\documentclass[structabstract]{aa}  

\usepackage{graphicx}
\usepackage{txfonts}
\usepackage{natbib}
\usepackage{color}
\definecolor{red}{rgb}{1.0, 0.0, 0.0}
  \def\cT2{c_T^2}

\newcommand{\overlim}[1]{{\buildrel{#1}\over\longrightarrow\;}}

\begin{document}

   \title{Consistent dust and gas models for protoplanetary disks}
   \subtitle{II. Chemical networks and rates}

   \author{I. Kamp\inst{1}
   \and
   W.-F. Thi\inst{2}
   \and
   P. Woitke\inst{3}
   \and
   C. Rab\inst{4}
   \and
   S. Bouma\inst{1}
   \and
   F. M\'{e}nard\inst{5}
       }

         \institute{
             Kapteyn Astronomical Institute, University of Groningen, Postbus 800,
             9700 AV Groningen, The Netherlands
             \and
             Max Planck Institute for Extraterrestrial Physics, Giessenbachstrasse, 85741 Garching, Germany                     
             \and
             SUPA, School of Physics \& Astronomy, University of St. Andrews, North Haugh, St. Andrews, KY16 9SS, UK
             \and
             University of Vienna, Department for Astrophysics, T\"{u}rkenschanzstr.17, 1180 Vienna, Austria
             \and
             University of Grenoble Alpes, CNRS, IPAG, F-38000 Grenoble, France
 }

   \date{Received ; accepted }

 
\abstract
   {}
    {We define a small and large chemical network which can be used for the quantitative simultaneous analysis of molecular emission from the near-IR to the submm. We revise reactions of excited molecular hydrogen, which are not included in UMIST, to provide a homogeneous database for future applications.}
    {We use the thermo-chemical disk modeling code ProDiMo and a standard T Tauri disk model to evaluate the impact of various chemical networks, reaction rate databases and sets of adsorption energies on a large sample of chemical species and emerging line fluxes from the near-IR to the submm wavelength range.}
    {We find large differences in the masses and radial distribution of ice reservoirs when considering freeze-out on bare or polar ice coated grains. Most strongly the ammonia ice mass and the location of the snow line (water) change. As a consequence molecules associated to the ice lines such as N$_2$H$^+$ change their emitting region; none of the line fluxes in the sample considered here changes by more than 25\% except CO isotopologues, CN and N$_2$H$^+$ lines. The three-body reaction N+H$_2$+M plays a key role in the formation of water in the outer disk. Besides that, differences between the UMIST 2006 and 2012 database change line fluxes in the sample considered here by less than a factor 2 (a subset of low excitation CO and fine structure lines stays even within 25\%); exceptions are OH, CN, HCN, HCO$^+$ and N$_2$H$^+$ lines.  However, different networks such as OSU and KIDA 2011 lead to pronounced differences in the chemistry inside 100~au and thus affect emission lines from high excitation CO, OH and CN lines. H$_2$ is easily excited at the disk surface and state-to-state reactions enhance the abundance of CH$^+$ and to a lesser extent HCO$^+$. For sub-mm lines of HCN, N$_2$H$^+$ and HCO$^+$, a more complex larger network is recommended.}
    {More work is required to consolidate data on key reactions leading to the formation of water, molecular ions such as HCO$^+$ and N$_2$H$^+$ as well as the nitrogen chemistry. This affects many of the key lines used in the interpretation of disk observations. Differential analysis of various disk models using the same chemical input data will be more robust than the interpretation of absolute fluxes.}


   \maketitle

\section{Introduction}

Observations often detect a multitude of simple molecules which have bright emission lines in protoplanetary disks such as CO, HCO$^+$, HCN, CN, N$_2$H$^+$, H$_2$CO, CH$^+$ \citep[e.g.][]{Thi2004,Dutrey2007,Oeberg2010,Thi2011b,Qi2013}. Even though studying differences in the molecular content of disks is interesting in its own right, molecules are frequently used as tracers of disk properties, such as outer gas radius \citep[e.g.][]{Panic2009}, the position of the CO ice line \citep[e.g.][]{Qi2013b}, the ionization degree \citep[e.g.][]{Qi2003}, the irradiation by X-rays \citep[e.g.][]{Aikawa2001} and the deuterium fractionation \citep[e.g.][]{Ceccarelli2005b}.

Due to the importance of molecular lines for protoplanetary disk research, several studies have focussed on the size of chemical networks and the uncertainties in chemical rates. \citet{Semenov2004} find that the midplane and the ionized surface layer can be described using very small networks, while the intermediate layer, where most of the ion-molecule chemistry happens, requires large networks with of the order of 100 species. \citet{Vasyunin2004} varied the rate constants within the uncertainties using a Monte Carlo approach and conditions representative of diffuse and dark clouds. In dark clouds, they find abundance changes of less than 0.5~dex for simple species such as N$_2$H$^+$, HCO$^+$, while HCN can change up to 1~dex. Interestingly, CO is  among the most robust species. \citet{Vasyunin2008} expanded this study to protoplanetary disk conditions. Again, CO is the most robust species while HCN, N$_2$H$^+$, and HCO$^+$ column densities can typically vary by a factor 2.5-3. Local changes can, however, be larger than this.

Thermo-chemical disk models often use a single network throughout the entire disk. When comparing such models to a set of observational data including molecular emission lines, we often rely in a first step on simple molecules such as CO and HCN. For those molecules, the chemistry is simple, resulting in robust model predictions. The chemical network used in this step should be small and fast to solve in order to allow the computation of larger model grids or the use of evolutionary search strategies to derive basic properties of the disk such as the dust mass, the mass of CO gas, radial extent of the disk, and the amount of flaring. In a second step, based on the previously found disk model, species with a more complex chemistry can be studied using larger chemical networks. It is important however to note that the chemical network is only one aspect in the interpretation of line observations. The other one, which is not addressed in this work, is the calculation of the excitation of the molecule which can be limited by the availability of molecular data (mainly collision cross sections), the complexity of IR and UV pumping or resonance scattering for optically thick lines \citep[e.g.][]{Cernicharo2009,Bethell2011,Kamp2013,Thi2013}.

We develop in this paper simple rules for the construction of chemical networks that avoid artificial sinks and ensure links between various sub-networks such as the carbon, oxygen and nitrogen chemistry. We use a standard T Tauri disk model from \citet{woitke2016} to study the impact of the size of the network, different chemical databases and ice adsorption scenarios on the species mass and emission from a representative sample of atomic and molecular lines (Sect.~\ref{Sect:models}). \citet{woitke2016} show that time-dependent chemistry has little effect on the resulting line fluxes at ages beyond $0.5$~Myr. Hence, we focus here entirely on stationary chemistry. In Section~\ref{Sect:results}, we discuss how changes in the UMIST database and the use of other databases affect the disk chemistry and line emission (Sect.~\ref{Sect:resultsUMISTintime}, \ref{Sect:resultsDatabases}). We investigate the role of three-body reactions for water chemistry (Sect.~\ref{Sect:resultsDatabases}) and how the composition of the grains (bare or polar ices) affects the various ices reservoirs and emission lines connected to them (Sect.~\ref{Sect:resultsEads}). We study the importance of reactions with excited molecular hydrogen (Sect.~\ref{Sect:resultsH2EXC}) and we end with assessing which emission lines require the use of larger chemical networks (Sect.~\ref{Sect:resultsSizenetwork}).

\section{The modeling}
\label{Sect:models}

\subsection{The disk model}
\label{Sect:diskmodel}

We choose for this study a parametrized disk structure representative of a typical T Tauri star. The full model is described in \citet{woitke2016}. 
Table~\ref{tab:diskparameters} repeats only the most important stellar and disk parameters.
 
\begin{table}[h]
\caption{Basic model parameters for the standard T Tauri disk.}
\begin{tabular}{lll}
\hline
\hline
 Quantity       & Symbol & Value \\
\hline
Stellar mass & $M_\ast$ & 0.7~M$_\odot$\\
Effective temperature & $T_{\rm eff}$ & 4000~K\\
Stellar luminosity & $L_\ast$ & 1.0~L$_\odot$\\
FUV excess & $f_{\rm UV}$ & 0.01\\
                       & $p_{\rm UV}$ & 1.3\\
Cosmic ray ionization rate & $\zeta_{\rm CR}$ & $1.7\,10^{-17}$~s$^{-1}$ \\
Chemical heating efficiency & $\gamma^{\rm chem}$ & 0.2 \\
\hline
Disk gas mass$^1$ & $M_{\rm gas}$ & 0.033~M$_\odot$\\
dust-to-gas mass ratio & $\delta$ & 0.01 \\
Inner disk radius & $R_{\rm in}$ & 0.07~au\\
Outer disk radius$^2$ & $R_{\rm out}$ & 700~au\\
Tapered edge radius & $R_{\rm taper}$ & 100~au\\
Radial column density power index & $\epsilon$ & 1.0\\
Reference radius & $R_0$ & 100~au \\
Scale height at $R_{0}$ & $H_0$ & 10.0~au \\
Disk flaring power index & $\beta$ & 1.15 \\
\hline
Minimum dust particle radius & $a_{\rm min}$ & 0.05 $\mu$m\\
Maximum dust particle radius & $a_{\rm max}$ & 3000.0 $\mu$m \\
Dust size distribution power index & $a_{\rm pow}$ & 3.5\\
\hline
\end{tabular}
\label{tab:diskparameters}
\tablefoot{(1) The disk mass is a factor 3.3 higher than in the original \citet{woitke2016} model. (2) The outer radius is defined as the radius where the surface density column drops to $N_{\rm \langle H\rangle, ver}=10^{20}$~cm$^{-2}$.} 
\end{table}

To calculate the two dimensional physical and chemical structure, we use the radiation thermo-chemical disk code ProDiMo \citep{Woitke2009,Kamp2010,Thi2011a}. The disk structure is set up using a tapered edge and mildly flaring geometry ($\beta\!=\!1.15$). It extends from 0.07 to 700~au (characteristic radius at 100~au) and contains a gas mass of $0.033~M_\odot$. The dust grain opacities are calculated using hollow spheres and a mixture of 60\% silicates and 15\% amorphous carbon with 25\% vacuum \citep{Min2016a} and we use the canonical dust-to-gas mass ratio of 0.01.

We use here a model series where we vary the base set of reaction rates, the set of adsorption energies, and the size of the chemical network using stationary chemistry. The reaction rate databases are UMIST2012 \citep{McElroy2013b}, UMIST2006 \citep{UMIST2006}, OSU (Ohio State University chemical network from Eric Herbst)
and KIDA2011 \citep{Wakelam2012}. Three sets of adsorption energies are taken from \citet{Aikawa1996}, \citet{Garrod2006}, and UMIST2012 \citep{McElroy2013b}. The adsorption energies will be discussed in more detail in Sect.~\ref{Sect:resultsEads}. The rules for compiling the small and large chemical network are provided in the next subsection. Table~\ref{tab:modelseries} describes the entire model series.

Since the disk chemistry and heating/cooling balance are intimately coupled, we fixed the gas temperature structure to that of the reference model 1 \citep[UMIST2012, adsorption energies from][]{Aikawa1996}. This allows us to interpret changes in the chemical structure and emitted line fluxes entirely from the various chemical input data sets. The coupling of heating/cooling and chemical equations is highly non-linear, so the impact of our approximation can only be checked from additional models. We calculated a single additional model where we used the KIDA database and re-computed the gas temperature self-consistently. Our models --- discussed in more detail in the next section --- show that KIDA produces less water compared to UMIST in the warm surface layer stretching to a few au at gas temperatures higher than 200~K. Indeed, the additional model shows that the gas temperature in that layer increases slightly since the water cooling is diminished with respect to the UMIST reference model. However, the effects are very subtle, if the overall gas temperature distribution is considered.

\begin{table}
\caption{Series of disk models: the columns denote the set of adsorption energies $E_{\rm ads}$, chemical network and rate database used.}
\begin{tabular}{l|lll}
\hline
model & $E_{\rm ads}$ &network size, mode & base rates \\
\hline
\hline\\[-2mm]
model 1 & Aikawa & small, steady state & UMIST2012 \\
 & & & $+$ CL reactions \\
model 2 & Aikawa & small, steady state & UMIST2006 \\
model 3 & Aikawa & small, steady state & OSU \\
model 4 & Aikawa & small, steady state & KIDA2011 \\[1mm]
\hline\\[-2mm]
model 1a & Aikawa & small, steady state & UMIST2012 \\[1mm]
\hline\\[-2mm]
model 5 & GH06 & small, steady state & UMIST2012 \\
 & & & $+$ CL reactions \\
model 6 & UMIST2012 & small, steady state & UMIST2012 \\
 & & & $+$ CL reactions \\
model 7 & $T$-dependent & small, steady state & UMIST2012 \\
 & & & $+$ CL reactions \\[1mm]
\hline\\[-2mm]
model 8 & UMIST2012 & large, steady state & UMIST2012 \\
 & & & $+$ CL reactions \\
 \hline
\end{tabular}
\label{tab:modelseries}
\end{table}

\subsection{The chemical network}
\label{Sect:chemicalnetwork}

We follow two approaches here: (1) provide a robust and fast standard that enables to deal with simple species (robust tracers) such as C, O, Ne, CO, CN, CO$_2$, OH, and H$_2$O, (2) provide a consistent standard that can be used as a starting point for further investigation of the chemistry of more complex species such as HCN, H$_2$C$_2$, HCO$^+$, N$_2$H$^+$. So far, in the literature no chemistry standard has been defined for disks in the context of multi-wavelength fitting of observational data. Instead many different species lists are used and we only start to understand the impact of the species choices and/or presence of ices and electron sinks for the abundance of specific low abundance tracers such as HCO$^+$ ($ \epsilon\!\sim\!10^{-10}$) \citep[][Rab et al.\ in preparation]{Kamp2013}. 

\begin{table*}[!htbp]
\caption{Selection of elements and chemical (gas$+$ice) species in the small network.}
\label{tab:standard-species}
\vspace*{1mm} 
\begin{tabular}{|c|p{11cm}|r|}
\hline
12 elements & H, He, C, N, O, Ne, Na, Mg, Si, S, Ar, Fe & \\
\hline
(H)         & H, H$^+$, H$^-$, {\bf H$_2$}, H$_2^+$, H$_3^+$, H$_2^{\rm exc}$          &  7 \\
(He)        & He, He$^+$,                                                           &  2 \\
(C-H)       & C, C$^+$, C$^{++}$, CH, CH$^+$, {\bf CH$_2$}, CH$_2^+$, 
              CH$_3$, CH$_3^+$, {\bf CH$_4$}, CH$_4^+$, CH$_5^+$,                     & 12 \\
(C-N)       & CN, CN$^+$, {\bf HCN}, HCN$^+$, HCNH$^+$                               & 5 \\  
(C-O)       & {\bf CO}, CO$^+$, HCO, HCO$^+$, {\bf CO$_2$}, CO$_2^+$, HCO$_2^+$,         & 7 \\
(N-H)       & N, N$^+$, N$^{++}$, NH, NH$^+$, NH$_2$, NH$_2^+$, {\bf NH$_3$}, NH$_3^+$, NH$_4^+$   & 9 \\
(N-N)       & {\bf N$_2$}, N$_2^+$, HN$_2^+$,                                                   & 3 \\
(N-O)       & NO, NO$^+$,                                                             & 2 \\  
(O-H)       & O, O$^+$, O$^{++}$, OH, OH$^+$, {\bf H$_2$O}, H$_2$O$^+$, H$_3$O$^+$,     & 8 \\
(O-O)       & {\bf O$_2$}, O$_2^+$,                                                  & 2 \\
(O-S)       & SO, SO$^+$, {\bf SO$_2$}, SO$_2^+$, HSO$_2^+$                          & 5 \\
(S-H)       & S, S$^+$, S$^{++}$,                                                  & 3 \\
(Si-H)      & Si, Si$^+$, Si$^{++}$, SiH, SiH$^+$, SiH$_2^+$,                     & 6 \\    
(Si-O)      & {\bf SiO}, SiO$^+$, SiOH$^+$,                                    & 3 \\    
(Na)        & Na, Na$^+$, Na$^{++}$,                                           &  3 \\
(Mg)        & Mg, Mg$^+$, Mg$^{++}$,                                           &  3 \\
(Fe)        & Fe, Fe$^+$, Fe$^{++}$,                                           &  3 \\
(Ne)        & Ne, Ne$^+$, Ne$^{++}$,                                           &  3 \\
(Ar)        & Ar, Ar$^+$, Ar$^{++}$,                                           &  3 \\
ice         & CO\#, H$_2$O\#, CO$_2$\#, CH$_4$\#, NH$_3$\#, SiO\#, SO$_2$\#, O$_2$\#, HCN\#, N$_2$\#  & 10 \\
\hline
 species    & total                                                           & 100 \\
\hline
\end{tabular}
\tablefoot{{\bf Neutral more stable} molecules are indicated in bold font and ices are indicated by a trailing \#.}
\end{table*}

\begin{table*}[!htbp]
\caption{Selection of elements and chemical (gas$+$ice) species in the large network.}
\label{tab:species}
\vspace*{1mm} 
\begin{tabular}{|c|p{11cm}|r|}
\hline
13 elements & H, He, C, N, O, Ne, Na, Mg, Si, S, Ar, Fe, PAH & \\
\hline
(H)         & H, H$^+$, H$^-$, {\bf H$_2$}, H$_2^+$, H$_3^+$, H$_2^{\rm exc}$          &  7 \\
(He)        & He, He$^+$,                                                     &  2 \\
(He-H)      & HeH$^+$,                                                       &  1 \\
(C-H)       & C, C$^+$, C$^{++}$, CH, CH$^+$, CH$_2$, CH$_2^+$, 
              CH$_3$, CH$_3^+$, {\bf CH$_4$}, CH$_4^+$, CH$_5^+$,                     & 12 \\
(C-C)       & {\bf C$_2$}, C$_2^+$, C$_2$H, C$_2$H$^+$, {\bf C$_2$H$_2$}, C$_2$H$_2^+$, 
              C$_2$H$_3$, C$_2$H$_3^+$, {\bf C$_2$H$_4$}, C$_2$H$_4^+$, C$_2$H$_5$, 
              C$_2$H$_5^+$, & \\
            & {\bf C$_3$}, C$_3^+$, C$_3$H, C$_3$H$^+$, {\bf C$_3$H$_2$},  
              C$_3$H$_2^+$, C$_3$H$_3^+$,  & \\
            & {\bf C$_4$}, C$_4^+$, C$_4$H$^+$, & 23 \\
(C-N)       & CN, CN$^+$, {\bf HCN}, HCN$^+$, HCNH$^+$, HNC, H$_2$CN, OCN, OCN$^+$, & 9 \\  
(C-O)       & {\bf CO}, CO$^+$, HCO, HCO$^+$, & \\
            & {\bf CO$_2$}, CO$_2^+$, HCO$_2^+$, C$_2$O, C$_2$O$^+$, HC$_2$O$^+$, & \\
            & {\bf H$_2$CO}, H$_2$CO$^+$, CH$_3$O, H$_3$CO$^+$, CH$_2$OH, & \\
            & {\bf CH$_3$OH}, CH$_3$OH$^+$, CH$_3$OH$_2^+$,                            & 18 \\
(C-S)       & {\bf CS}, CS$^+$, HCS, HCS$^+$, {\bf H$_2$CS}, H$_2$CS$^+$, H$_3$CS$^+$, & \\
            & {\bf OCS}, OCS$^+$, HOCS$^+$,        & 10 \\
(N-H)       & N, N$^+$, N$^{++}$, NH, NH$^+$, NH$_2$, NH$_2^+$, {\bf NH$_3$}, 
              NH$_3^+$, NH$_4^+$,                                              & 10 \\
(N-N)       & {\bf N$_2$}, N$_2^+$, HN$_2^+$,                                        & 3 \\
(N-O)       & NO, NO$^+$, {\bf NO$_2$}, NO$_2^+$, {\bf HNO}, HNO$^+$, H$_2$NO$^+$,  & 7 \\  
(N-S)       & NS, NS$^+$, HNS$^+$                                              & 3 \\  
(O-H)       & O, O$^+$, O$^{++}$, OH, OH$^+$, {\bf H$_2$O}, H$_2$O$^+$, H$_3$O$^+$,    & 8 \\
(O-O)       & {\bf O$_2$}, O$_2^+$, O$_2$H$^+$,                                              & 3 \\
(O-S)       & SO, SO$^+$, {\bf SO$_2$}, SO$_2^+$, HSO$_2^+$,                           & 5 \\
(S-H)       & S, S$^+$, S$^{++}$, HS, HS$^+$, {\bf H$_2$S}, H$_2$S$^+$, H$_3$S$^+$,    & 8 \\
(Si-H)      & Si, Si$^+$, Si$^{++}$, SiH, SiH$^+$, SiH$_2$, SiH$_2^+$, SiH$_3$, 
              SiH$_3^+$, {\bf SiH$_4$}, SiH$_4^+$,  SiH$_5^+$,                        & 12 \\    
(Si-C)      & {\bf SiC}, SiC$^+$, HCSi$^+$,                                               & 3 \\    
(Si-N)      & {\bf SiN}, SiN$^+$, HNSi$^+$,                                               & 3 \\    
(Si-O)      & {\bf SiO}, SiO$^+$, SiOH$^+$,                                               & 3 \\    
(Si-S)      & {\bf SiS}, SiS$^+$, HSiS$^+$,                                               & 3 \\    
(Na)        & Na, Na$^+$, Na$^{++}$,                                           &  3 \\
(Mg)        & Mg, Mg$^+$, Mg$^{++}$,                                           &  3 \\
(Fe)        & Fe, Fe$^+$, Fe$^{++}$,                                           &  3 \\
(Ne)        & Ne, Ne$^+$, Ne$^{++}$,                                           &  3 \\
(Ar)        & Ar, Ar$^+$, Ar$^{++}$,                                           &  3 \\
(PAH)       & PAH-, PAH, PAH$^+$, PAH$^{++}$, PAH$^{+++}$,                      &  5 \\
ice         & all neutral gas species except noble gases have ice counterparts         & 64 \\
\hline
 species    & total                                                           & 235 \\
\hline
\end{tabular}
\tablefoot{Neutral more stable molecules are indicated in bold font.}
\end{table*}

Table~\ref{tab:standard-species} shows the selection used for many years in the disk modeling for the Herschel open time key program GASPS (Gas Evolution in Planetary Systems, PI: B.\ Dent); we subsequently refer to this as the small network. The abundances of the robust tracers listed above should be calculated with sufficient accuracy and this will be tested in Sect.~\ref{Sect:resultsSizenetwork}.

Table~\ref{tab:species} details the selection of chemical species in the large network. We cover the most important C/N/O chemistry and a simple S and Si chemistry. Other elements (Na, Mg, Fe, Ne, Ar) are represented only by their atoms and ions.  Detailed PAH charging is used, as well as a large selection of ice species. The selection of species is largely based on the chemical networks of Prasad \& Huntress (1980)\nocite{Prasad1980a} (C-H, C-C chemistry), van Dishoeck (1990)\nocite{vanDishoeck1990} (C-H chemistry) Sternberg \& Dalgarno (1995)\nocite{Sternberg1995} (Si-chemistry, S-H chemistry, N-Chemistry, O-H chemistry), Agundez et al.\ (2008)\nocite{Agundez2008} (high temperature C-H, C-C chemistry), Hily-Blant et al.\ (2013)\nocite{Hily-Blant2013} (N-chemistry). The size of our network is controlled by a combination of species becoming less reactive or saturated. We apply the following rules to ensure the completeness of the chemical network used:
\begin{itemize}
\item Negative ions/molecules have been omitted for the time being except H$^-$.
\item We include for all atoms/molecules the positively charged counterpart (for elements also double charged). In some cases (HeH, HNS, HSO, CH$_3$O), the neutral one is missing since it is not present/has no reactions in UMIST (e.g.\ unstable molecule or other reasons).
\item C-H chemistry processes via H$_2$ addition reactions up to CH$_5^+$, which is the maximum hydrogenation possible. CH$_5^+$ can then recombine dissociatively to give the closed-shell molecule CH$_4$. We proceed similarly for Si-H chemistry, thus stopping at SiH$_5^+$, and for O-H chemistry, thus stopping at H$_3$O$^+$.
\item We identify the neutral more stable species to be H$_2$, CH$_2$, CH$_4$, C$_2$, C$_3$, C$_4$, C$_2$H$_2$ (acetylene), C$_2$H$_4$ (ethylene), C$_3$H$_2$ (cyclopropenylidene), HCN (hydrogen cyanide), CO, CO$_2$, H$_2$CO (formaldehyde), CH$_3$OH (methanol), CS (carbon monosulfide), H$_2$CS (thioformaldehyde), NH$_3$ (ammonia), NO$_2$, HNO (nitroxyl), N$_2$, H$_2$O, SO$_2$, H$_2$S (hydrogen sulfide), OCS (carbonyl sulfide), O$_2$, SiH$_4$ (silane), SiC (silicon carbide), SiN (silicon nitride), SiO (silicon monoxide), SiS (silicon sulfide). For those molecules, we ensure that the respective positive ion and the protonated ion are included. The exception is HNO$_2^+$, which is not included in UMIST (HNO$_2$ is included as species in UMIST, but has no reactions).
\item We decided to keep the isomers CH$_3$O and CH$_2$OH to study the gas phase formation of methanol. We also keep the isomers HNC and HCN since they are both observed. However, we only include the ion HCN$^+$ and subsequent hydrogenation.
\item We included several species that link the chemical networks with each other, especially for the heavier elements such as S, N, and Si. An interesting example is the radical H$_2$CN (amidogen). It is formed by collisions between N and C-H chains and forms a CN bond. This connects the C-H, C-C chemistry with the nitrogen chemistry.
\item Neutral atoms/molecules (including radicals) except noble gases can freeze out.
\end{itemize}

\subsection{Reaction rates}
\label{Sect:Databases}

ProDiMo selects from the UMIST2012 database all reactions among the species defined in the chemical network above. However, in some cases, we add additional reactions and/or overwrite UMIST reactions following the procedures described in Appendix~\ref{App:diffUMIST}-\ref{App:X-rays}. Alternatively, we also use the UMIST2006, the OSU and the KIDA 2011 databases.

\subsection{Element abundances}

Table~\ref{Tab:elements} describes the selection of elements and their respective abundances. These are very similar to the low-metal abundances used in the literature e.g.\ \citet{Lee1998}. All following models adopt these low metal abundances.

\begin{table}
\caption{Elements, their abundances on the scale $\log n_{\rm H}=12$ and their masses in amu.}
\begin{tabular}{crr|crr}
\hline
element & \hspace{-1mm}$12+\log\epsilon$ & $m$ [amu] & element &  \hspace{-1mm}$12+\log\epsilon$  & $m$ [amu] \\
\hline
\hline
H   & 12.00  & 1.0079 & Na  & 3.36  & 22.990 \\
He & 10.98 & 4.0026 & Mg  & 4.03  & 24.305 \\
C  &  8.14 &  12.011 & Si  & 4.24 &  28.086 \\
N  &  7.90  & 14.007 & S  &  5.27 &  32.066 \\
O  &  8.48  & 15.999 & Ar &  6.08 &  39.948 \\
Ne  & 7.95 &  20.180 & Fe  & 3.24 &  55.845 \\
\hline
\end{tabular}
\label{Tab:elements}
\end{table}

\begin{table}
\caption{Lines used to analyse flux changes related to chemistry.}
\begin{tabular}{lllll}
\hline
species \hspace*{-3mm} & designation \hspace*{-4mm} & $E_{\rm up}$~[K] & $A_{ij}$~[s$^{-1}$] & $\lambda$ [$\mu$m] \\
\hline
\hline
        CO & J=2-1 & 16.60 & 6.910(-7) & 1300.40 \\
 $^{13}$CO & J=2-1 & 15.87 & 6.038(-7) &1360.23 \\  
C$^{18}$O & J=2-1 & 15.81 & 6.011(-7) & 1365.42 \\  
        CO & J=3-2  & 33.19 & 2.497(-6) & 866.96 \\ 
 $^{13}$CO & J=3-2 & 31.73 & 2.181(-6) & 906.84 \\
    C$^{18}$O & J=3-2 & 31.61 & 2.172(-6) & 910.31 \\  
        CO  & J=18-17 & 944.97 & 5.695(-4) & 144.78 \\
        CO   & J=36-35 & 3668.78 & 3.638(-3) & 72.84 \\  
        CO   & v=1-0 J=3-4 & 3116.70 & 1.950(1) & 4.699950 \\  
        CO   & v=1-0 J=35-36  & 6523.52 & 1.407(1) & 5.040484\\  
        CO   & v=2-1 J=3-4 & 6162.10 & 3.745(1) & 4.758863\\  
      OI   &$^3$P$_1$-$^3$P$_2$  & 227.712 & 8.91(-5) & 63.18 \\
      OI  & $^3$P$_0$-$^3$P$_1$ & 326.579 & 1.750(-5) & 145.53 \\  
      OI   & $^1$D$_2$-$^3$P$_2$ & 22830.18 \hspace*{-3mm} & 6.535(-3) & 0.63003 \\  
     CII  & $^2$P$_{3/2}$-$^2$P$_{1/2}$ & 91.21 & 2.300(-6) & 157.74 \\ 
      CI  & $^3$P$_1$-$^3$P$_0$ & 23.620 & 7.880(-8) & 609.14 \\  
      CI &  $^3$P$_2$-$^3$P$_1$ & 62.462 & 2.650(-7) & 370.42 \\  
    NeII   & $^2$P$_{1/2}$-$^3$P$_{3/2}$ & 1122 & 8.59(-3) & 12.815 \\ 
    NeIII   & $^3$P$_1$-$^3$P$_2$ & 924.98 & 5.97(-3) & 15.555 \\ 
    SII   & $^2$D$_{5/2}$-$^4$S$_{3/2}$ & 21420 & 3.338(-4)& 0.67164 \\ 
    SIII   & $^3$P$_2$-$^3$P$_1$ & 1199.904  \hspace*{-3mm}& 2.07(-3) & 18.716 \\ 
    ArII   & $^2$P$_{1/2}$-$^2$P$_{3/2}$ & 2059.72 & 5.3(-2) & 6.985 \\ 
    ArIII   & $^3$P$_0$-$^3$P$_1$ & 2259.2 & 5.19(-3) & 21.816 \\ 
    FeII   & $^6$D$_{9/2}$-$^6$D$_{7/2}$ & 553.6 & 2.13(-3) & 25.988 \\ 
    SiII   & $^2$P$_{1/2}$-$^2$P$_{3/2}$ & 413.21 & 2.132(-4) & 34.807 \\ 
        OH   & $^2\Pi_{1/2}$ 7/2$^+$-5/2$^-$ & 617.9 & 1.012 & 71.22 \\ 
        OH   & $^2\Pi_{1/2}$ 7/2$^-$-5/2$^+$ & 617.6 & 1.014 & 71.17 \\  
        OH   & $^2\Pi_{1/2}$ 1/2$^-$-$^2\Pi_{3/2}$ 3/2$^+$  \hspace*{-3mm}& 181.9 & 3.606(-2) & 79.11 \\ 
        OH   & $^2\Pi_{1/2}$ 1/2$^+$-$^2\Pi_{3/2}$ 3/2$^-$  \hspace*{-3mm}& 181.7 & 3.598(-2) & 79.18 \\  
        OH  & $^2\Pi_{3/2}$ 5/2$^-$-3/2$^+$ & 120.7 & 1.388(-1) & 119.23 \\  
       OH  & $^2\Pi_{3/2}$ 5/2$^+$-3/2$^-$ & 120.5 & 1.380(-1) & 119.44 \\  
     o-H$_2$O \hspace*{-3mm}& 1$_{10}$-1$_{01}$  & 61.0 & 3.458(-3) & 538.29 \\  
     o-H$_2$O \hspace*{-3mm} & 2$_{12}$-1$_{01}$ & 114.4 & 5.593(-2) & 179.53 \\  
     o-H$_2$O \hspace*{-3mm} & 4$_{23}$-3$_{12}$ & 432.2 & 4.838(-1) & 78.74 \\  
     o-H$_2$O  \hspace*{-3mm}&  8$_{18}$-7$_{07}$& 1070.7 & 1.751 & 63.32 \\  
     p-H$_2$O \hspace*{-3mm} & 1$_{11}$-0$_{00}$ & 53.4 & 1.842(-2) & 269.27 \\  
     p-H$_2$O \hspace*{-3mm} & 4$_{13}$-4$_{04}$ & 396.4 & 3.726(-2) & 187.110 \\  
     p-H$_2$O \hspace*{-3mm} & 3$_{22}$-2$_{11}$ & 296.8 & 3.524(-1) & 89.988 \\  
      o-H$_2$   & v=0-0 S(1) J=3-1 & 1015 & 4.76(-10) & 17.034 \\  
      p-H$_2$   & v=0-0 S(2) J=4-2 & 1682 & 2.754(-9) & 12.278 \\  
      p-H$_2$   & v=0-0 S(4) J=6-4 & 3474 & 2.642(-8) & 8.025 \\  
      o-H$_2$   & v=0-0 S(9) J=11-9 & 10262 & 4.898(-7) & 4.694 \\  
      o-H$_2$   & v=2-1 S(1) J=3-1 & 12550 & 4.977(-7) & 2.248 \\  
      p-H$_2$   & v=1-0 S(0) J=2-0 & 6472 & 2.526(-7) & 2.223 \\  
      o-H$_2$   & v=1-0 S(1) J=3-1 & 6952 & 3.471(-7) & 2.122 \\  
        CN & N=2-1 J=5/2-3/2 & 16.34 & 1.143(-4) & 1321.380 \\  
        CN  & N=5-4 J=11/2-9/2 & 81.64 & 2.027(-3) &  528.78 \\  
       HCN & J=3-2 & 25.52 & 8.3559(-4) &  1127.521 \\ 
       HCN & J=4-3 & 42.53 & 2.054(-3) &  845.66 \\ 
       CH$^+$  & J=2-1 & 120.195 & 4.760(-2) & 179.594 \\  
       CH$^+$  & J=4-3 & 400.086 & 0.3781 & 90.011 \\  
       CH$^+$  & J=5-4 & 599.524 & 0.7346 & 72.137 \\  
     HCO$^+$ & J=1-0 & 4.28 & 4.2512(-5) & 3361.334 \\  
      HCO$^+$ & J=3-2 & 25.68 & 1.4757(-3) & 1120.478 \\  
     HCO$^+$ & J=4-3 & 42.80 & 3.6269(-3) &  840.380 \\  
      N$_2$H$^+$ & J=3-2 & 26.83 & 1.2586(-3) & 1072.558 \\  
      \hline
\end{tabular}
\tablefoot{The notation $x(-y)$ stands for $x\,10^{-y}$.}
\label{Tab:linelist}
\end{table}

\subsection{The line list}
\label{Sect:linelist}

Table~\ref{Tab:linelist} describes the list of lines used to analyse how changes in disk chemistry propagate into observable line fluxes. The atomic and molecular data is collected from LAMDA \citep{Schoier2005}, NIST and CHIANTI \citep{Chianti1997}. Line fluxes are calculated using level populations from statistical equilibrium and a simplified 2D escape probability approach. Detailed radiative transfer tests show that line fluxes from escape probability are typically off by less than 50\% except for close to edge on disk geometries and/or lines where a significant fraction of total emission originates from the inner rim \citep[e.g.][]{Woitke2009b,Antonellini2015}. 

Details on collision cross sections and collision partners can be found in a series of papers: atoms, ions and H$_2$ \citep{Woitke2009}, CH$^+$ \citep{Thi2011b}, double-ionized species \citep{Aresu2012}, CO \citep{Thi2013}, H$_2$O \citep{Kamp2013}. The collision data for the remaining molecules is taken from the LAMDA database. The CN collision partners are He and e; the HCO$^+$ collision partner is H$_2$; the HCN collision partners are H$_2$, He and e; the OH collision partners are ortho- and para-H$_2$.

\section{Results}
\label{Sect:results}

\subsection{The base model}

The physical properties of the reference model are described in detail in \citet{woitke2016} and we summarize here a few key features relevant for the chemical studies. The model reaches total hydrogen number densities of $10^{14}\!-\!10^{16}$~cm$^3$ in the midplane inside 1~au. The dust temperature decreases from $\sim\!1500$~K at the inner rim to 100~K at $\sim\!1$~au. Gas and dust are thermally well coupled below $A_{\rm V}\!\sim\!1$ (towards the disk midplane). At the disk surface above $A_{\rm V}\!=\!1$, the gas temperature reaches values up to several 1000~K. Only in the outer disk atmosphere beyond 100~au and below $z/r\!=\!0.4$ (corresponding to $\sim\!20^{\circ}$ opening angle), the gas temperature drops below that of the dust.

\begin{figure*}[!htbp]
\includegraphics[width=5.9cm]{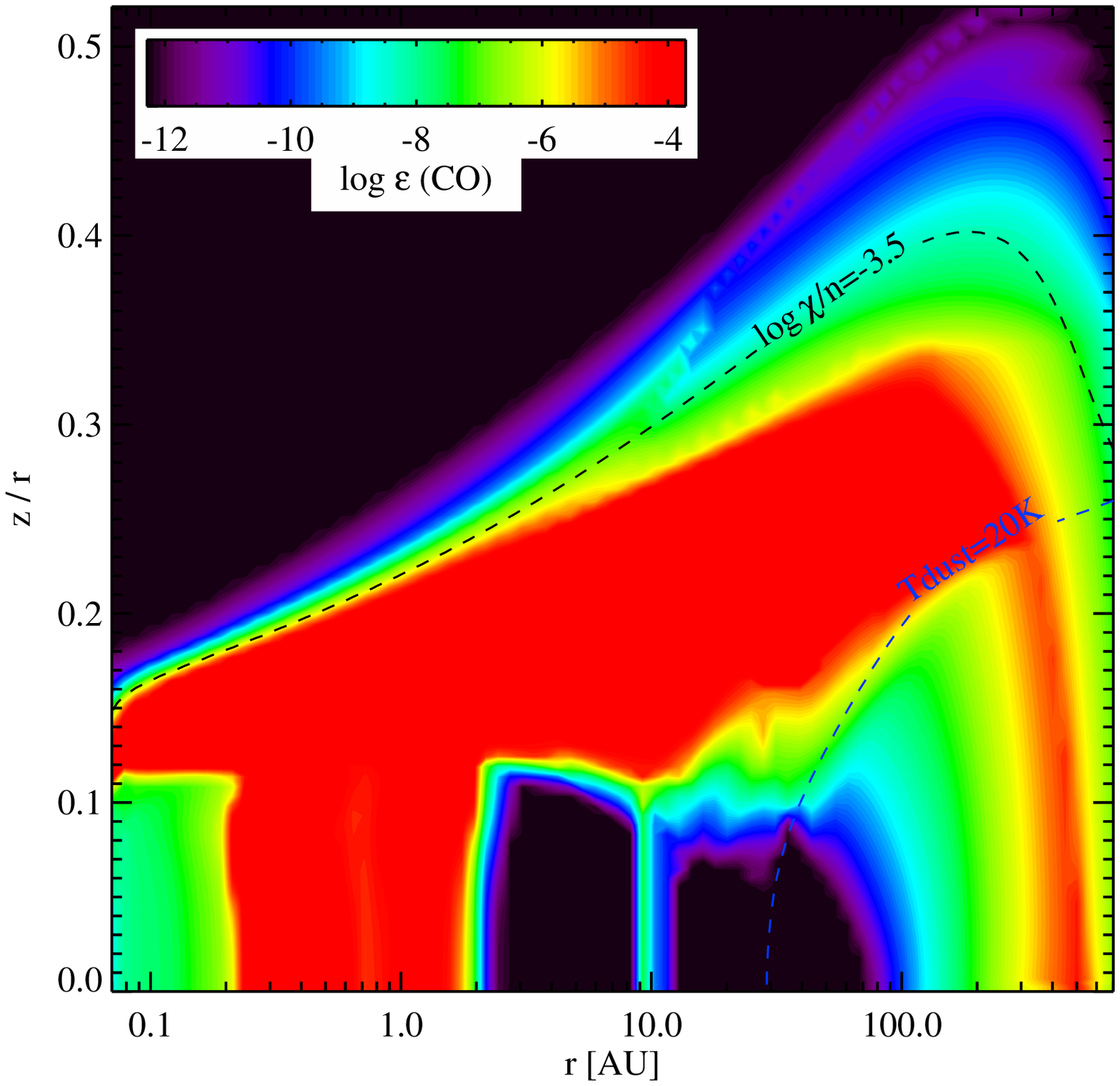}
\includegraphics[width=5.9cm]{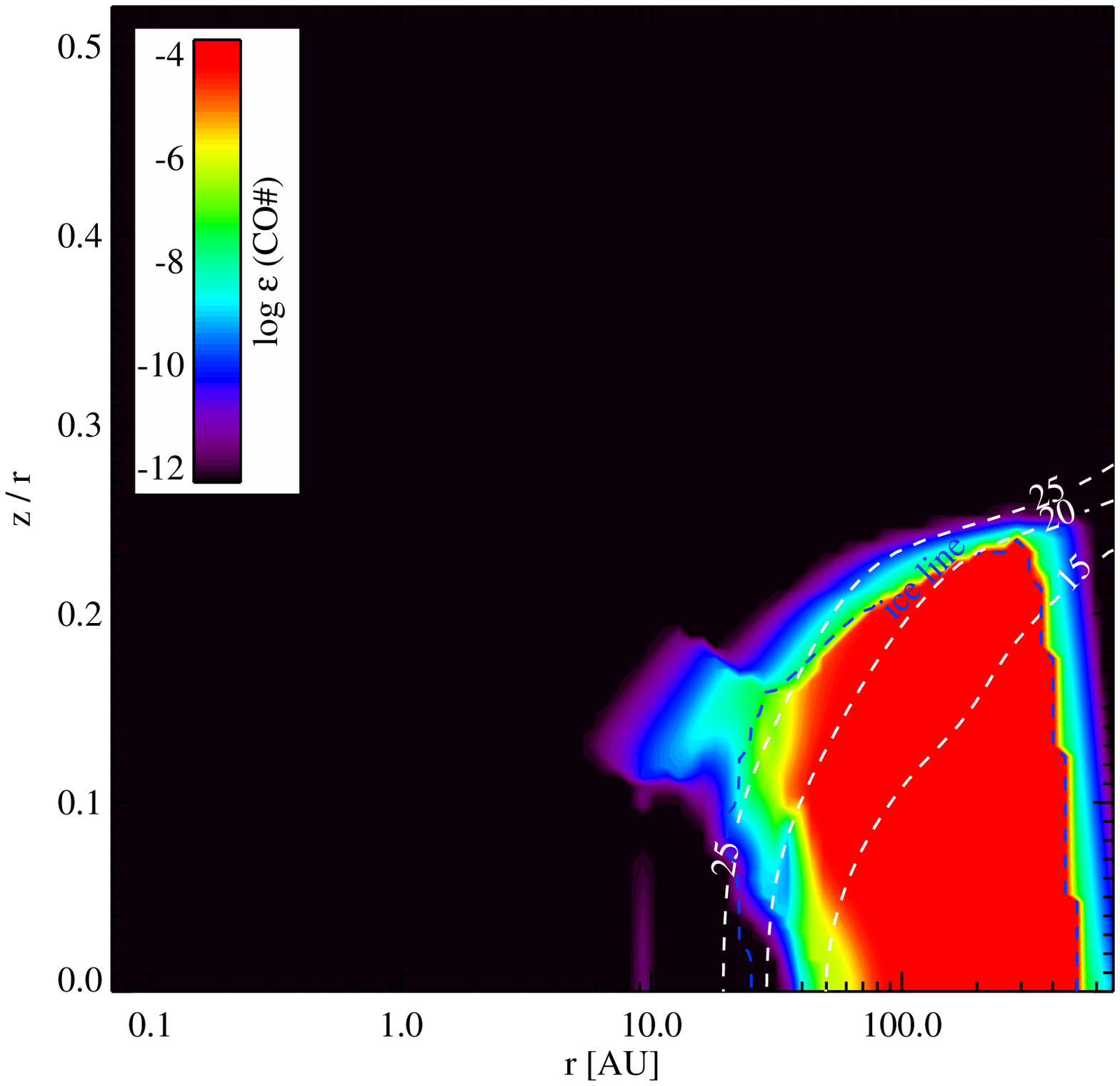}
\includegraphics[width=5.9cm]{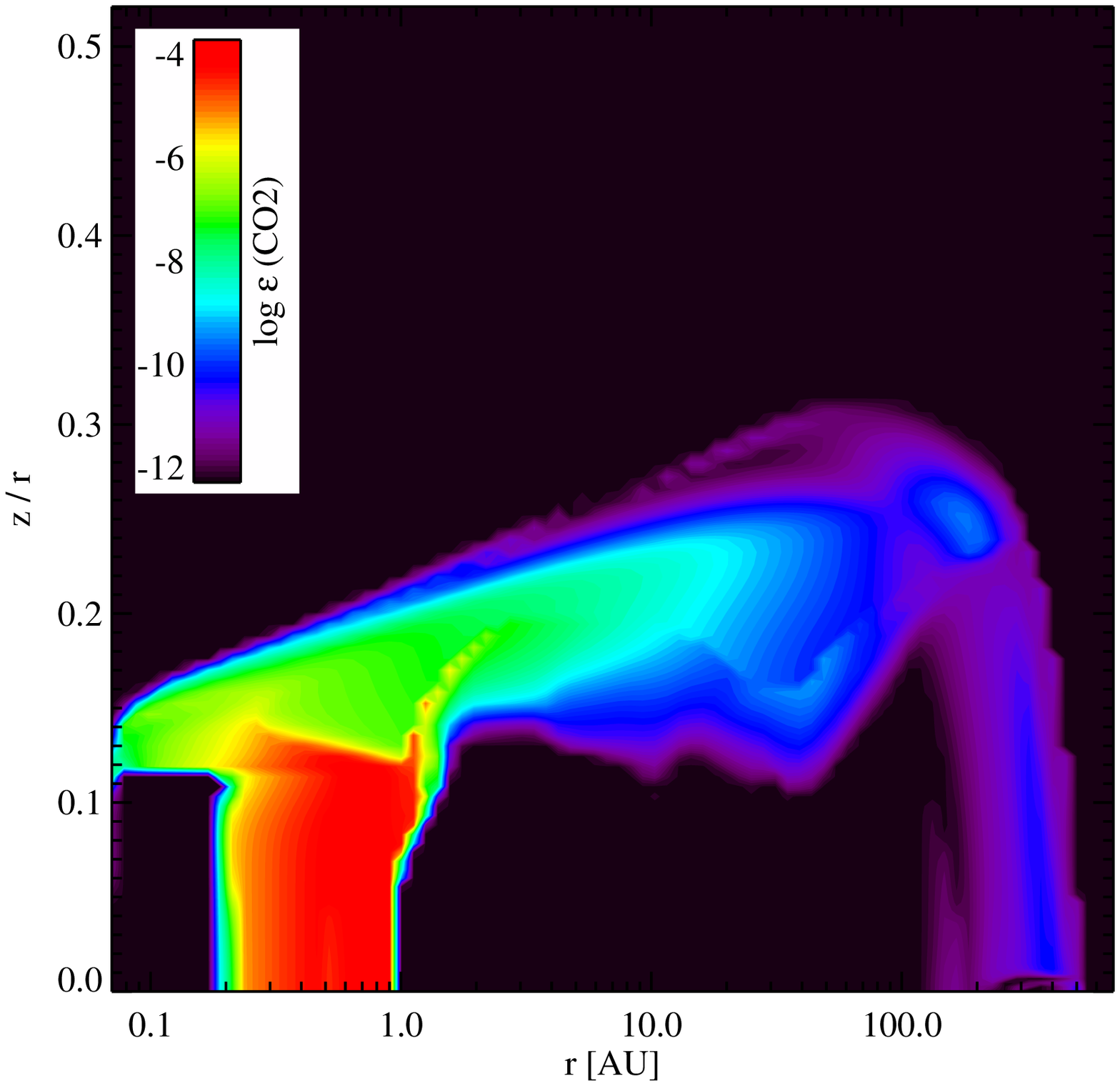}
\includegraphics[width=5.9cm]{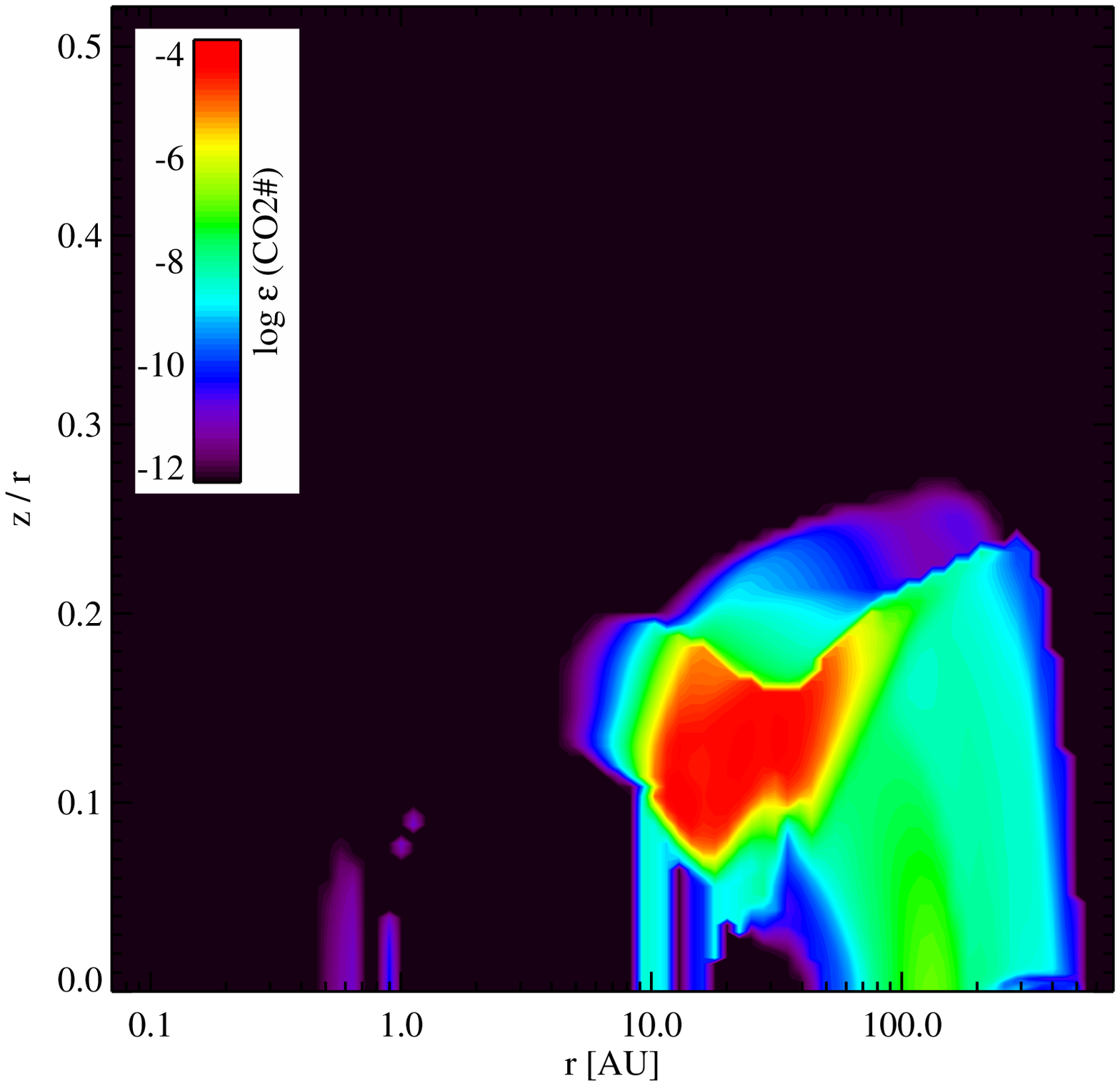}
\includegraphics[width=5.9cm]{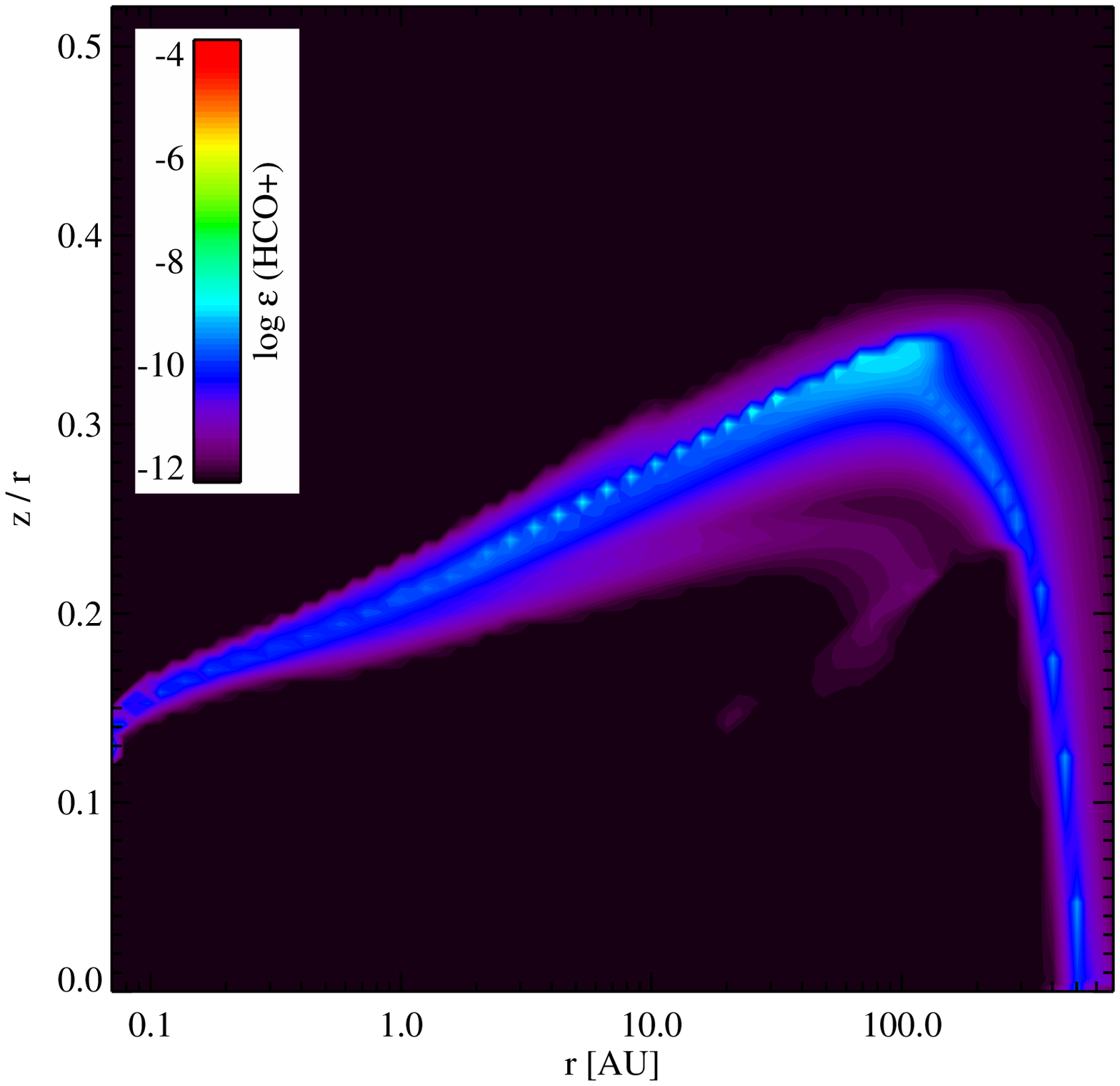}
\includegraphics[width=5.9cm]{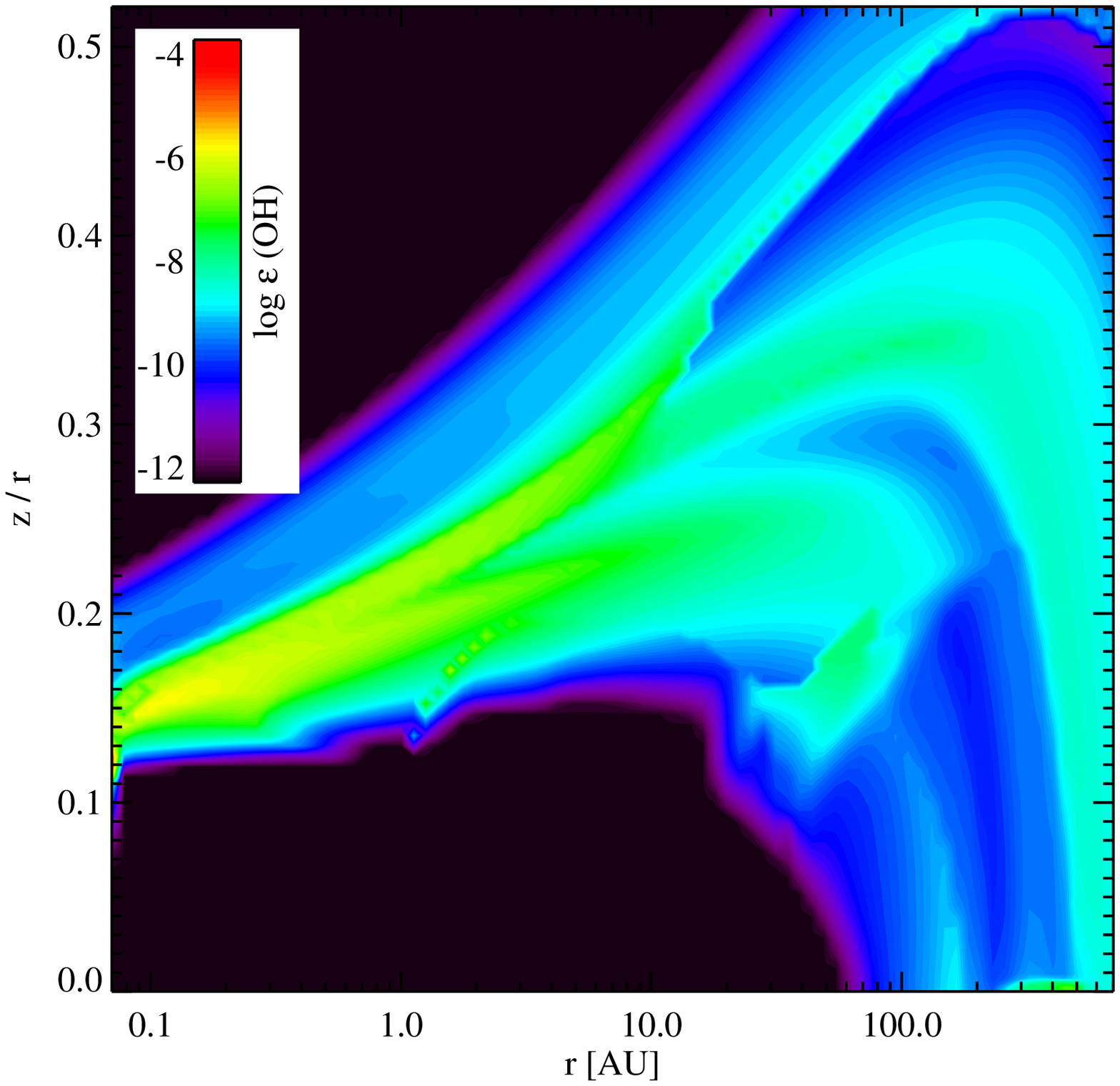}
\includegraphics[width=5.9cm]{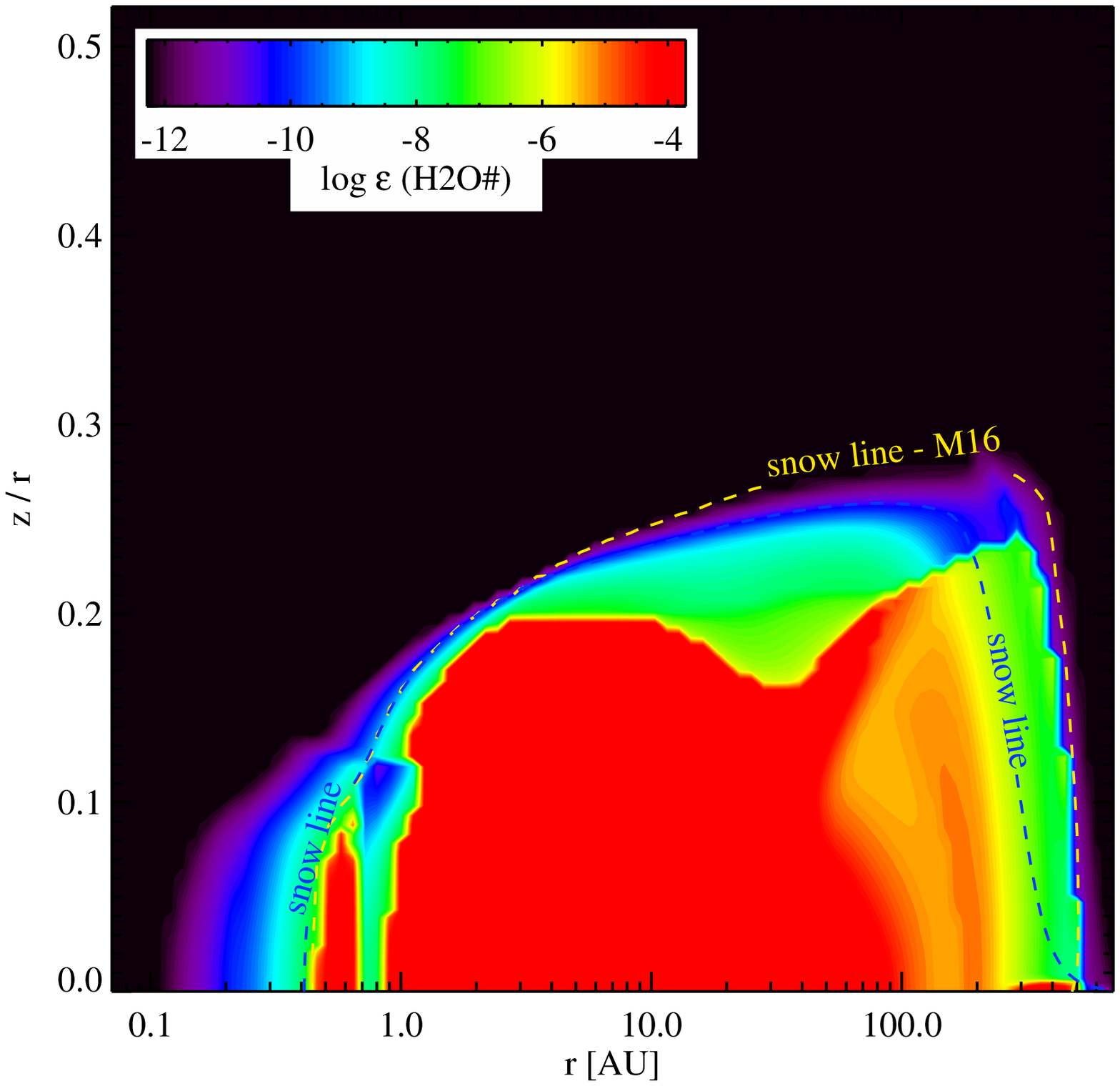}
\includegraphics[width=5.9cm]{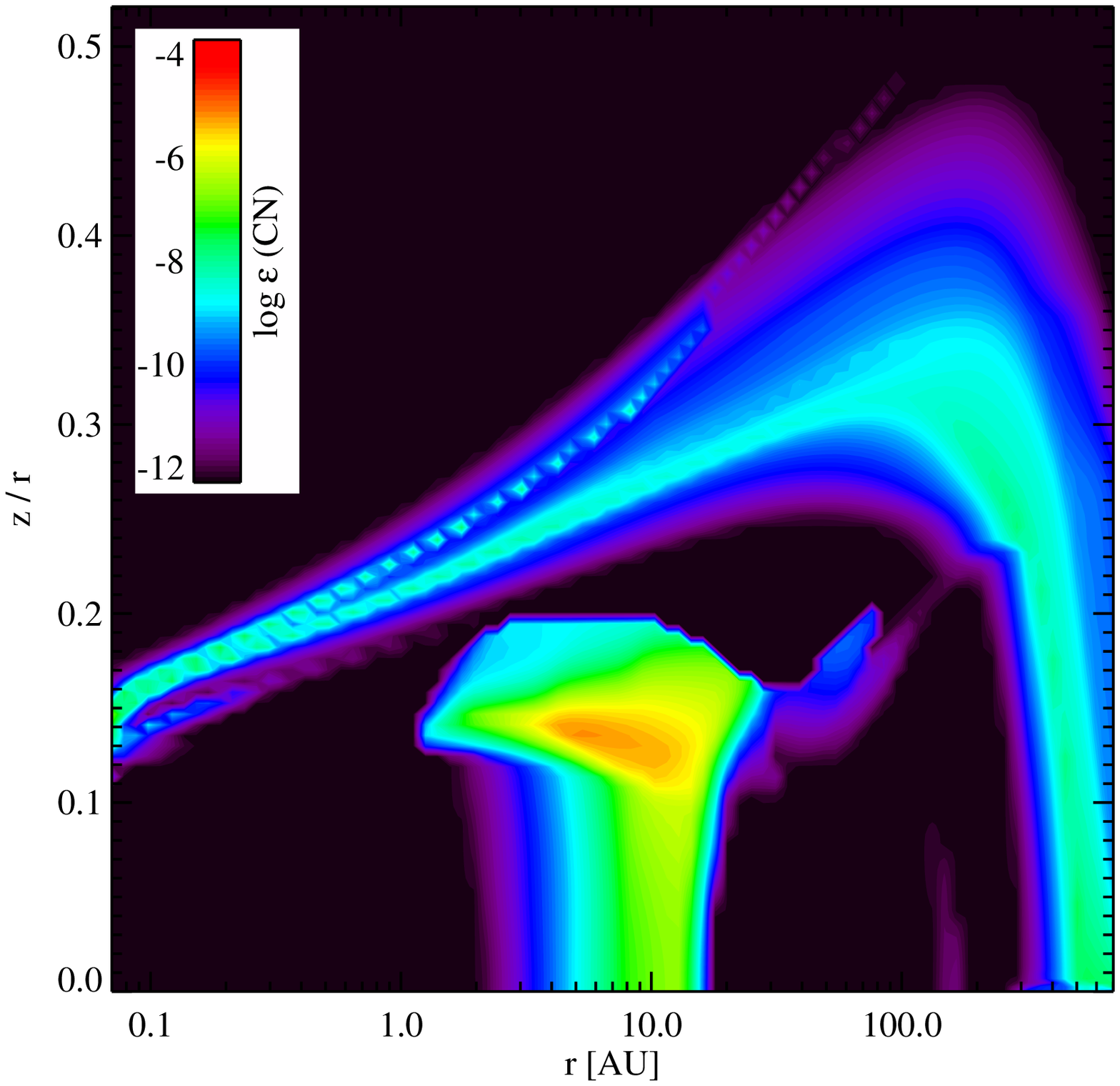}
\includegraphics[width=5.9cm]{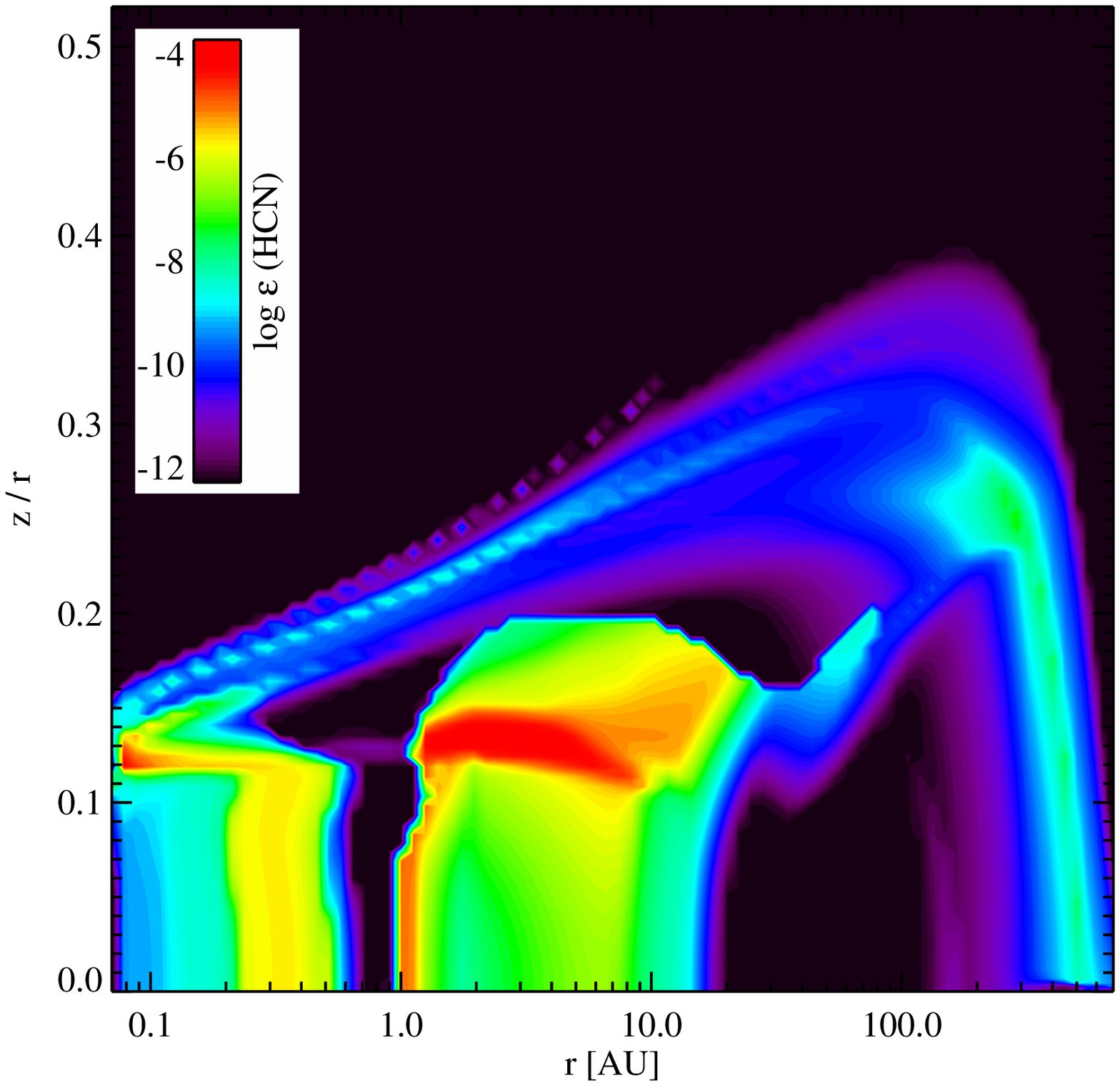}
\includegraphics[width=5.9cm]{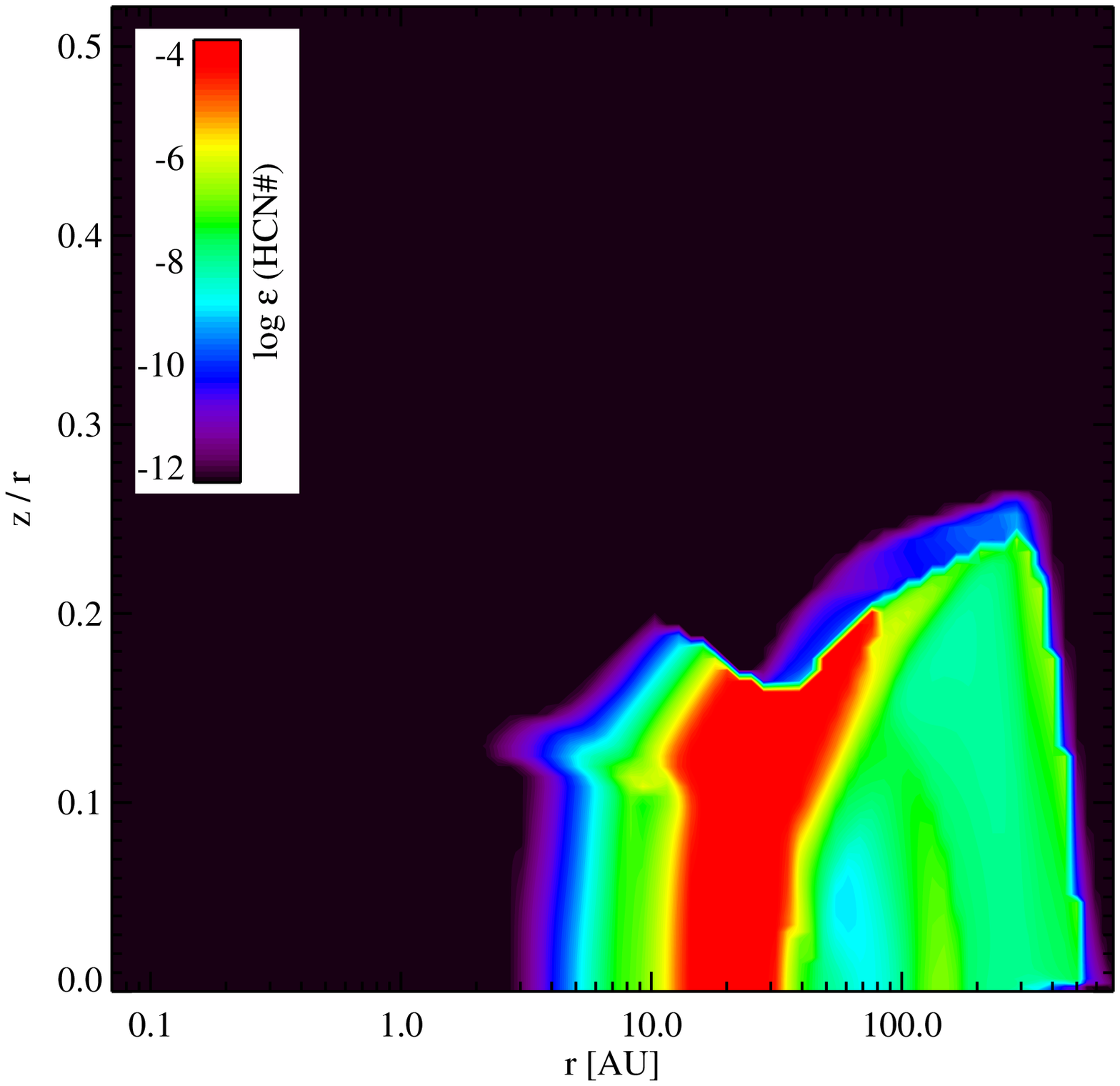}
{\hspace*{3mm}\includegraphics[width=5.9cm]{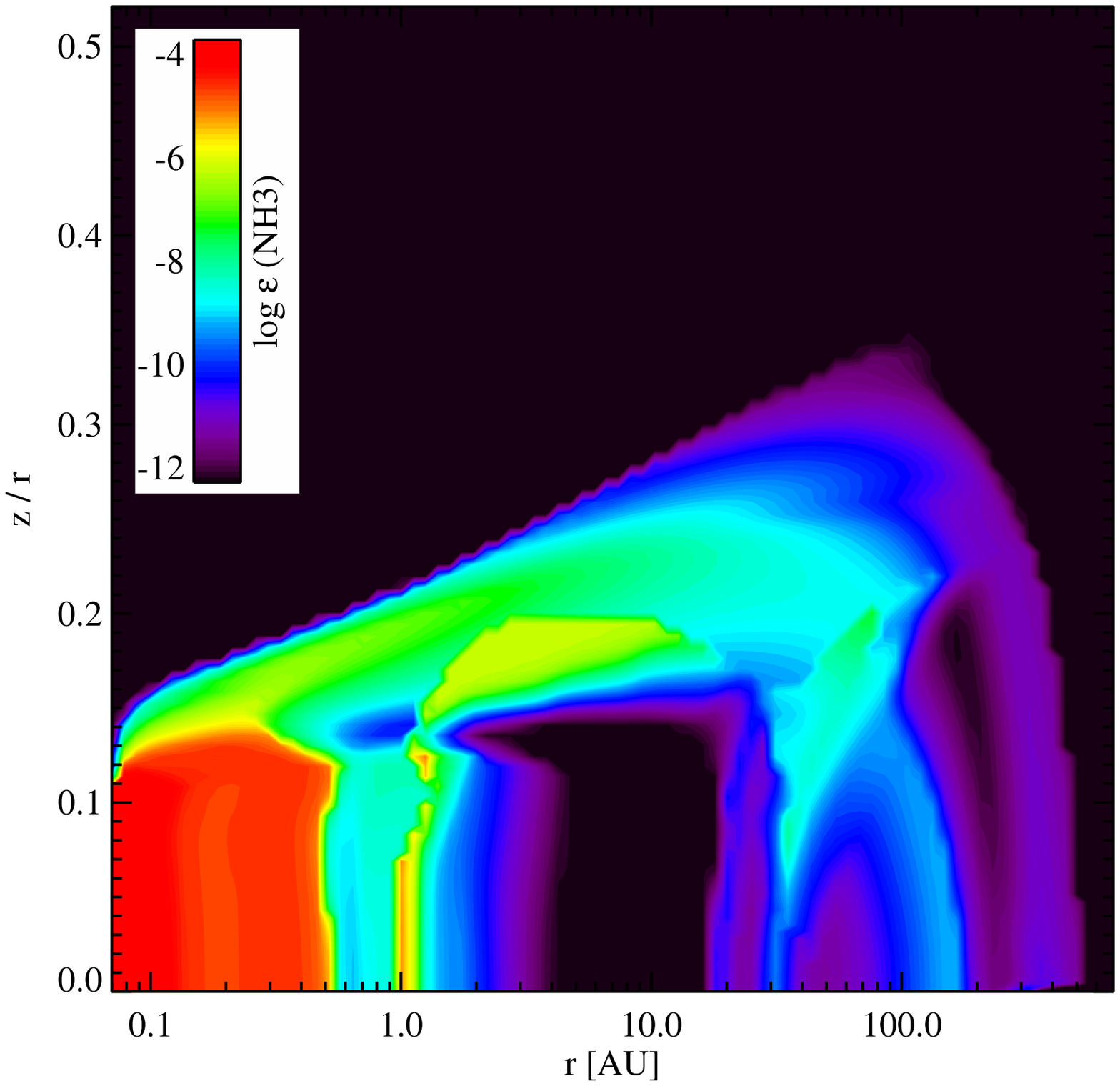}}
{\hspace*{3mm}\includegraphics[width=5.9cm]{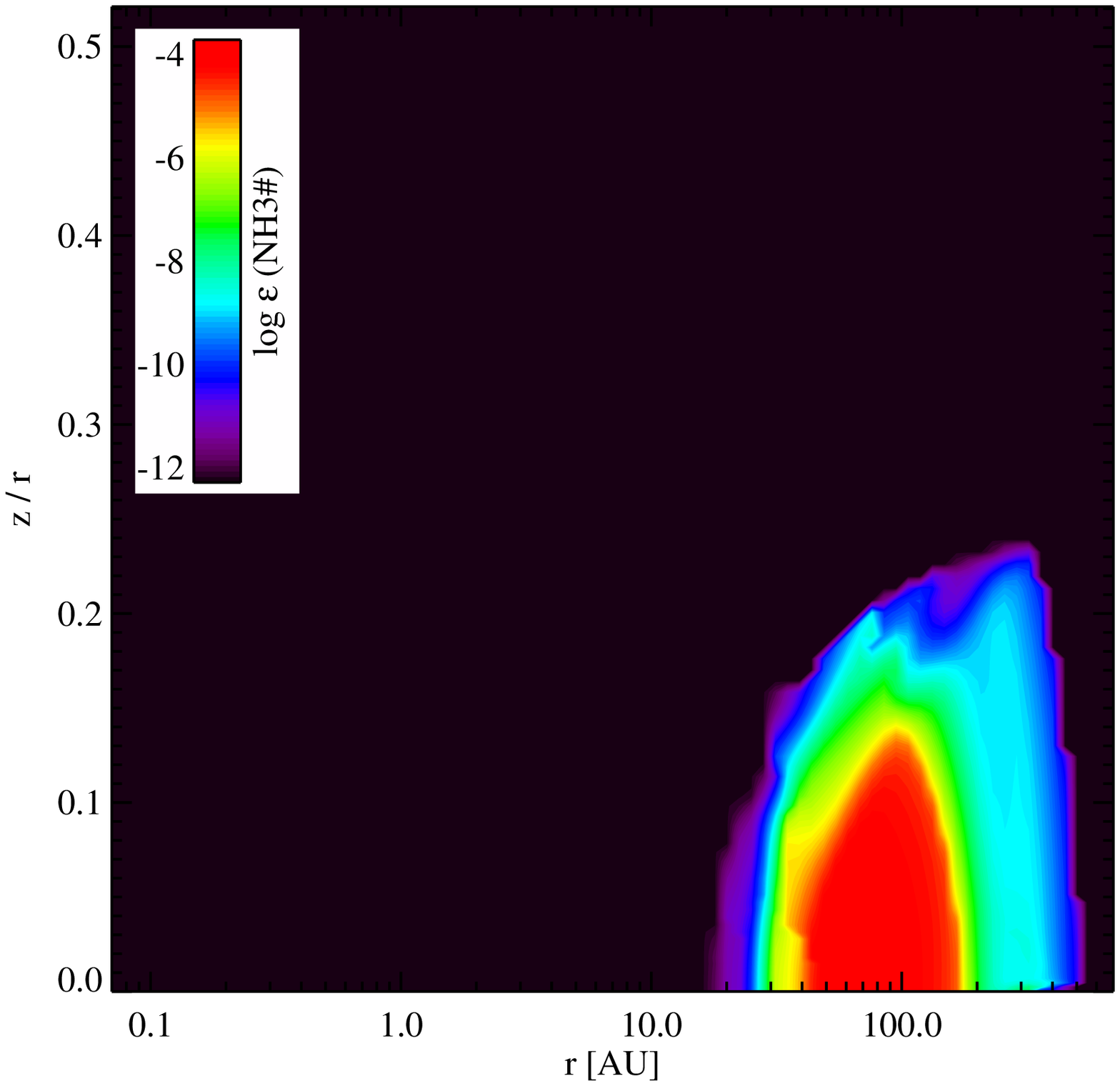}}
\caption{Distribution of key species abundances in the base model: CO, CO\#, CO$_2$, CO$_2$\#, HCO$^+$, OH, H$_2$O\#, CN, HCN, HCN\#, NH$_3$, NH$_3$\#. For CO, the black contour shows the PDR parameter $\log \chi/\langle n_{\rm H} \rangle\!=\!-3.5$ and the blue contour $T_{\rm dust}\!=\!20$~K where CO starts to freeze out on dust grains. For CO\#, the white dashed contours show dust temperatures of 15, 20 and 25~K and the blue dashed line shows the CO ice line estimate from rate equilibrium \citep{Antonellini2016}. For water ice, two approximations of the snow line are indicated: (1) Estimate based on the local density, dust temperature and radiation field \citep[][yellow dashed]{Min2016} and (2) Estimate from rate equilibrium \citep[][blue dashed]{Antonellini2016}.}
\label{fig:base-model}
\end{figure*}

The abundance distribution --- using the small network --- for the key species CO, CO\#, CO$_2$, CO$_2$\#, HCO$^+$, OH, H$_2$O\#, CN, HCN, HCN\#, NH$_3$, NH$_3$\# is shown in Fig~\ref{fig:base-model} for the reference model (model1; note that the trailing \# denotes the ice form of this species). The CO surface is reasonably well described using the PDR parameter $\log \chi/\langle n_{\rm H} \rangle$. For values larger than $-3.5$, CO is efficiently photo dissociated\footnote{Self-shielding is taken into account using the approach described in \citet{Woitke2009}.} and has abundances below $\log n_{\rm CO}/\langle n_{\rm H}\rangle\!=\!\log \epsilon_{\rm CO}\!=\!-8$. Here, $\langle n_{\rm H}\rangle$ is the total hydrogen number density. The CO ice line is reasonably well described by the $T_{\rm dust}\!=\!20$~K line, but a rate equilibrium approach works even better \citep[][white dashed line]{Antonellini2016}. The disk shows a ring of high CO$_2$ abundance inside 1~au. The CO$_2$ ice is sandwiched between the water and CO ice reservoirs. HCO$^+$ only resides in a very thin layer below the C$^+$/C/CO transition when the small chemical network is used.

The OH molecule constitutes the first step in the neutral-neutral chemical pathway to water formation. It is concentrated in the surface layers of the inner disk ($r\!<\!10$~au) where gas temperatures are between 200 and 2000~K. Just below the OH reservoir, inside 0.5~au, densities are high enough to efficiently form water with an abundance of $10^{-4}$. Beyond the snow line, water freezes out onto the cold dust grains. The water ice reservoir is outlined well by a rate equilibrium approach \citep[][yellow dashed line]{Antonellini2016} or using the water vapor pressure together with the local radiation field \citep[][white dashed line]{Min2016}. 

The disk model contains only minute amounts of CN in the disk atmosphere ($\log \epsilon_{\rm CN}\!<\!-8$). Instead we identify two large HCN reservoirs with $\log \epsilon_{\rm HCN}\!\sim\!-4$, a narrow ring  around $0.2$~au and a broader ring between 1 and 5~au. These two reservoirs sit below the $A_{\rm V}\!=\!1$ surface where $T_{\rm gas}\!=\!T_{\rm dust}$. An additional lower abundance reservoir ($\log \epsilon_{\rm HCN}\!\sim\!-8$) can be found in the outer disk atmosphere beyond 100~au. The most stable nitrogen bearing molecule, NH$_3$ is only found in a very narrow ring close to the inner rim of the disk. In this particular model, NH$_3$ ice plays a minor role as a nitrogen reservoir. 

Some of these results will depend on the details of the chosen disk model, on the set of adsorption energies used and also on the size of the chemical network. The impact of the latter two will be discussed in the subsequent sections.

\clearpage

\subsection{Chemical rates from UMIST2006 to 2012}
\label{Sect:resultsUMISTintime}

\begin{figure}[!htbp]
\includegraphics[width=9cm]{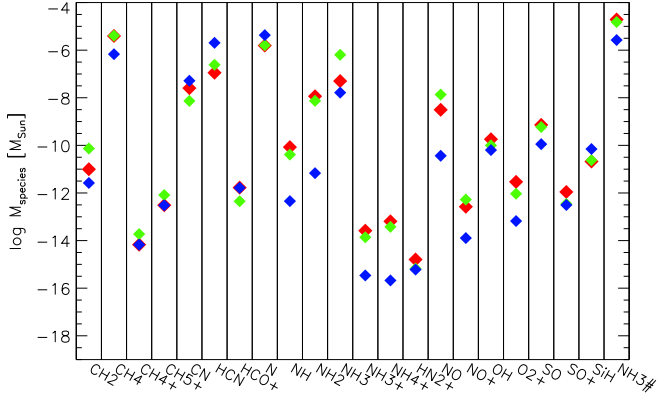}
\vspace*{-8mm}
\caption{Differences in species masses between three different sets of reactions: UMIST2012 plus collider reactions from UMIST2006 (red, model 1), UMIST2006 (green, model 2) and UMIST2012 (blue, model 1a).}
\label{fig:Reactions-change-in-speciesmass}
\end{figure}

The revision of the UMIST database in 2012 reveals major differences in species masses especially for nitrogen bearing species. The main reason is the missing collider reactions with respect to the UMIST2006 rate file. Fig.~\ref{fig:Reactions-change-in-speciesmass} shows this effect for the species that change by more than a factor three and have absolute masses above $10^{-15}$~M$_\odot$. Species not shown here vary by less than a factor three. 

\begin{figure}[!htbp]
\includegraphics[width=4.45cm]{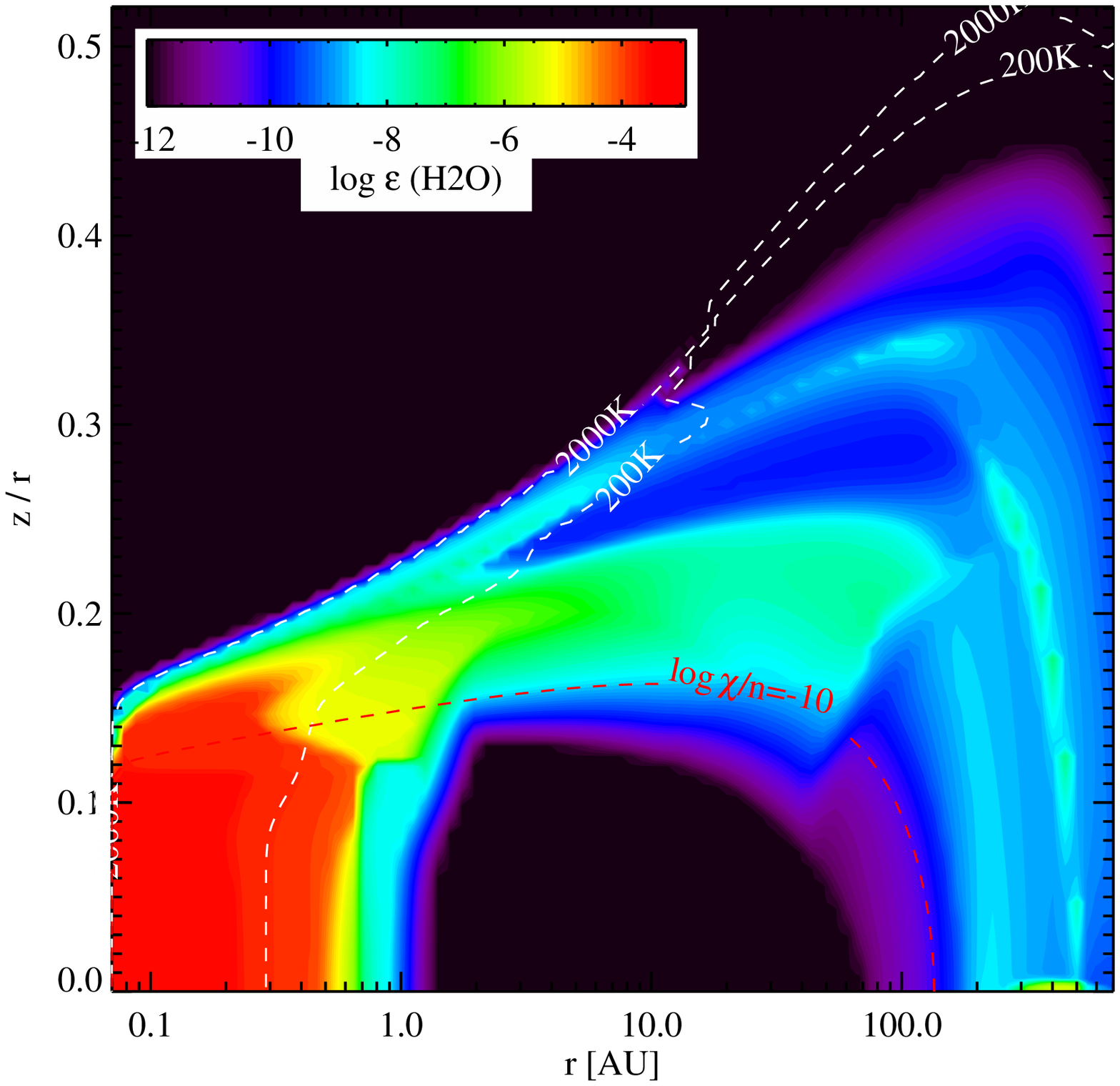}
\includegraphics[width=4.45cm]{model1_OH.ps}
\includegraphics[width=4.45cm]{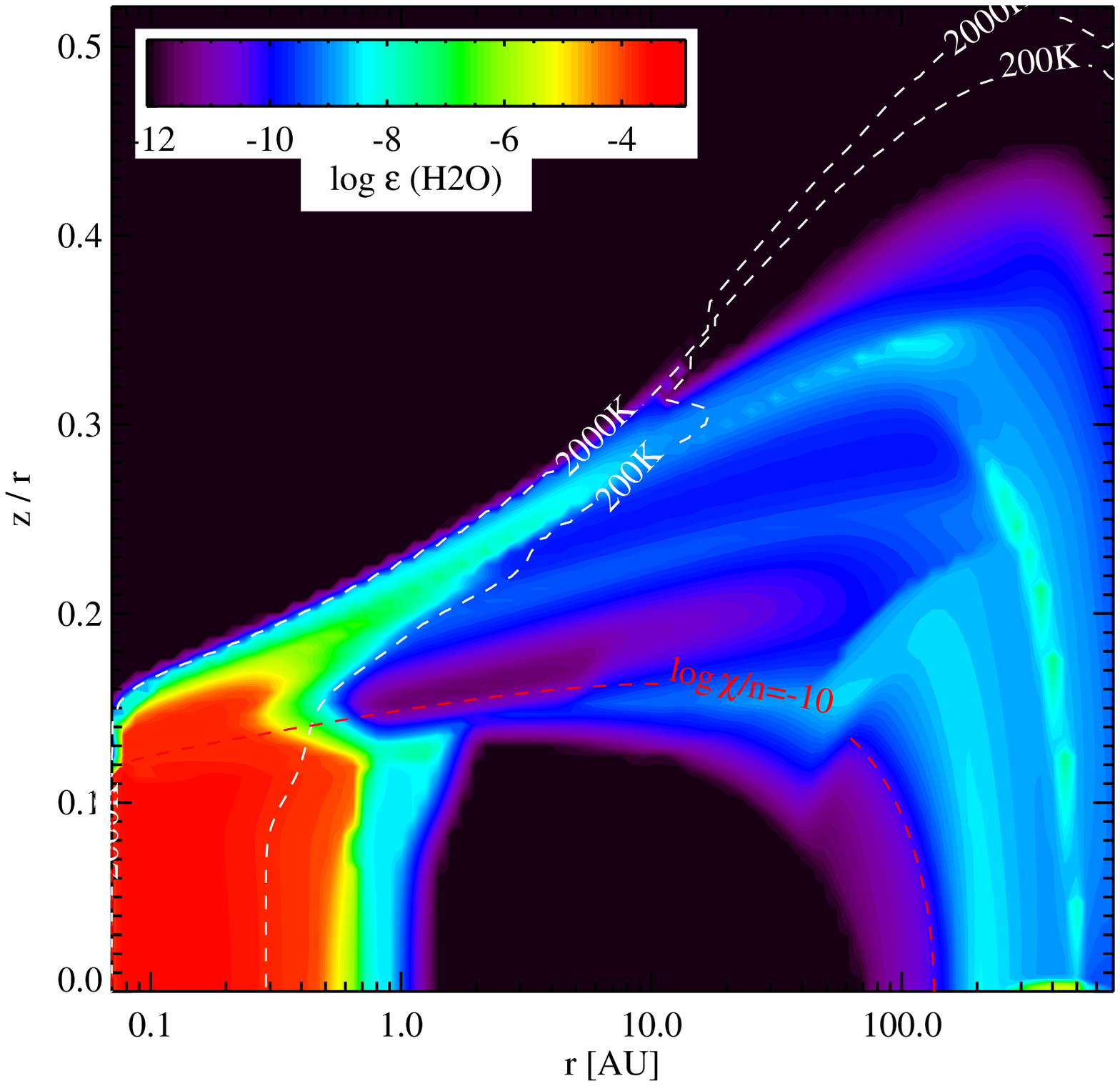}
\includegraphics[width=4.45cm]{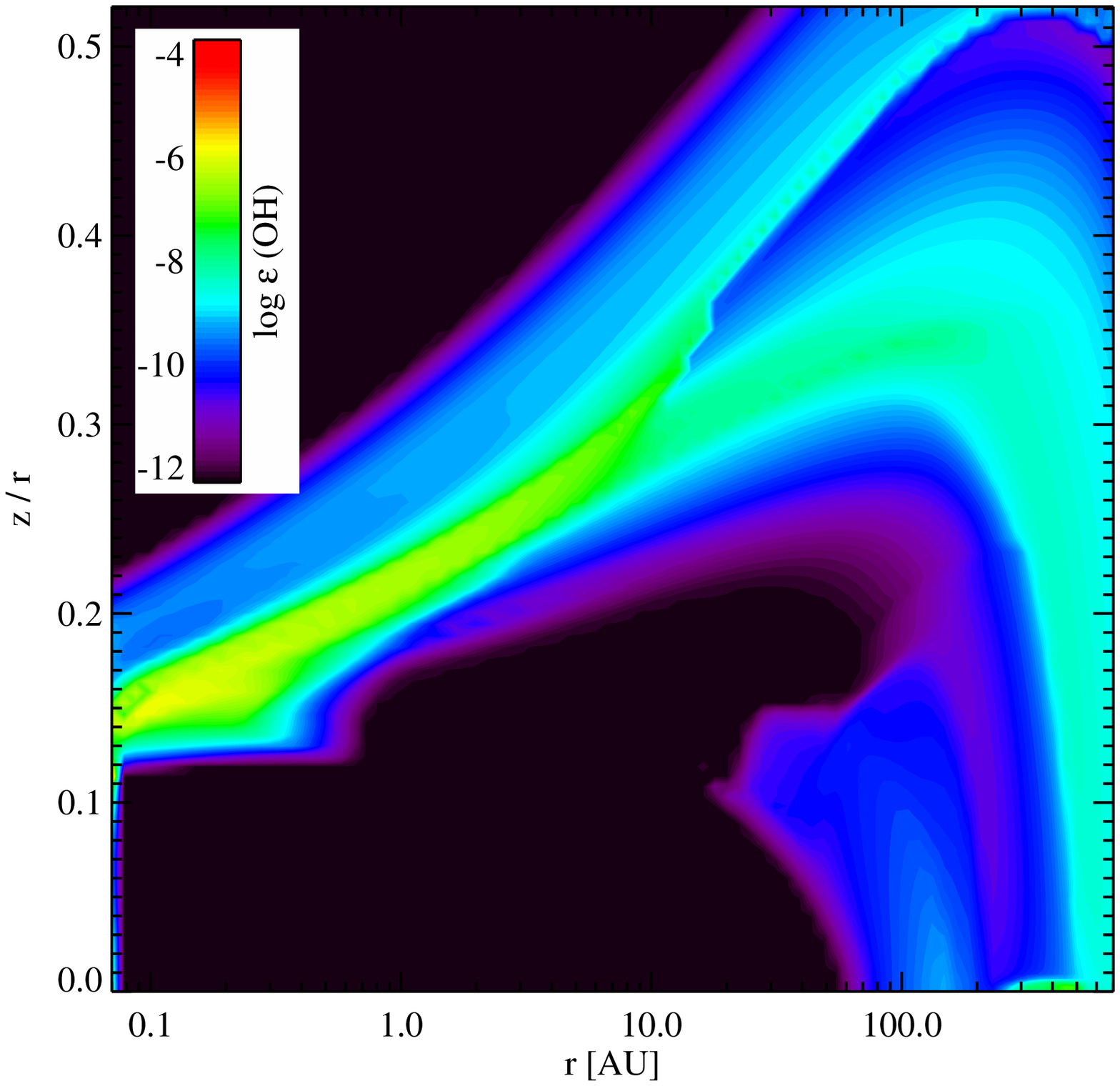}
\caption{Comparison of water and OH abundances in the UMIST2012 model with (top, model 1) and without collider reactions (bottom, model 1a).}
\label{fig:H2O-collider}
\end{figure}

\begin{figure*}[!htbp]
\begin{center}
\includegraphics[width=16cm]{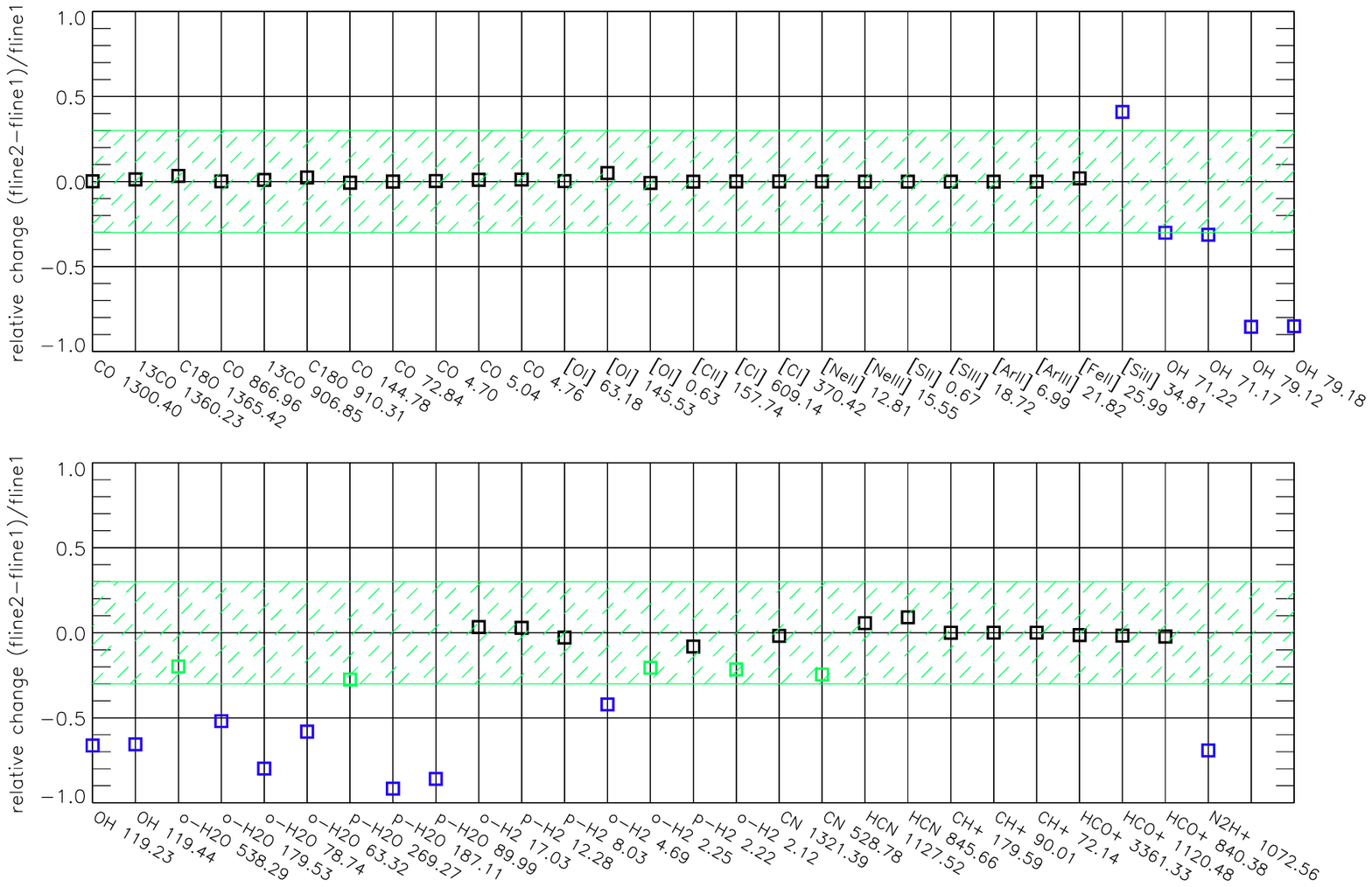}
\caption{Comparison of line fluxes for two sets of reactions: UMIST2012 plus CL reactions from UMIST2006 (fline1, model 1) and UMIST2012 (fline2, model 1a). Black and green squares denote differences of less than 25\% and less than a factor two respectively, blue squares and red triangles denote differences larger than a factor three and ten respectively.}
\label{fig:chem-lines-1}
\end{center}

\begin{center}
\includegraphics[width=16cm]{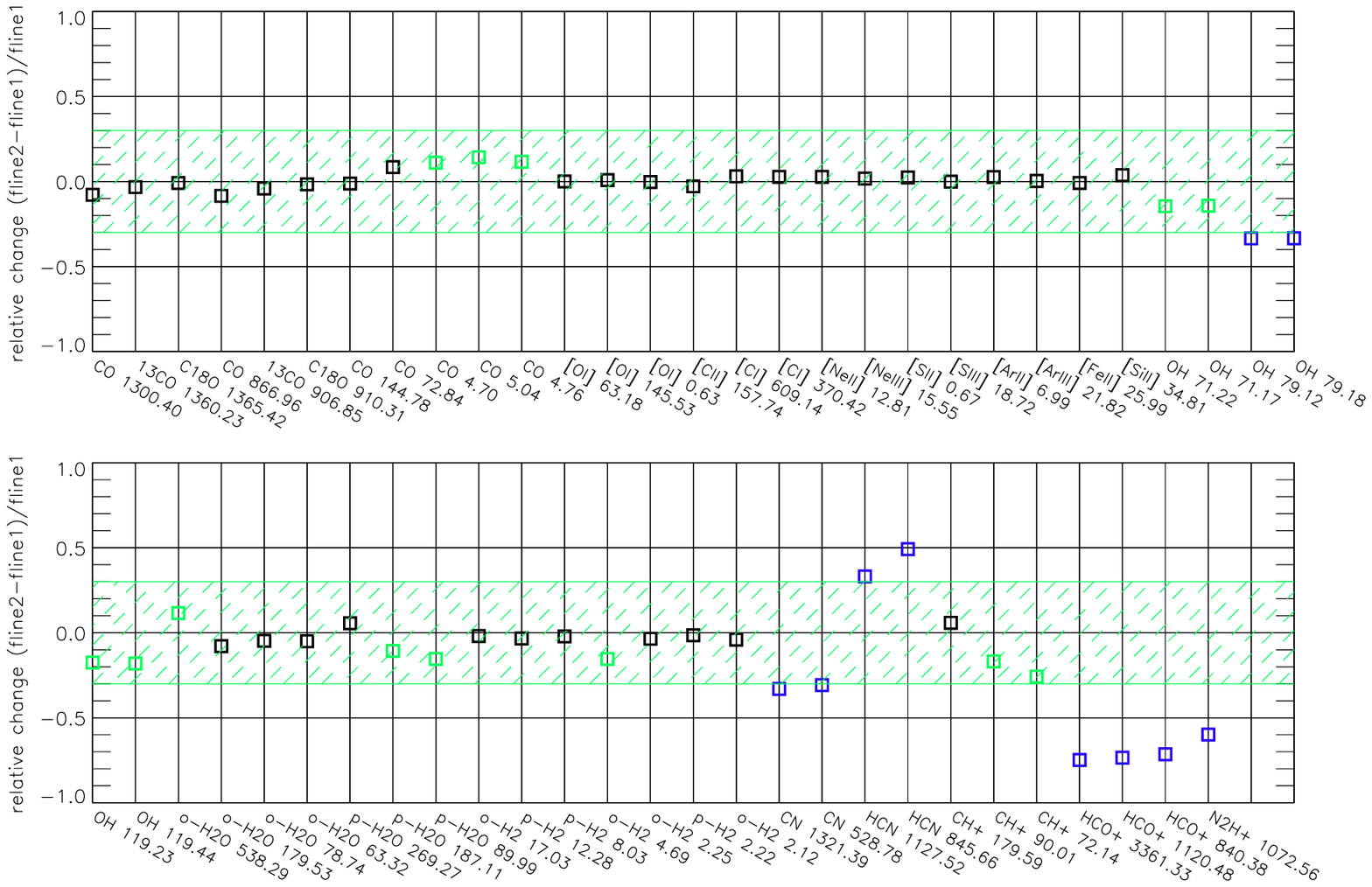}
\caption{Comparison of line fluxes for two sets of reactions: UMIST2012 plus CL reactions from UMIST2006 (fline1, model 1) and UMIST2006 (fline2, model 2). Black and green squares denote differences of less than 25\% and less than a factor two respectively, blue squares and red triangles denote differences larger than a factor three and ten respectively.}
\label{fig:chem-lines-2}
\end{center}
\end{figure*}

In the case of UMIST2012 without the collider reactions, water and OH abundances in the surface of the outer disk change by orders of magnitude; in fact, the entire water vapor reservoir on top of the water ice reservoir disappears (Fig.~\ref{fig:H2O-collider}). This is also reflected in the water and OH line fluxes changing by a factor 3-10 (Fig.~\ref{fig:chem-lines-1}). The rates were not deliberately omitted, but simply not re-assessed in UMIST2012, hence the UMIST2006 collider reactions should be used (Millar, private communication). Adding the collider reactions brings back the water and OH reservoir and also leads to a match of the water and OH line fluxes to within a factor 2-3 (Fig.~\ref{fig:chem-lines-2}). The three-body (collider) reaction opening the water formation pathway is
\begin{equation}
{\rm N} + {\rm H_2} + {\rm M} \rightarrow {\rm NH_2}
\end{equation}
with a reaction rate of $10^{-26}$~cm$^6$~s$^{-1}$ \citep{Avramenko1966}. The rate is constant over the temperature range $564 - 796$~K according to NIST. Since we extrapolate rates outside the temperature range, it gets applied also in the somewhat cooler disk surface regions ($20-300$~K). This rate stems from a very old measurement and definitely needs to be revisited. NH$_2$ subsequently reacts with oxygen to form NH and OH. Both radicals react further to form water \citep{Kamp2013}. The more classical neutral-neutral pathway identified e.g. by \citet{Glassgold2009} 
\begin{equation}
{\rm O} \overlim{\rm H_2} {\rm OH} \overlim{\rm H_2} {\rm H_2O} 
\end{equation}
acts at higher gas temperatures ($T_{\rm gas}\!\gg\!200$~K) closer to the star.

CN, OH and HCO$^+$ show differences in species mass of up to 0.5~dex between UMIST2006 and UMIST2012 (plus collider reactions). Lines of these species are frequently observed in the far-IR and submm wavelength range and their predicted line fluxes can differ by up to a factor 2.5 for CN and OH and up to a factor six for HCO$^+$, with UMIST2012 giving systematically higher fluxes (Fig.~\ref{fig:chem-lines-2}). Throughout the remainder of this paper, we use `UMIST2012' as a replacement for `UMIST2012 rate database including the collider reactions'.

\subsection{Chemical reaction databases}
\label{Sect:resultsDatabases}

We test the impact of different sets of reaction rates on the overall disk chemistry and appearance. Three databases are investigated: UMIST2012 (model 1), OSU (model 3) and KIDA (model 4). In all cases, we use the small chemical network and the adsorption energies of \citet{Aikawa1996}. We also keep the physical and thermal structure of the underlying disk constant. 

\begin{figure}[!htbp]
\includegraphics[width=4.4cm]{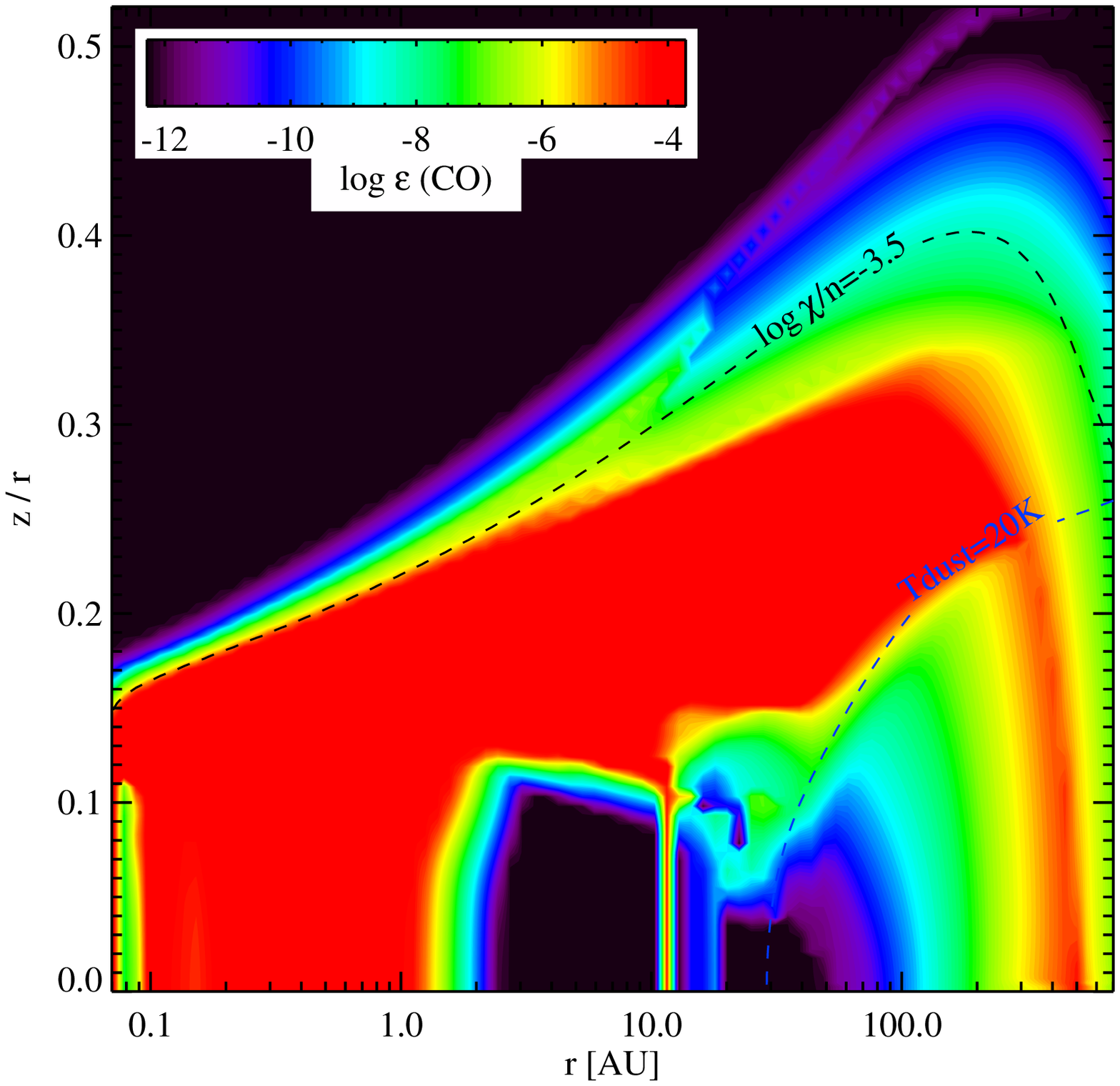}
\includegraphics[width=4.4cm]{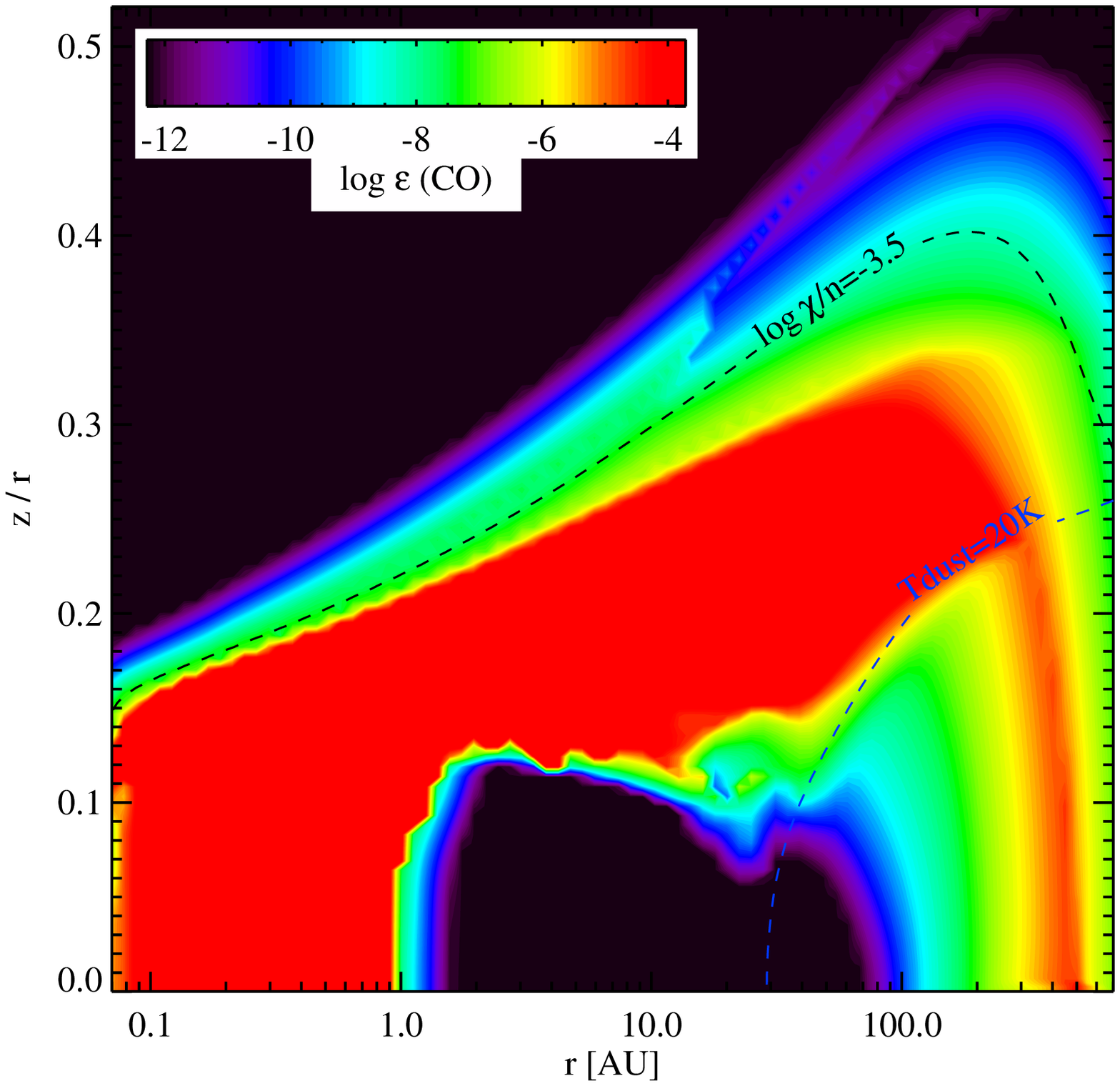}
\includegraphics[width=4.4cm]{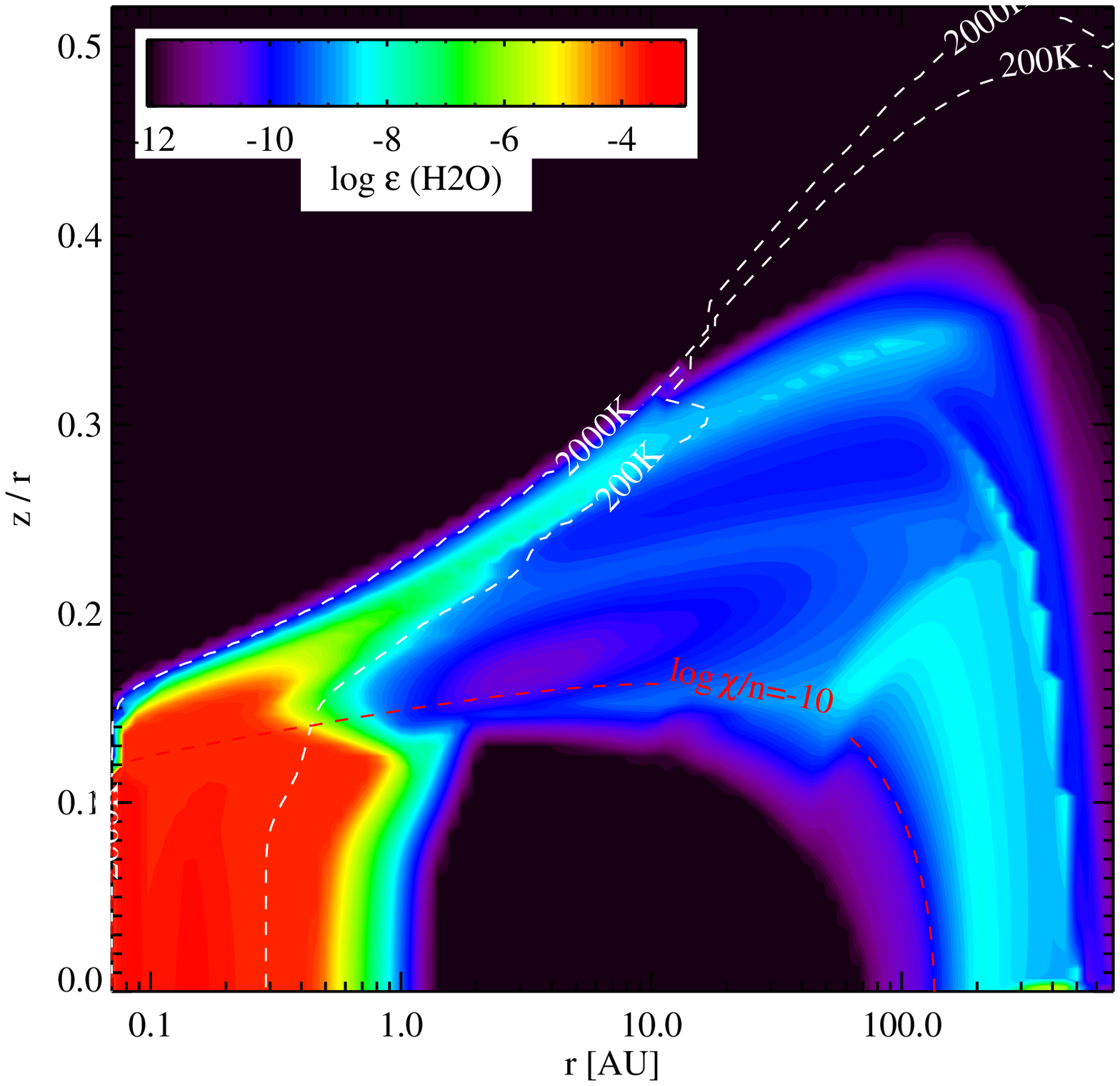}
\includegraphics[width=4.4cm]{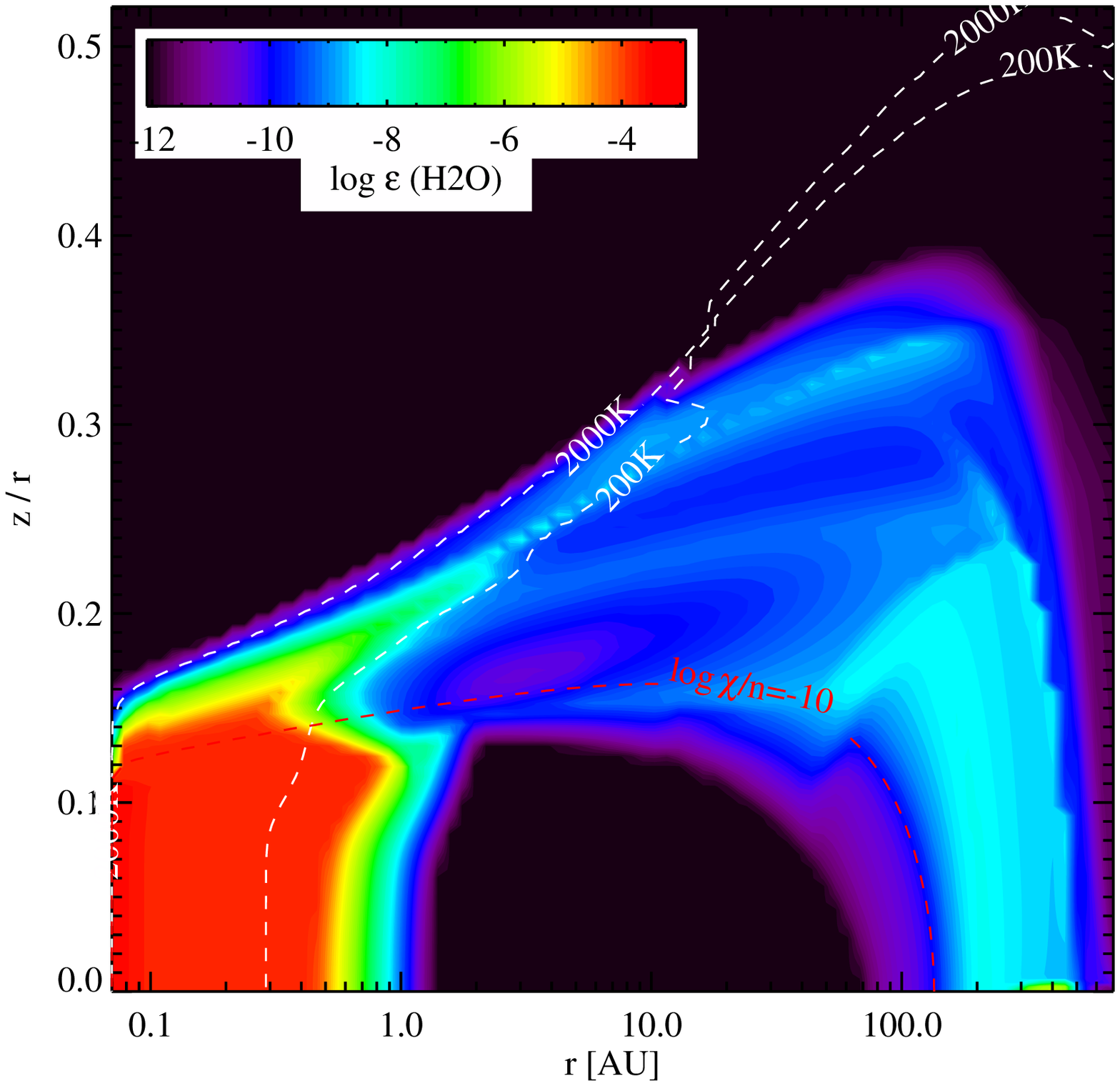}
\includegraphics[width=4.4cm]{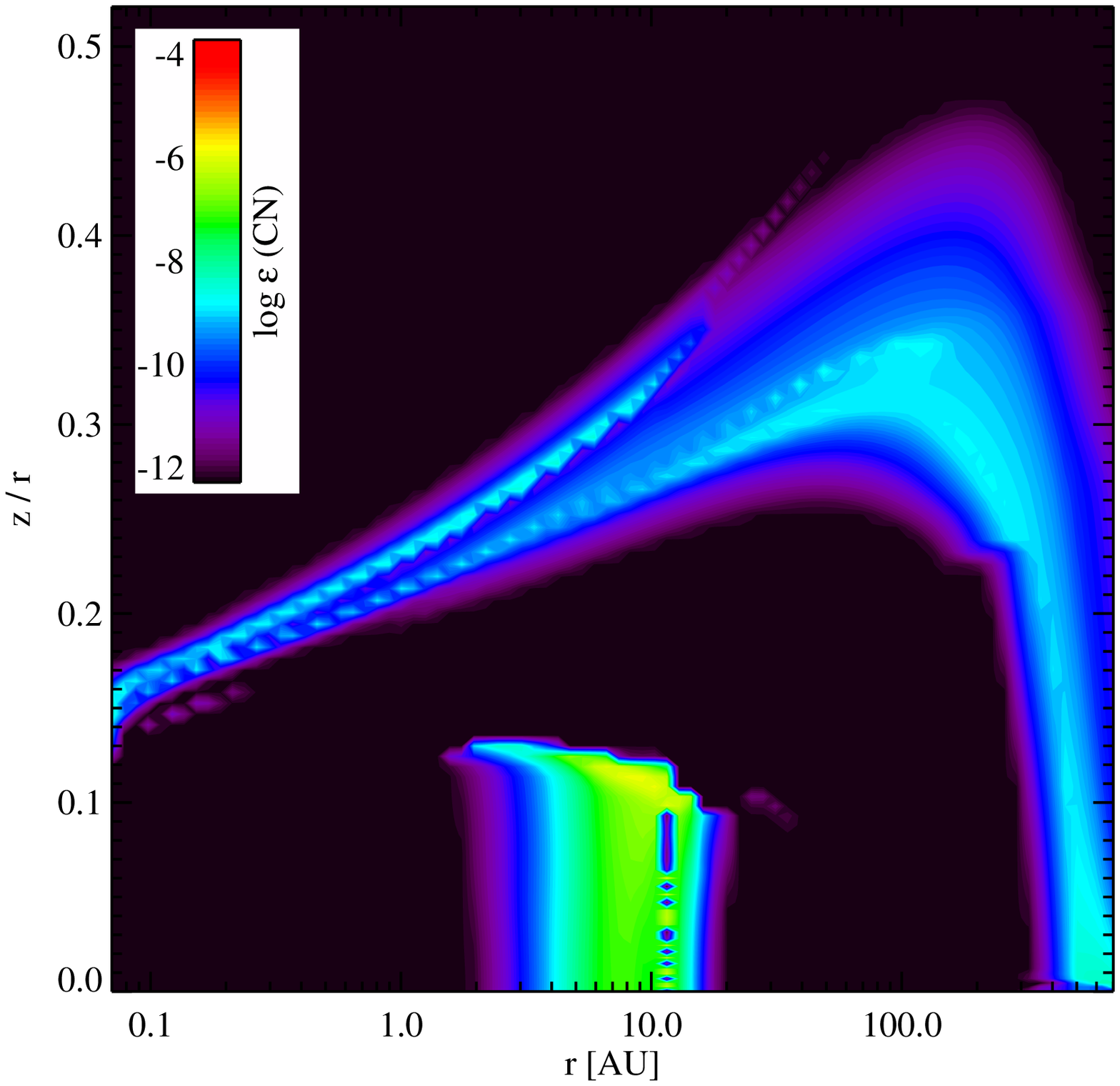}
\includegraphics[width=4.4cm]{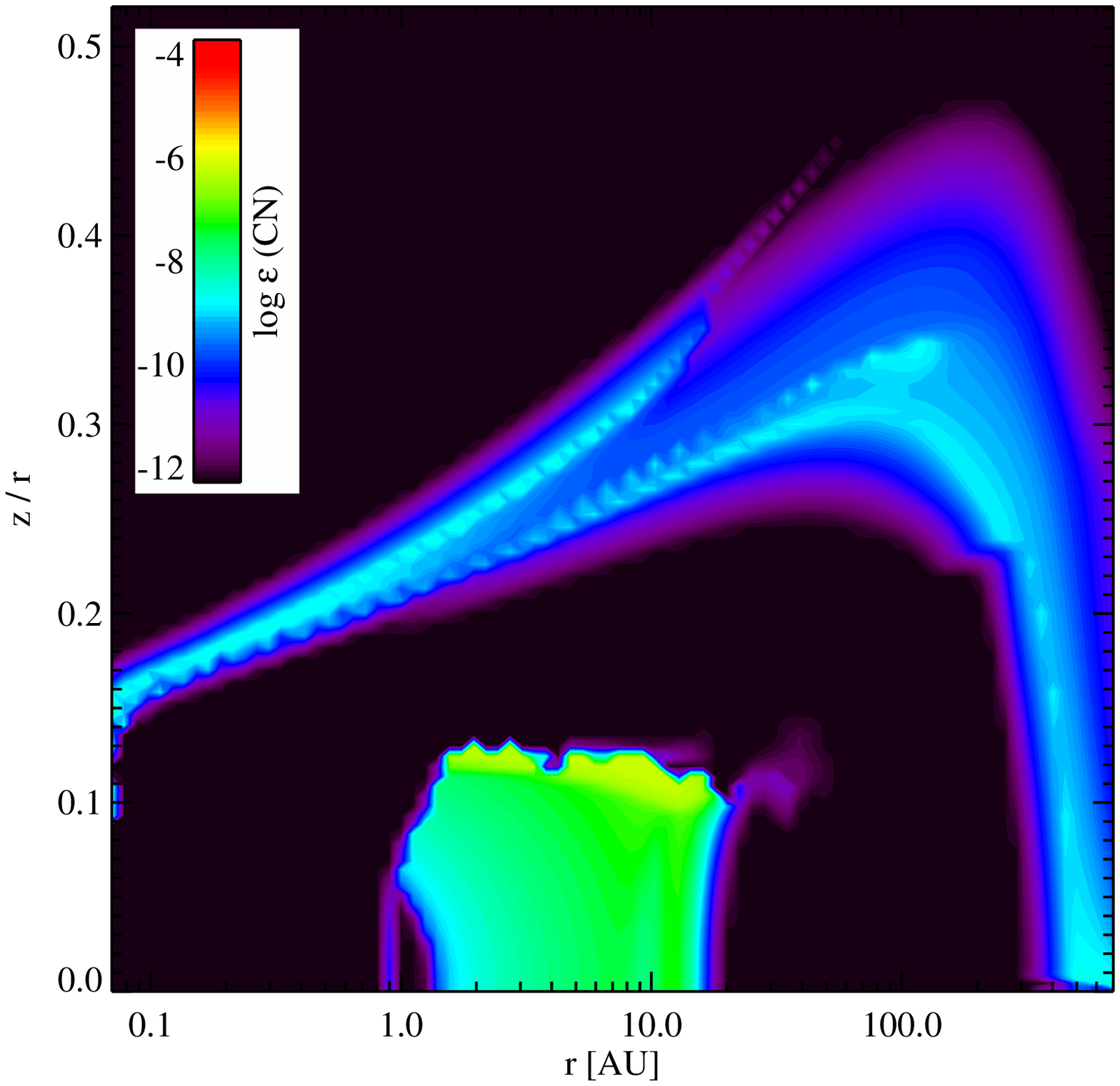}
\includegraphics[width=4.4cm]{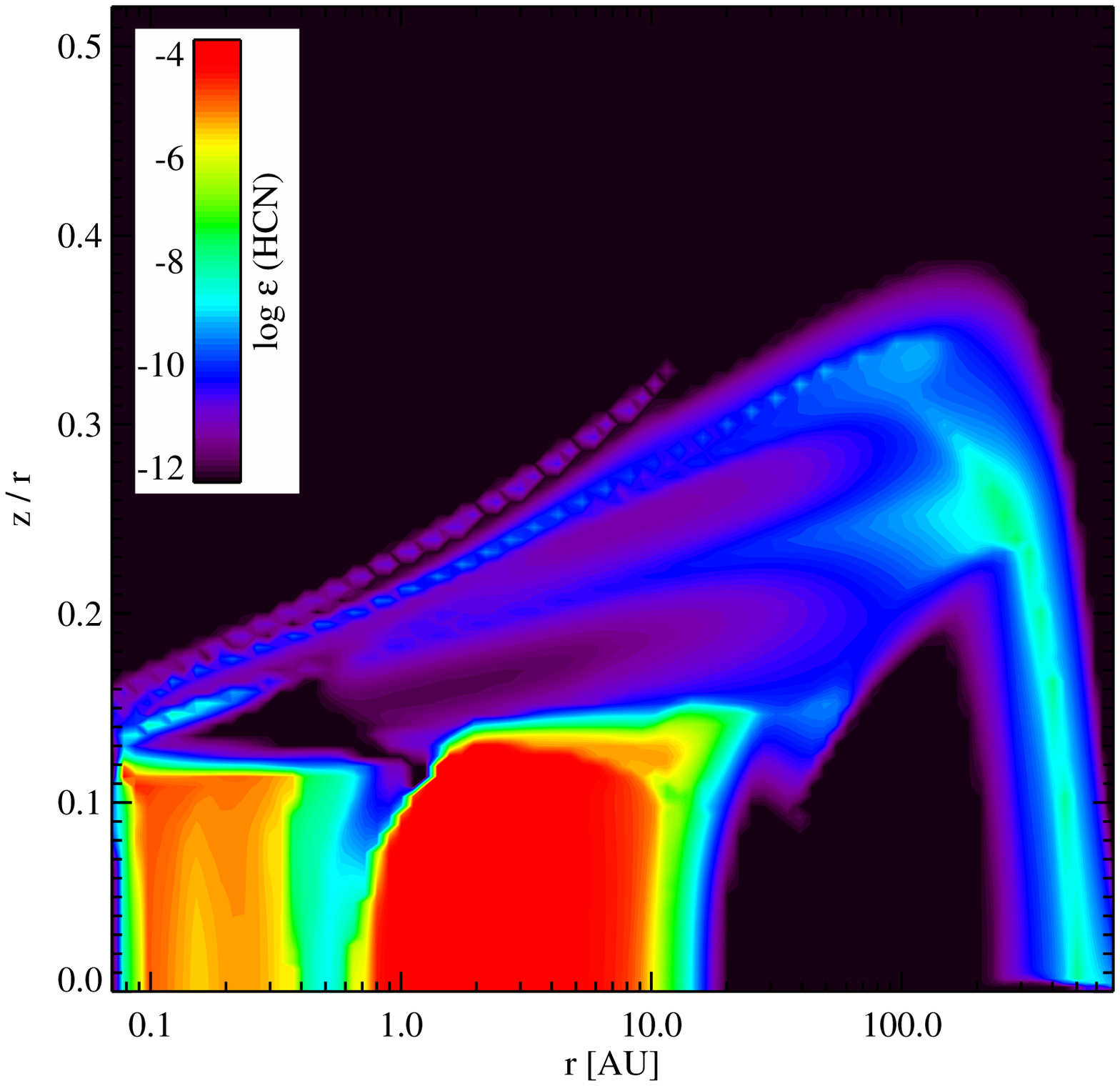}
{\hspace*{1mm}\includegraphics[width=4.4cm]{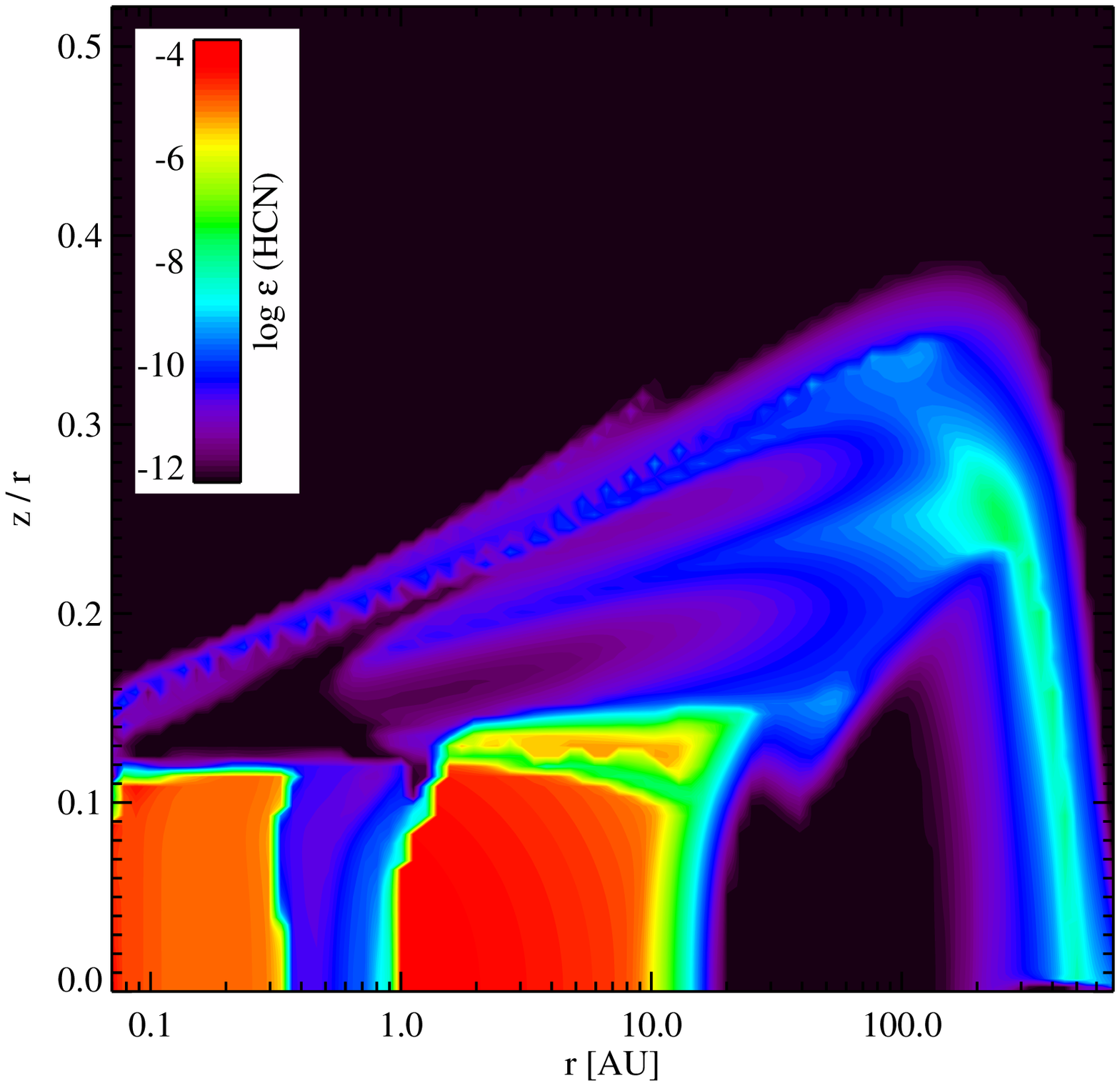}}
\caption{Distribution of key species abundances using OSU (left) and KIDA (right): CO, H$_2$O, CN, HCN. Contours are the same as those in Fig.~\ref{fig:base-model} and \ref{fig:H2O-collider}.}
\label{fig:Chemistrydiffdatabases}
\end{figure}

Given that these databases have been compiled with different focus, it is not surprising that almost one third of the species masses change by more than 0.5~dex. In addition, collider reactions are not a priori included in these databases; hence water and OH are affected in the same way as described in Sect.~\ref{Sect:resultsUMISTintime}. Fig.~\ref{fig:Chemistrydiffdatabases} provides an overview of a few key species for OSU and KIDA and can be compared to Fig.~\ref{fig:base-model} and \ref{fig:H2O-collider}.

\begin{figure*}[!htbp]
\begin{center}
\includegraphics[width=16cm]{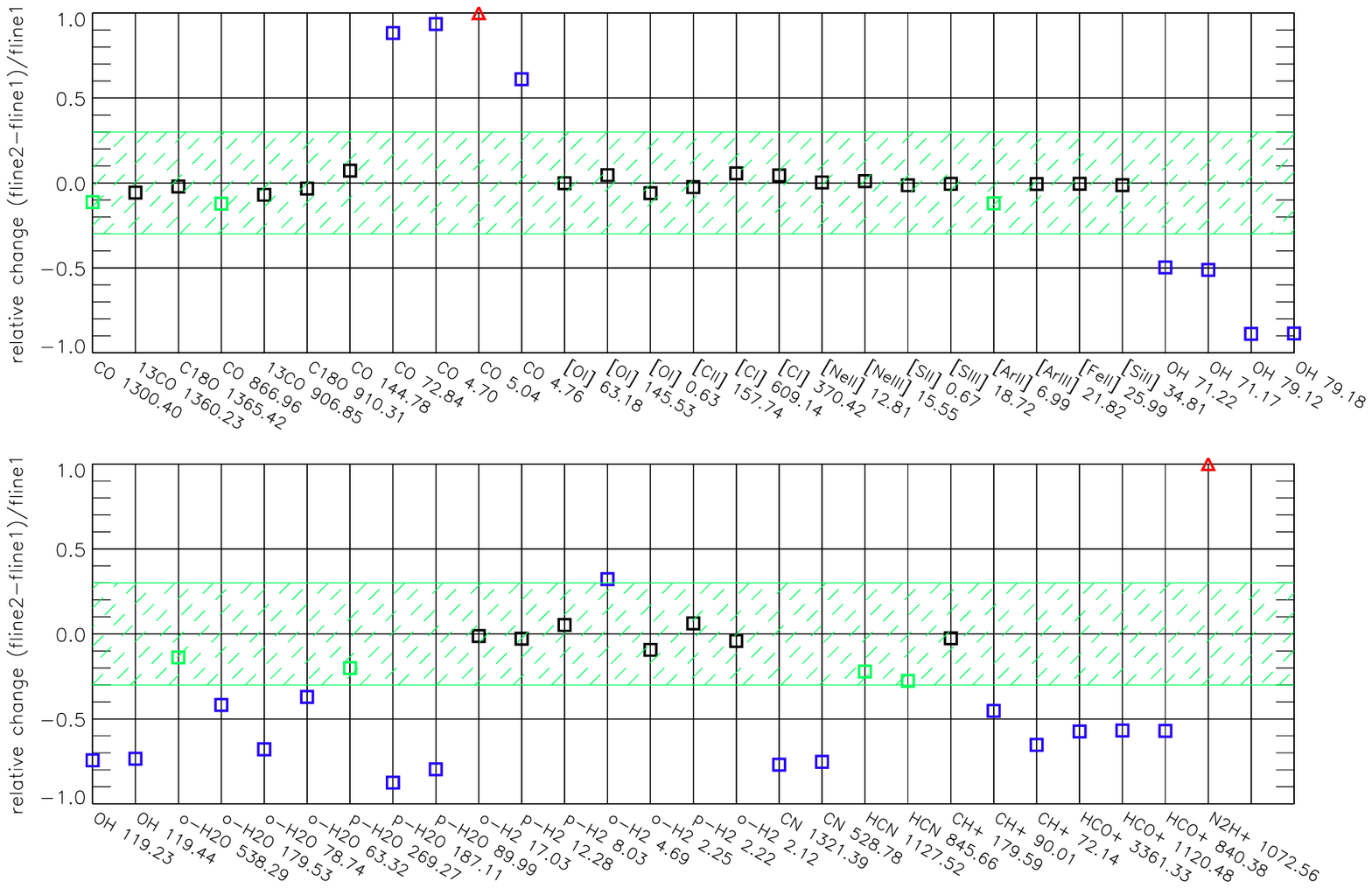}
\caption{Comparison of line fluxes for two databases: UMIST2012 (fline1, model 1) and OSU (fline2, model 3). Black and green squares denote differences of less than 25\% and less than a factor two respectively, blue squares and red triangles denote differences larger than a factor three and ten respectively.}
\label{fig:chem-lines-3}
\end{center}

\begin{center}
\includegraphics[width=16cm]{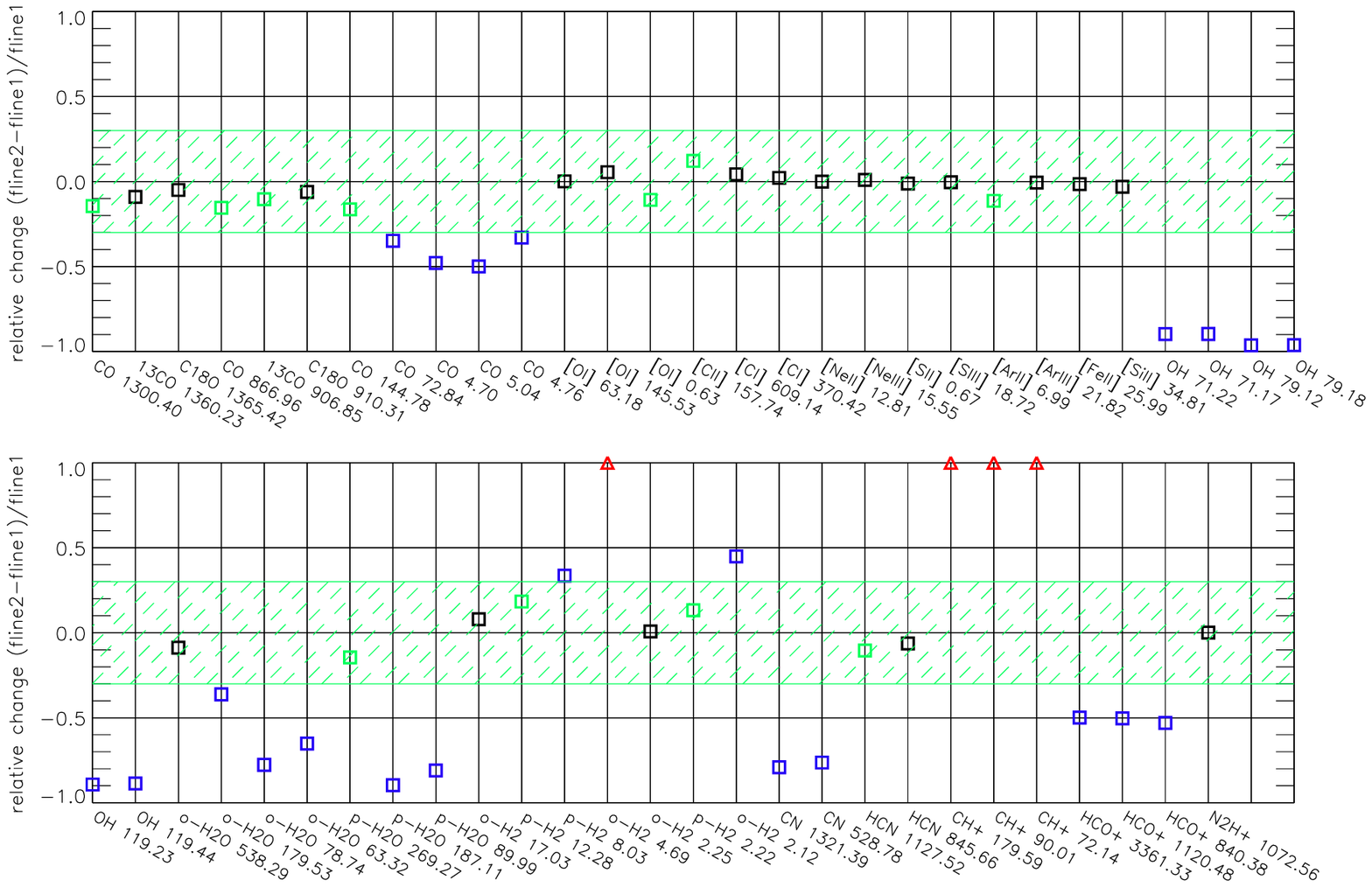}
\caption{Comparison of line fluxes for two databases: UMIST2012 (fline1, model 1) and KIDA (fline2, model 4). Black and green squares denote differences of less than 25\% and less than a factor two respectively, blue squares and red triangles denote differences larger than a factor three and ten respectively.}
\label{fig:chem-lines-4}
\end{center}
\end{figure*}

However, despite these large changes in the overall chemical structure, some line tracers stay very robust while others change by more than 1~dex (see Figs.~\ref{fig:chem-lines-3} and \ref{fig:chem-lines-4}): \\[-5mm]

\begin{itemize}
\item The low excitation CO lines hardly change. This is due to the simple molecular cloud like chemistry in the outer disk. However, there are significant changes in the high excitation rotational lines and ro-vibrational lines that originate in the inner 10~au where CO chemistry is driven by ion-molecule and in lower layers by neutral-neutral reactions. Interestingly, OSU gives systematically higher fluxes, while KIDA is lower than UMIST2012. This relates to the depth at which the C$^+$/C/CO transition is reached in those different networks. Self-shielding is treated in the same way for all three networks. The main CO formation reactions are via OH producing CO$^+$. The OH radical can react with H$_2$, N or C$^+$ and the latter forms CO$^+$ (OH + C$^+$ $\rightarrow$ CO$^+$ + H). Subsequent reactions with H and H$_2$ lead to CO (CO$^+$ + H $\rightarrow$ CO + H$^+$
and CO$^+$ + H$_2$ $\rightarrow$ HCO$^+$ + H followed by HCO$^+$ + e $\rightarrow$ CO + H). Reaction rates for these differ between the networks. Apparently, even small rate differences can lead to significant differences in line fluxes. The transition lies closest to the surface in the OSU model, while it lies deepest in the KIDA model (see Fig.~\ref{fig:base-model} and \ref{fig:Chemistrydiffdatabases}); note that the CO high excitation lines react very strongly to changes in temperature and the C$^+$/C/CO transition has a steep vertical temperature gradient.\\[-2mm]
\item The neutral and ionized atomic line fluxes are very robust and stay within a factor two across all networks; those species are dominated by photochemistry and their abundance (and line flux) directly reflects the choice of elemental abundances. \\[-2mm]
\item H$_2$ chemistry is driven by formation on dust and photodissociation; these processes are implemented outside the specific network. A crucial reaction in the inner disk destroying H$_2$ is collisions with atomic oxygen, leading to the formation of OH. The rate constants do not differ much in the three networks\\[2mm]
UMIST -- $A\!=\!3.14\,10^{-13}$, $B\!=\!2.7$, $C\!=\!3150$ (297-3532~K)\\[2mm]
OSU -- $A\!=\!3.44\,10^{-13}$, $B\!=\!2.67$, $C\!=\!3160$ (1-40\,000~K) \\[2mm]
KIDA -- $A\!=\!3.44\,10^{-13}$, $B\!=\!2.67$, $C\!=\!3160$ (10-280~K)\\[2mm]
However, for the reactions consuming OH, the rate coefficients are different in the three networks, propagating into the OH abundances in the surface layers inside 10~au. Emission lines of molecular hydrogen turn out to be mostly within a factor 3. Many of the H$_2$ lines discussed here originate in a thin surface layer limited in depth by the dust continuum. The high rotational line at 4.694~$\mu$m as well as the ro-vibrational lines are optically thin similar to what was found by \citet{Nomura2005}. This makes line flux predictions very sensitive to the exact placement of the H/H$_2$ transition in the disk model.  \\[-2mm]
\item HCN lines originating in the outer disk are also very robust; again, similar to the CO case, the chemistry here is largely molecular cloud chemistry.\\[-2mm]
\item CN outer disk abundances are lower in the OSU and KIDA disk models (see Fig.~\ref{fig:base-model} and \ref{fig:Chemistrydiffdatabases}) and the corresponding lines originating in the outer disk are systematically weaker for those networks compared to UMIST2012. One difference in the networks is the CN destruction reaction with oxygen, which is a factor $\sim\!2$ stronger in OSU and KIDA at low temperatures compared to UMIST2012.\\[-2mm]
\item OH and water lines differ within a factor 10 between the OSU/KIDA and UMIST2012. This is mainly due to the missing collider reactions that affect the outer lower reservoirs of these two molecules.\\ 
\item The largest differences (more than one order of magnitude) are seen in line fluxes of CH$^+$ between UMIST2012 and KIDA. This is due to differences in reaction rates leading to the formation and destruction of this radical (see Table~\ref{tab:CH+rates}).\\
\end{itemize}

\begin{table}
\caption{Comparison of rate coefficients for CH$^+$ formation and destruction reactions.}
\begin{tabular}{l|rrr|rrr}
\hline
Reaction      & \multicolumn{3}{c|}{KIDA 2011} & \multicolumn{3}{c}{UMIST 2012} \\
                     & $A$ & $B$ & $C$ & $A$ & $B$ & $C$ \\
                     \hline
                     \hline
C + H$_3^+$  &      2(-9) &0.0 & 0.0              &  2.00(-9) & 0.0 & 0.0 \\
C$^+$  + H$_2$  &      7.8(-10) & 0.0 & 4540.0    & 1.00(-10) & 0.0 & 4640.0 \\
CH$^+$ + H$_2$     &          1.20(-9) & 0.0 & 0.0                          &  1.20(-9) & 0.0 & 0.0 \\
CH$^+$ + e     &          7(-8) & -0.5 & 0.0                                        &  1.50(-7) & -0.42 & 0.0 \\
\hline
\end{tabular}
\tablefoot{The coefficients $A$, $B$ and $C$ have their usual meaning \citep[see e.g.][]{McElroy2013b}. The notation $x(-y)$ denotes $x\,10^y$.}
\label{tab:CH+rates}
\end{table}

\subsection{Adsorption energies}
\label{Sect:resultsEads}

After having seen differences arising from different networks, we focus now on adsorption energies that affect the gas/ice reservoirs in the disk. \citet{Collings2004} found for example that CO can be trapped in the polar water ice at much higher temperatures than in a non-polar CO ice. The dependence of chemical abundances on the specific grain surface --- SiO$_2$, polar, non-polar --- has already been noted by \citet{Bergin1995}. Here, we explore systematically the effects of using different sets of adsorption energies and explore a first simple model that illustrates the effect of surface dependent adsorption energies.

The thermal desorption rate of a species $i$ depends among other variables also on the adsorption energy $E_{\rm ads}(i)$ (expressed in K)
\begin{equation}
R = n_{i\#}  \nu_{\rm osc}(i) \exp\left(-\frac{E_{\rm ads}(i)}{T_{\rm dust}}\right) \,\,\, {\rm cm^{-3}}~{\rm s^{-1}}\,\,\, ,
\label{Eq:thermaldesorption}
\end{equation}
where $\nu_{\rm osc}(i)$ is the oscillation frequency of species $i$, $n_{i\#}$ the density of desorbable species $i$ on the grain surface\footnote{Details on the various thermal and non-thermal desorption processes can be found in \citet{Woitke2009}.}, and $T_{\rm dust}$ the temperature of the grain. The oscillation frequency depends only weakly on the adsorption energy of the species, thus making the exponential term in Eq.\,(\ref{Eq:thermaldesorption}) the dominant one.

\begin{figure}[!htbp]
\begin{center}
{\includegraphics[width=9cm]{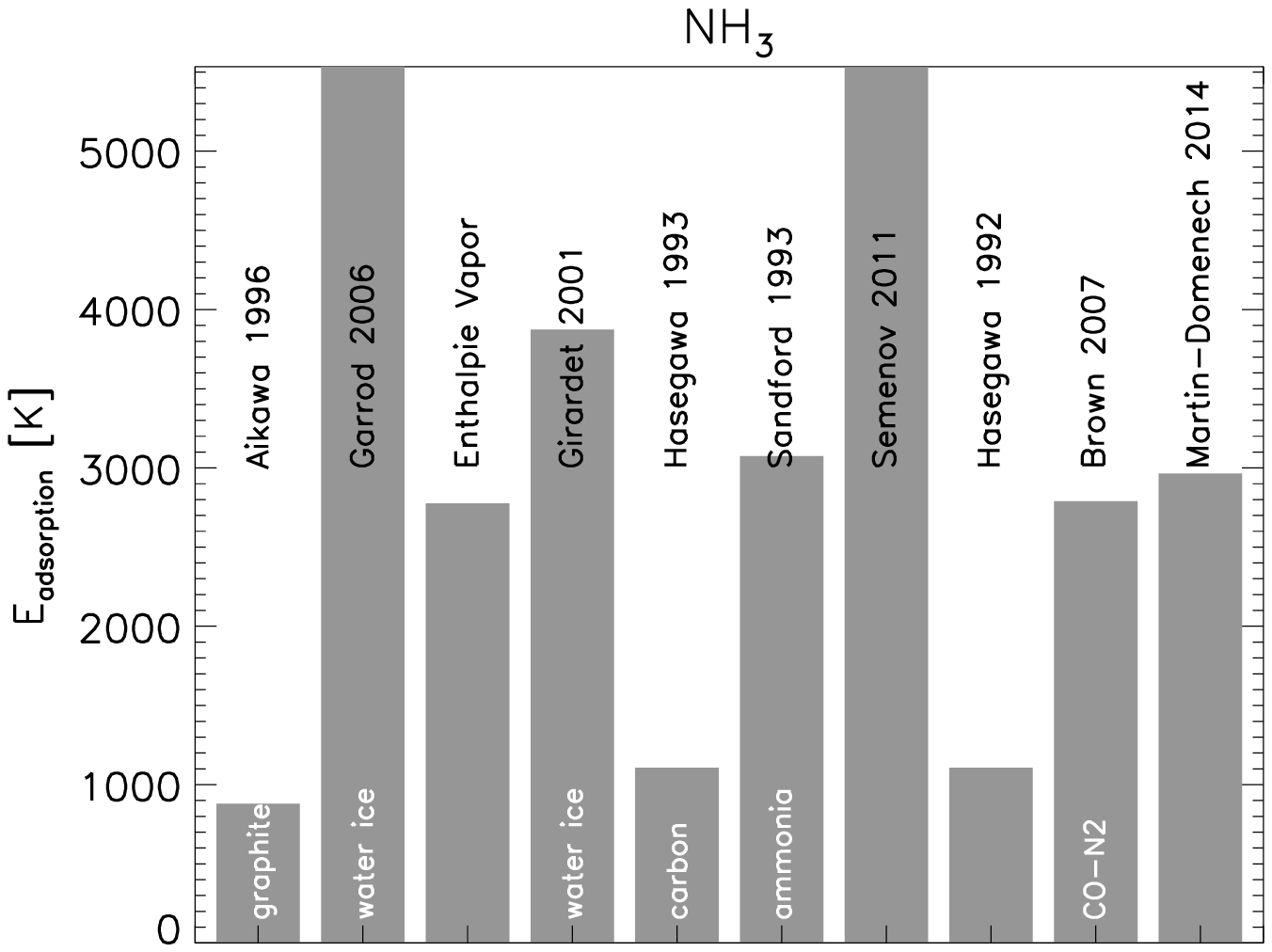}
\vspace*{-11mm}}
{\includegraphics[width=9cm]{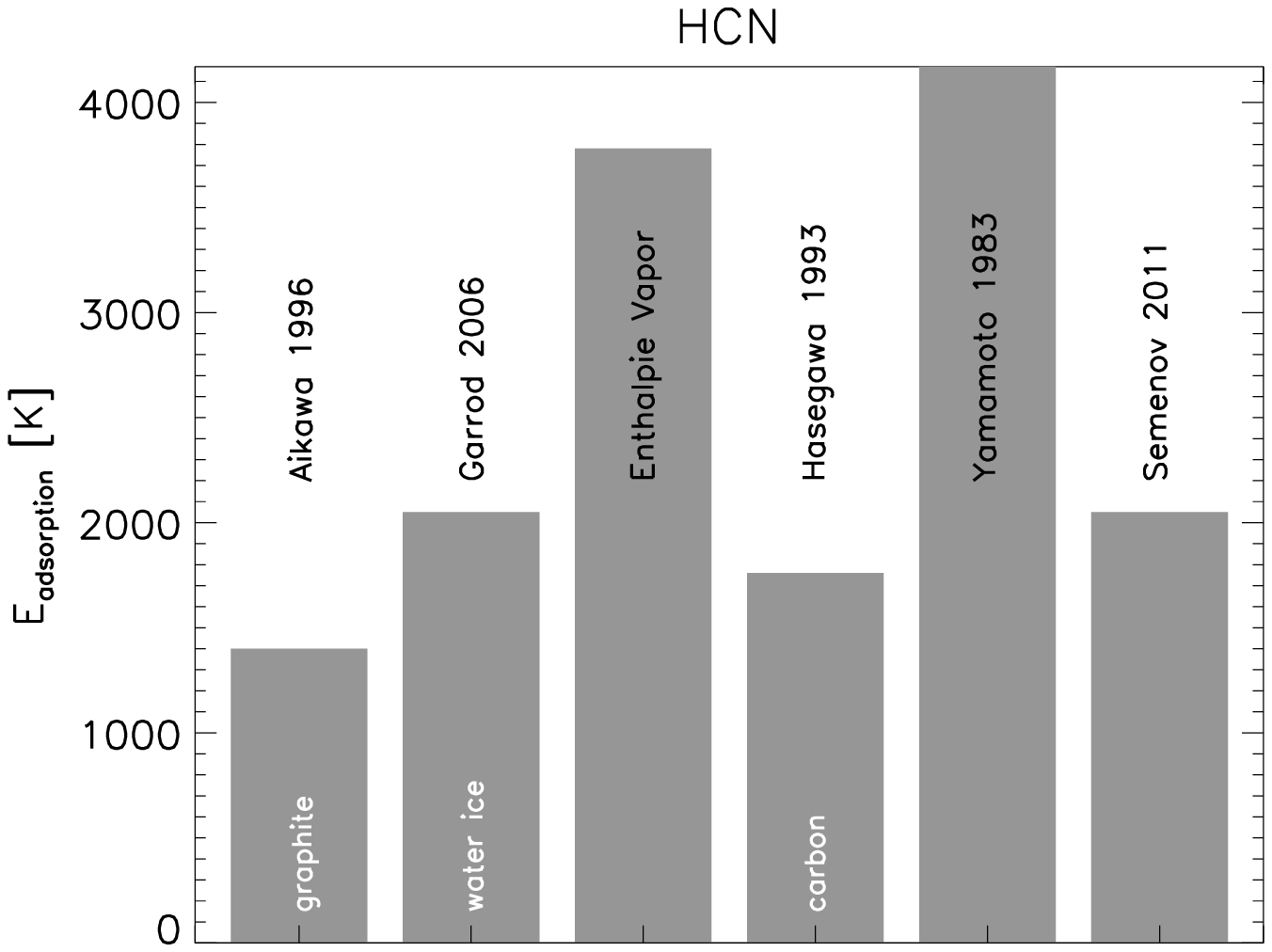}
\vspace*{-3mm}}
\caption{Comparison between various literature values for the adsorption energy of NH$_3$ and HCN. White text indicates the surface on which the adsorption energy was measured, so bare carbonaceous grains, ammonia ice, water ice or an ice mixture; black text are the references.}
\label{fig:Eads-differences}
\end{center}
\end{figure}

Adsorption energies measured in experiments differ largely depending on whether they are measured from ice on top of the same ice, ice mixtures or ice on bare graphite or silicate grains. Fig.~\ref{fig:Eads-differences} shows two examples for values collected for NH$_3$ and HCN ice. The set of adsorption energies from \citet{Aikawa1996} corresponds to bare carbonaceous or silicate surfaces. Alternatively, \citet{Garrod2006} compiled a set of adsorption energies that is valid for non-porous water ice surfaces. Fig.~\ref{fig:Eads-differences} shows that most values found in the literature indeed group around either the low bare grain value or the higher value on water ice. Values for the adsorption of a species on its own ice reside somewhere in between \citep[see e.g.\ NH$_3$ vapor enthalpy and NH$_3$ on ammonia ice,][]{Sandford1993}. UMIST2012 recommends a set of adsorption energies that largely agrees (within $30$\%) with \cite{Garrod2006}. The only exception among the species in common is HCOOCH$_3$ (GH06: $E_{\rm ads}\!=\!6300$~K, UMIST2012: $E_{\rm ads}\!=\!4000$~K). A more general overview of the uncertainties around adsorption energies and a critical review can be found in \citet{Cuppen2017}.
We compare in the following three sets of adsorption energies: \citet{Aikawa1996} (model 1), \citet{Garrod2006} (model 5) and UMIST2012 (model 6).

\begin{figure}[!htbp]
\vspace*{-8mm}
\includegraphics[width=9cm]{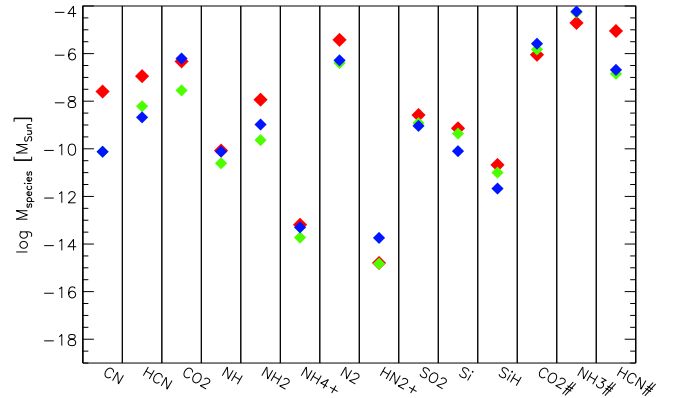}
\caption{Differences in species masses between three different sets of adsorption energies: \citet{Aikawa1996} (red, model 1), \citet{Garrod2006} (green, model 5) and \citet[][UMIST2012]{McElroy2013b} (blue, model 6). Shown are only species that differ by a factor three or more.}
\label{fig:Eads-change-in-speciesmass}
\end{figure}

Differences in the adsorption energies affect most species masses by less than a factor 2-3. However, a few species change by more than a factor 3, some even by an order of magnitude: CN, HCN, CO$_2$, NH$_2$, N$_2$, N$_2$H$^+$, Si, SiH, and HCN ice (Fig.~\ref{fig:Eads-change-in-speciesmass}). The extent of the various ice reservoirs changes from one set of adsorption energies to the other (see Fig.~\ref{fig:Eads-ices}). The most extreme case is NH$_3$ ice where the ice line moves from 40~au \citep[$E_{\rm ads}$ from][]{Aikawa1996} to 0.3~au \citep[$E_{\rm ads}$ from][]{Garrod2006}. Note that a significant change in the ice line does not have to lead to a significant change of the species mass and thus the two provide complementary information. We discuss in the following the processes behind the changes in the chemistry.\\[-5mm]

\begin{itemize}
\item {\it CO$_2$:} The snow line limits the radial extent of the CO$_2$ ring in the midplane of the disk, where water is not yet completely frozen onto the cold dust grains. The snow line changes from $\sim\!1$~au to $\sim\!0.3$~au for the two different values of $E_{\rm ads}$ for water, 4800~K and 5700~K. However, none of this affects the water ice reservoir since that is dominated by the mass in the outer disk.\\[-2mm]
\item {\it CN, HCN, NH$_2$, N$_2$:} The adsorption energy of NH$_3$ ice determines the radial ice line for this species. For $E_{\rm ads}\!=\!880$~K (Aikawa), the NH$_3$ ice reservoir extends from 40 to 200~au and nitrogen does not fully condense into NH$_3$ ice. For $E_{\rm ads}\!=\!5530$~K (GH06, UMIST2012), all nitrogen is bound in NH$_3$ ice between 0.3 and 200~au. With a low adsorption energy, sufficient nitrogen remains in the gas phase between 1 and 10~au to form CN, HCN, NH$_2$, and N$_2$. Fig~\ref{fig:Eads-change-in-speciesmass} shows that all these species have lower masses in case of the higher adsorption energy (GH06, UMIST2012), while the mass of NH$_3$ ice increases by a factor three.\\[-2mm]
\item {\it N$_2$H$^+$:} The ice line for several ice species, particularly also CO, shifts upward beyond 20~au, if the UMIST2012 adsorption energies are used instead of the \citet{Aikawa1996} ones. This increases the abundance of N-bearing species in the region which is oxygen poor. Since N$_2$H$^+$ resides in a thin layer at the disk surface, this extra reservoir causes an increase in mass by a factor $\sim\!10$. 
\item {\it Si, SiH:} The change in the mass of Si and SiH is related to CO$_2$. In models with a low adsorption energy for water, the CO$_2$ ring is extended out to a few AU. In this ring, CO$_2$ reacts with Si to form SiO which subsequently freezes out onto the cold dust grains, driving Si into SiO ice at the expense also of the SiH abundance. In the models with a high adsorption energy, Si remains atomic out to a few~au. The SiO that forms is efficiently destroyed by reactions with C$^+$ into Si$^+$ which is then subsequently neutralized by charge exchange with Mg and Na. \\[-2mm]
\end{itemize}

\begin{figure*}[!htbp]
{\includegraphics[width=9cm]{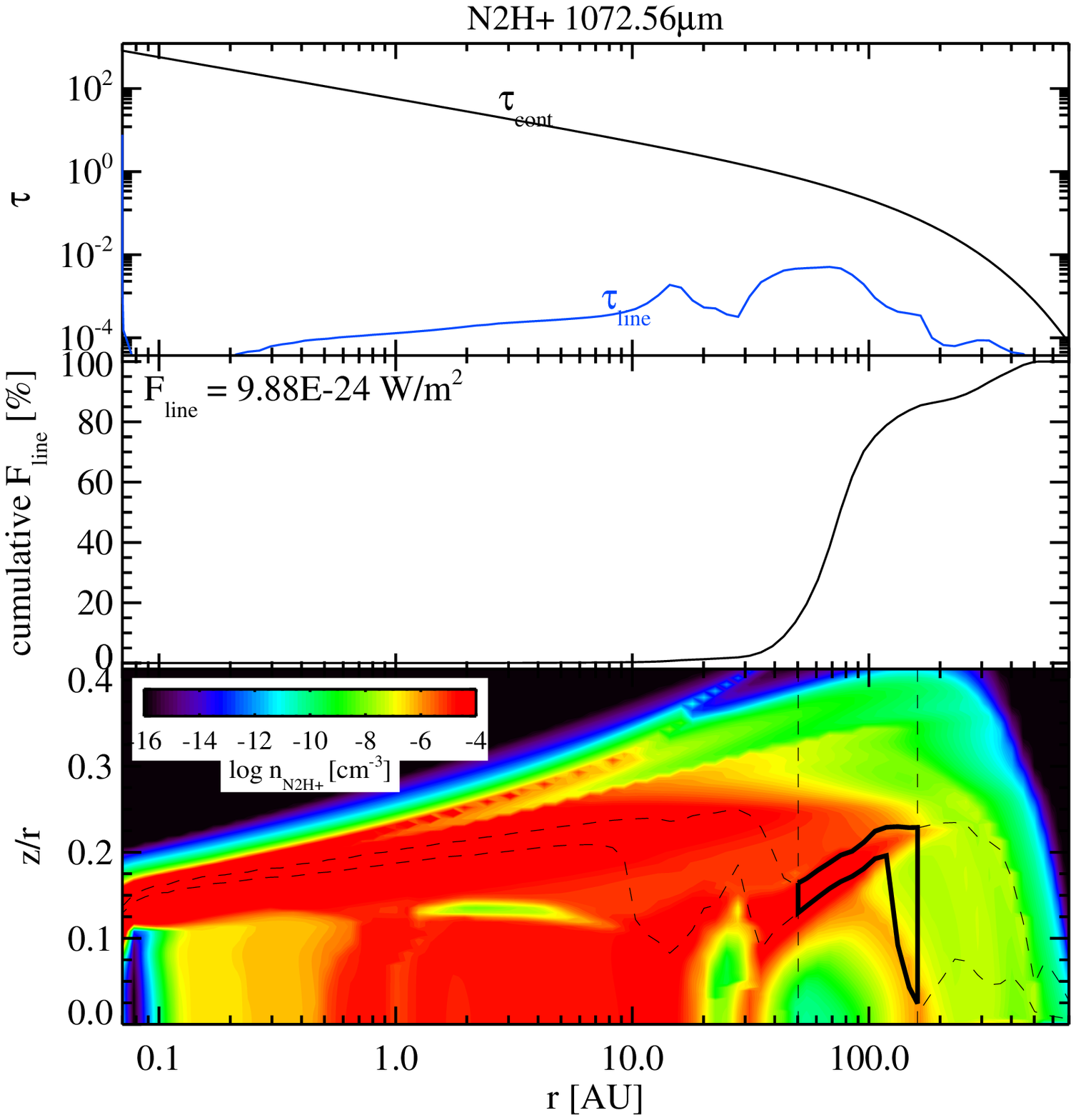}
\includegraphics[width=9cm]{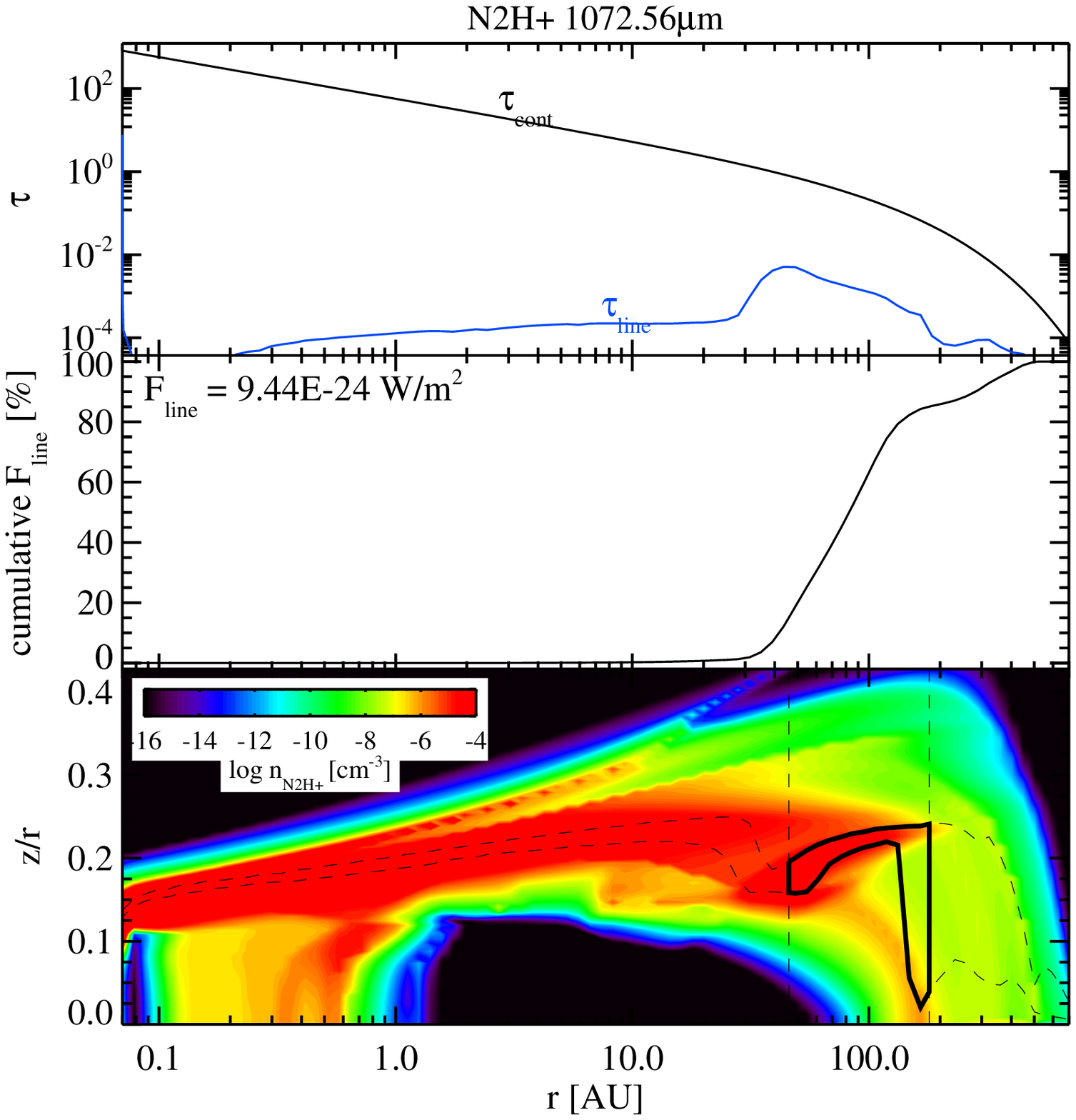}
\vspace*{3mm}}
\includegraphics[width=9cm]{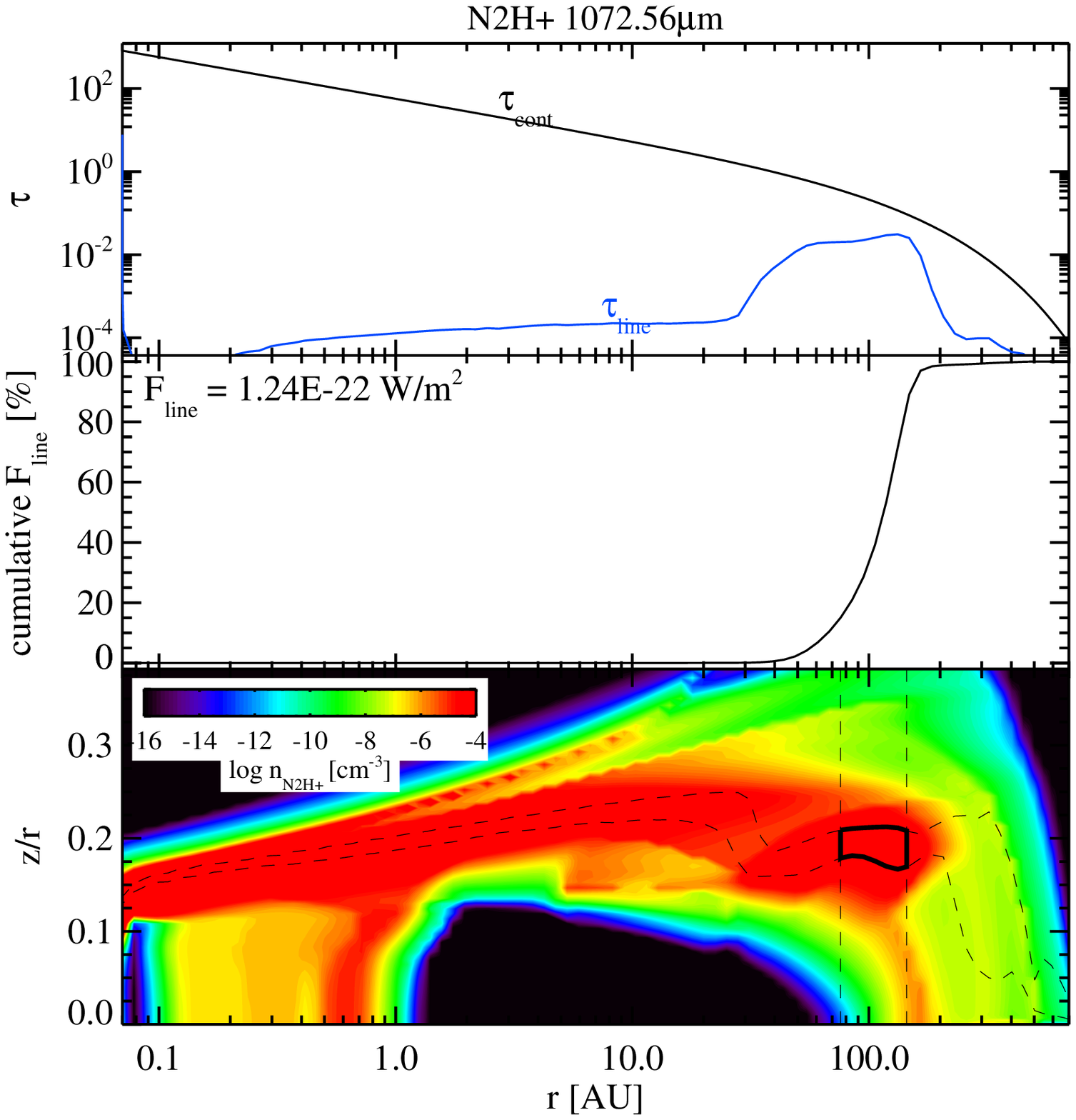}
\includegraphics[width=9cm]{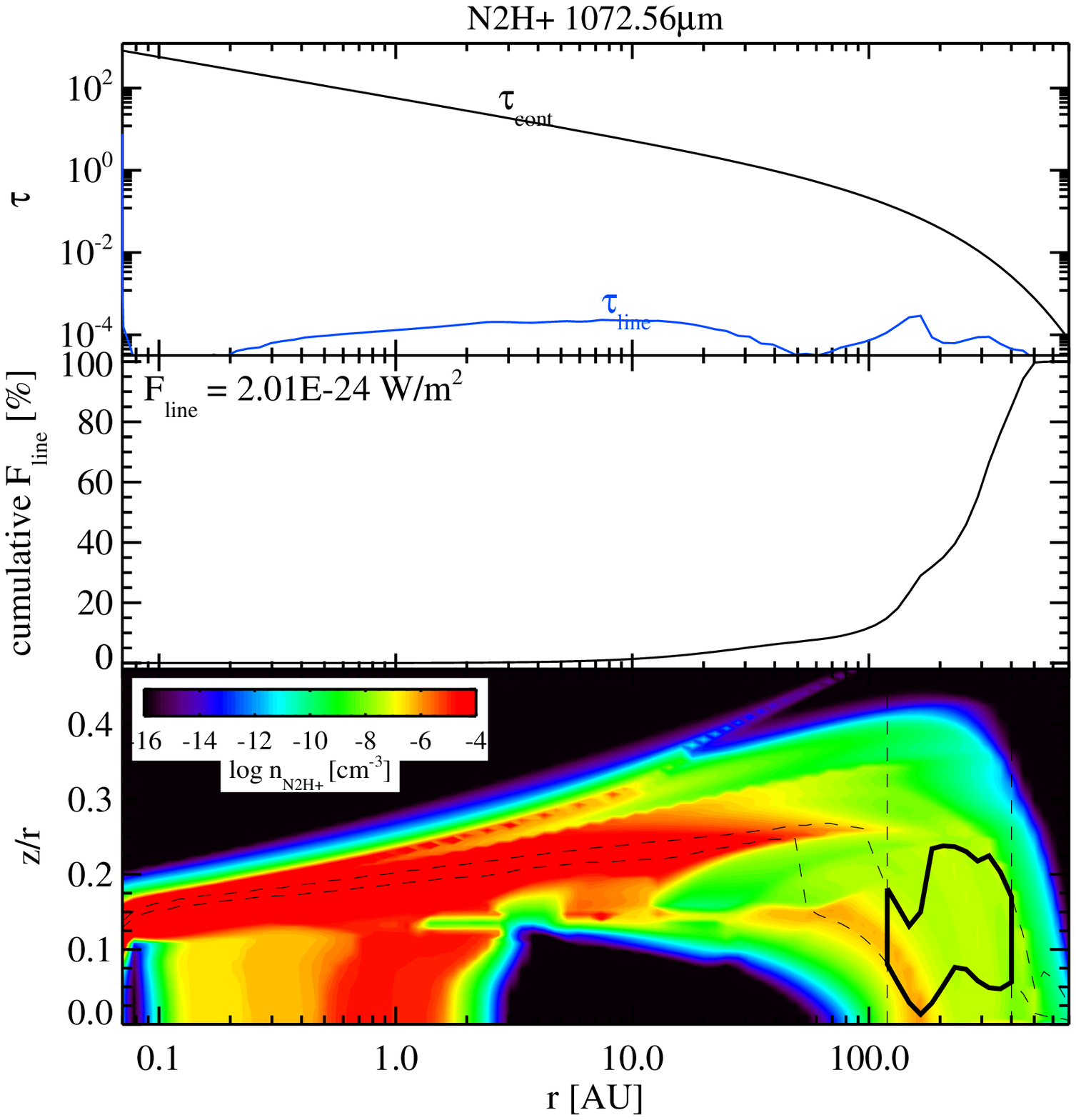}
\caption{N$_2$H$^+$ $J\!=\!3-2$ line (1072.56~$\mu$m) from vertical escape for the four models from top left to bottom right: The standard disk model using the set of adsorption energies from \citet{Aikawa1996} (model 1), \citet{Garrod2006} (model 5), UMIST2012 (model 6), $T$-dependent adsorption rates (model 7). The three panels show the optical depth in the line and the continuum, the cumulative line flux as a function of radius and the box (thick black line) in which 50\% of the line flux originates (15-75\% radially and vertically - dashed black lines) on the color background of the N$_2$H$^+$ density distribution.}
\label{fig:Eads-N2Hplusline}
\end{figure*}

\begin{figure}[!htbp]
\includegraphics[width=4.4cm]{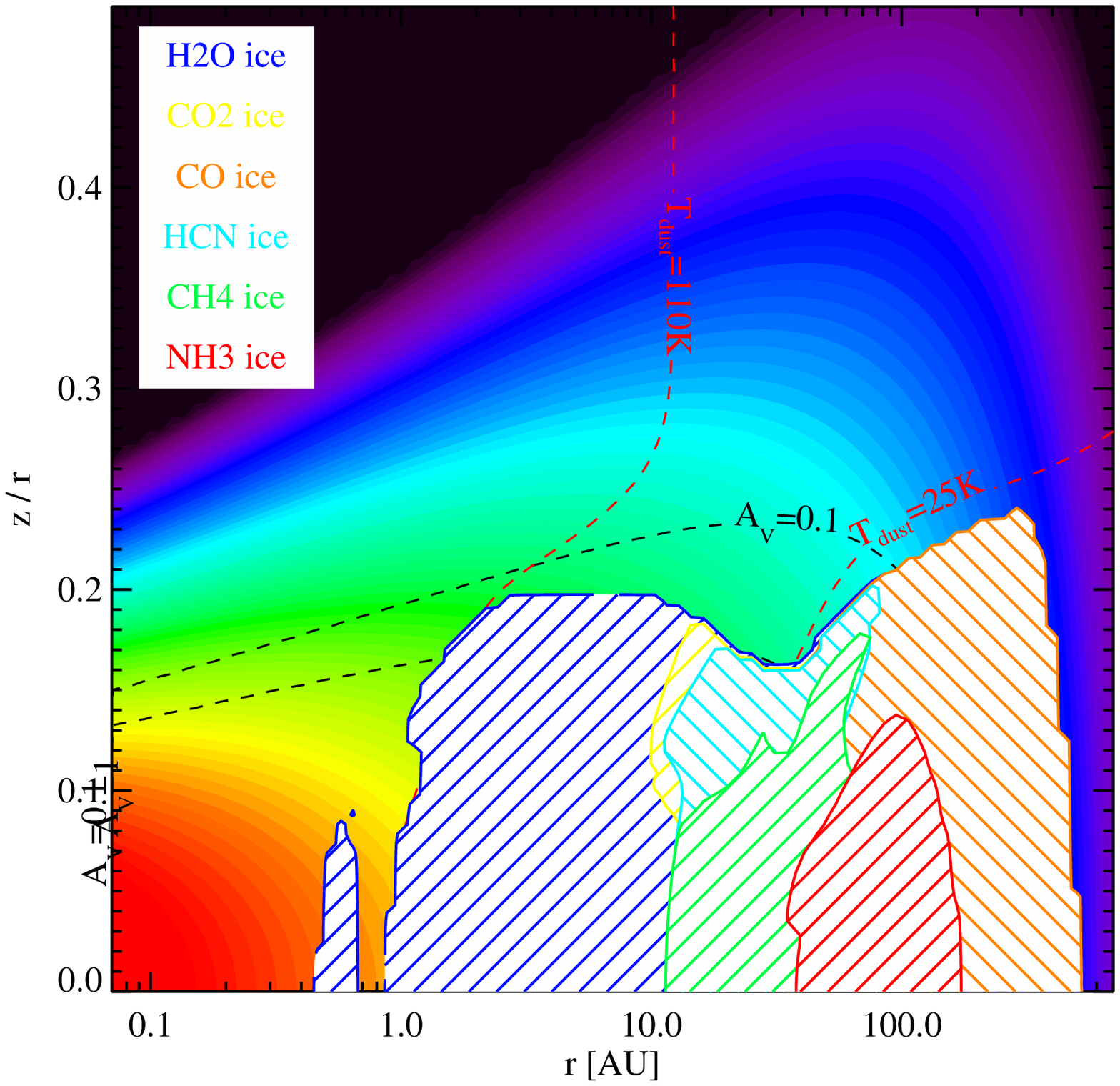}
\includegraphics[width=4.4cm]{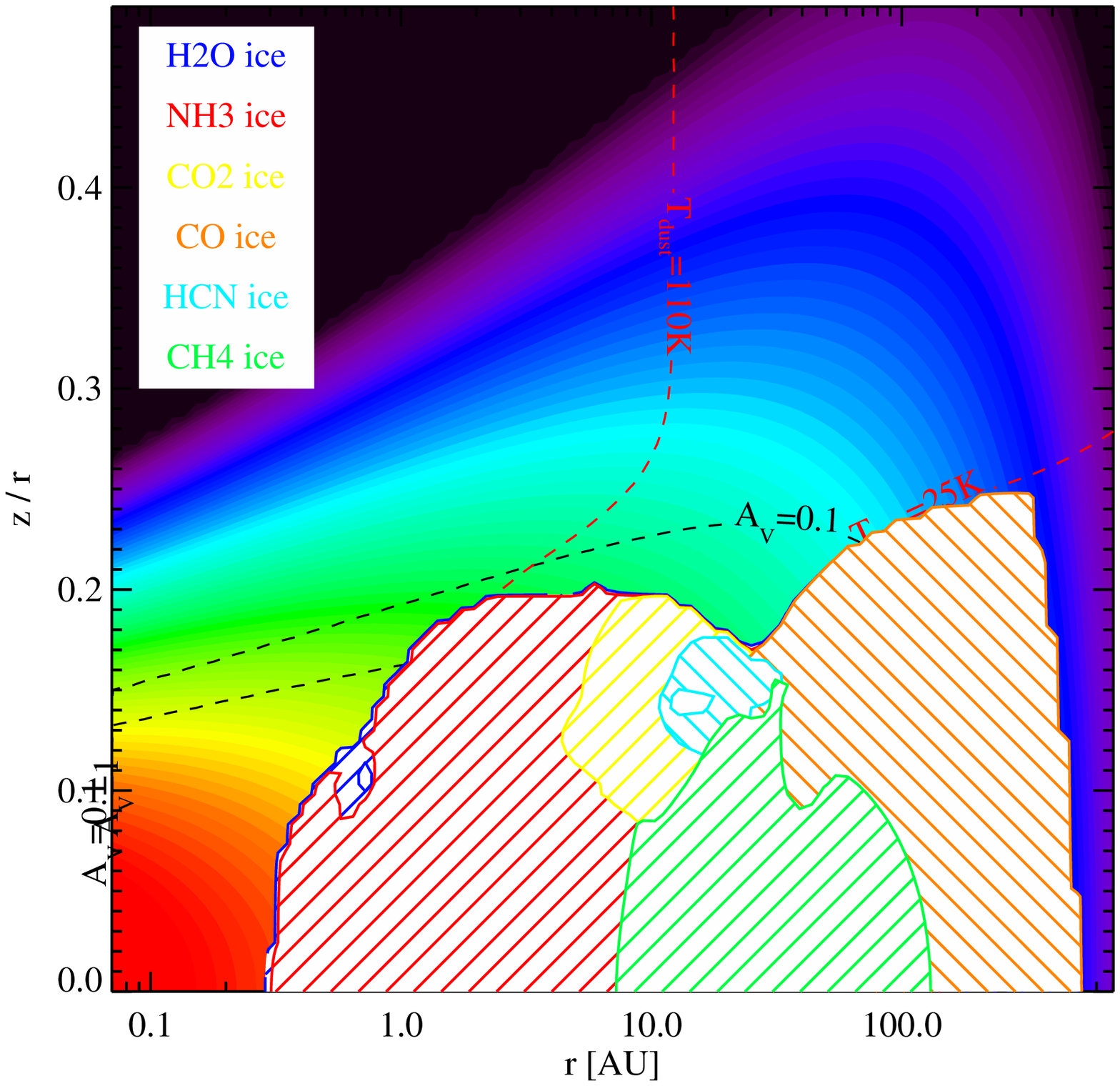}
\includegraphics[width=4.4cm]{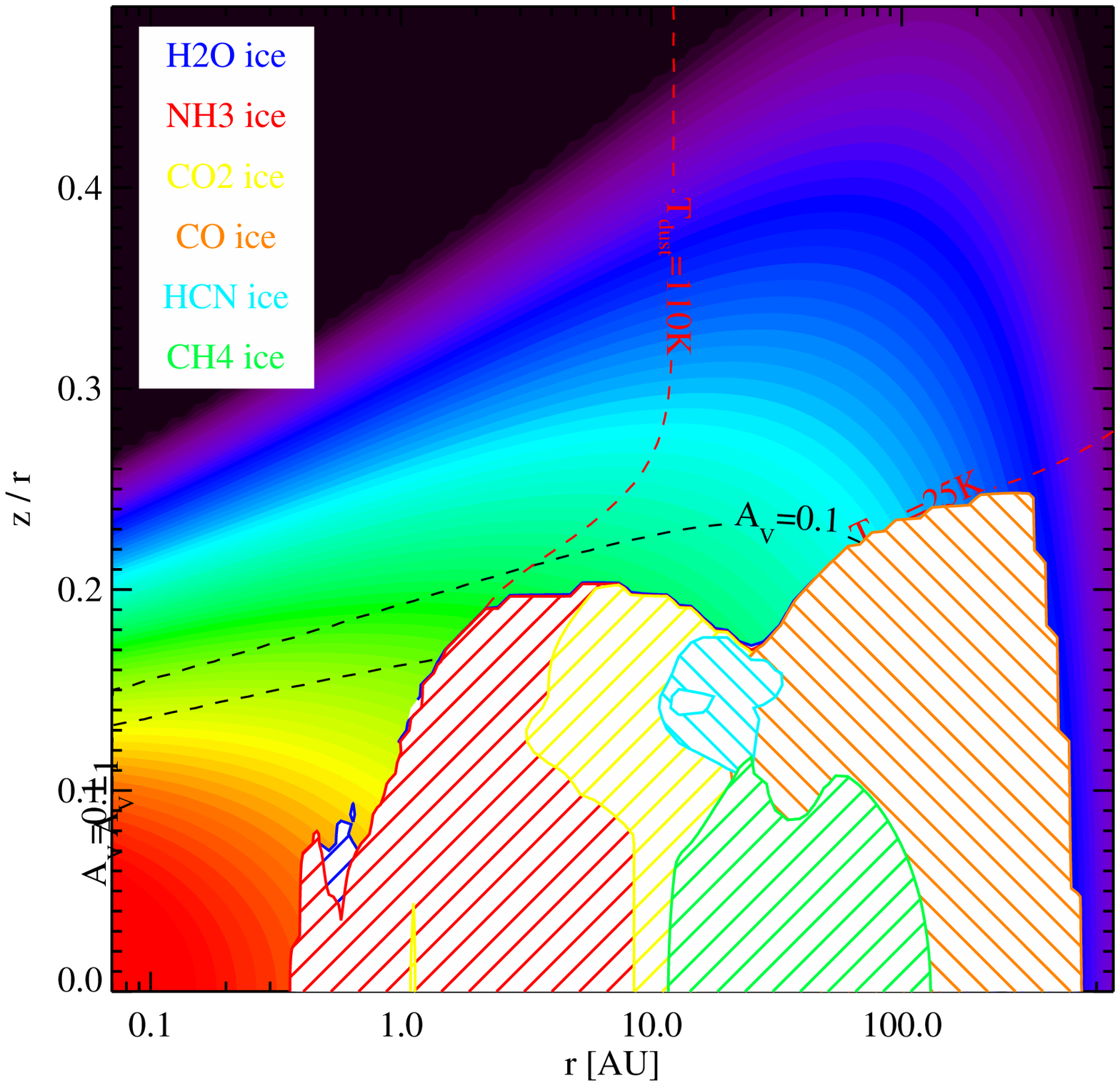}
{\hspace*{1mm}\includegraphics[width=4.4cm]{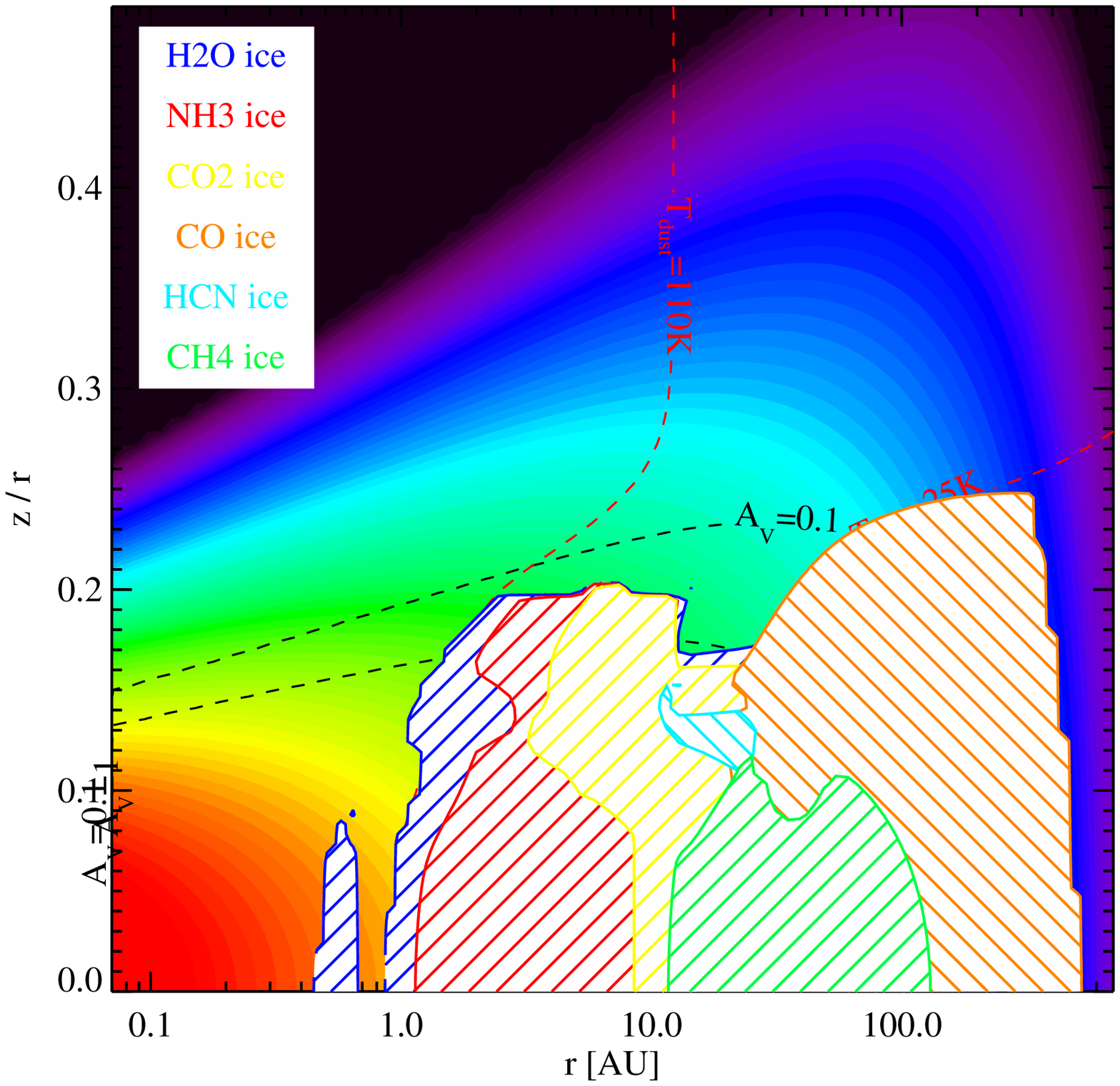}}
\caption{Distribution of ice reservoirs plotted on top of each other in order from top to bottom of legend. Note that some ices may be partially hidden behind others. The order but not the color scheme is changed for the upper left panel to make the ammonia ice visible. From top left to bottom right: The standard disk model using the set of adsorption energies from \citet{Aikawa1996} (model 1), \citet{Garrod2006} (model 5), UMIST2012 (model 6), $T$-dependent adsorption rates (model 7). The color scale in the background shows the total hydrogen number density in the disk model and the black dashed lines the $A_{\rm V}\!=\!0.1$ and 1 contours (minimum of radial and vertical $A_{\rm V}$).}
\label{fig:Eads-ices}
\end{figure}

Despite the fact that several gas species masses change drastically, none of the lines studied here is affected by more than a factor 2 except the N$_2$H$^+$ line (Fig.~\ref{fig:Eads-lines-1} and \ref{fig:Eads-lines-2}). Most of the lines are optically thick and/or originate in the disk surface and do not trace the chemistry changes occurring typically closer to the midplane. The only lines changing by a factor 2 are the optically thin lines of C$^{18}$O ($1365.42$ and $910.31~\mu$m), and the CN line ($528.78~\mu$m). The N$_2$H$^+$ line at $1072.56~\mu$m increases by more than a factor 10 if the UMIST2012 adsorption energies are used; this is related to the increase in N$_2$H$^+$ mass as shown in Fig.~\ref{fig:Eads-change-in-speciesmass} due to the change in N$_2$ absorption energy (GH06: 1000~K, UMIST: 790~K). Apart from the line flux changes, Fig.~\ref{fig:Eads-N2Hplusline} shows that the N$_2$H$^+$ density distributions and therefore also emitting regions in the disk change depending on the set of adsorption energies used.

\begin{table}
\caption{Adsorption energies used in the two temperature regimes: bare grains and polar ices.}
\begin{tabular}{l|rlrl}
\hline
 ice                    & \multicolumn{4}{c}{$E_{\rm ads}$ [K]} \\[1mm]
species & $T\!>\!110$~K & Ref. & $110 \! \le\!T\!\le\!10$~K & Ref. \\
\hline
\hline
CO          & 960 & A96 & 1150 & UMIST2012 \\
H$_2$O & 4800 & H09 & 5700  & GH06 \\
CO$_2$ & 2000 & A96 & 2990  & UMIST2012\\
CH$_4$ & 1360 & HH93 & 1090  & UMIST2012\\
NH$_3$ & 880 & A96 & 3874  & G01 \\
SO$_2$ & 2400 & A96 & 5330 & UMIST2012\\
O$_2$   & 960 & as CO & 1000 & UMIST2012\\
HCN      & 1400 & A96 & 2050 & UMIST2012\\
N$_2$   & 660 & scaled CO & 1870  & G01\\
\hline
\end{tabular}
\tablefoot{Abbreviations for references can be found in Table~\ref{Tab:diffUMIST2006}.}
\label{tab:T-dependent-Eads}
\end{table}

It is reasonable to assume that the adsorption energy of a specific molecule will depend on the surface property of the grain, i.e\ bare surfaces and/or the polarity of the ice. Hence, we ran an additional model in which we vary the adsorption energy as a function of temperature (model 7). For that we assume two temperature intervals: (1) bare grain surface values for $T\!>\!110$~K, and (2) polar ice values for 110~K\,$\le\!T\!\le\!10$~K. Table~\ref{tab:T-dependent-Eads} summarizes the values and regimes for the ices used in the small chemical network. The N$_2$ adsorption energy is now scaled by a factor 0.7 with respect to the one by \citet{Aikawa1996}; such a scaling has already been proposed by \citet{Bergin1997}. \citet{Ceccarelli2005} show that such a scaling is required to match H$_2$D$+$ observations and Rab et al.\ (in preparation) show that it matches typical N$_2$H$^+$ line fluxes from disks. Again, the disk density and thermal structure is kept constant.

\begin{figure}[!htbp]
\includegraphics[width=9cm]{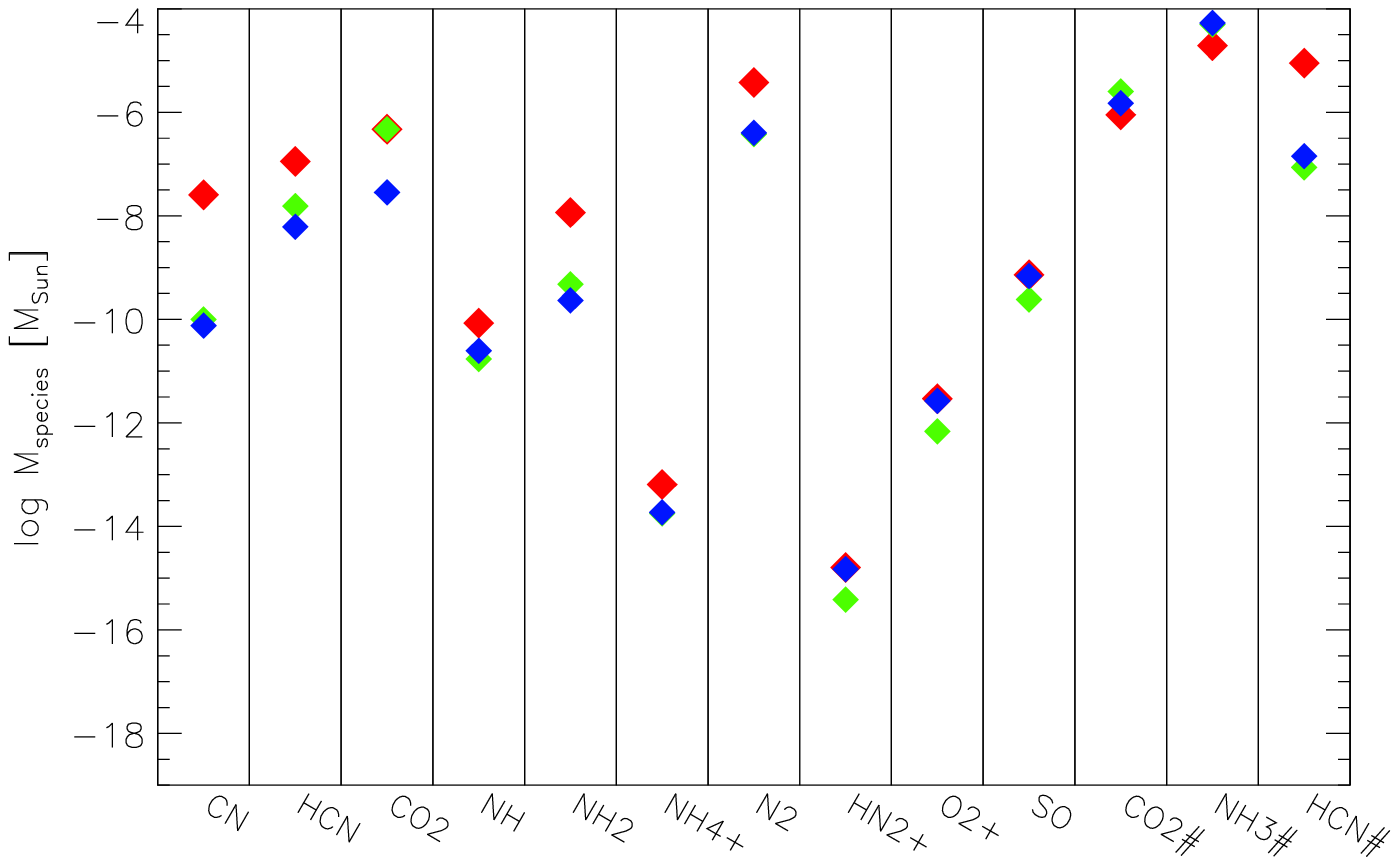}
\caption{Differences in species masses between $E_{\rm ads}$ of \citet[][red, model 1 - bare grains]{Aikawa1996}, temperature-dependent adsorption energies (green, model 7) and $E_{\rm ads}$ of \citet[][blue, model 5 - water ice]{Garrod2006}.}
\label{fig:EadsT-change-in-speciesmass}
\end{figure}

Fig.~\ref{fig:EadsT-change-in-speciesmass} shows the change in species masses with respect to the Aikawa adsorption energy set for those species that change by a factor three or more: CN, HCN, CO$_2$, NH, NH$_2$, NH$_4^+$, N$_2$, N$_2$H$^+$, O$_2^+$, SO, CO$_2$ ice, NH$_3$ ice and HCN ice. The species masses of the temperature-dependent case sometimes follow the bare grain case and sometimes the water ice surface case. Species with high abundances in the inner disk such as CO$_2$ stay close to the results from bare grains since this is in fact the adsorption energy that governs their behavior in the temperature-dependent case. Most other species stay close to the results from water ice surfaces since they are dominated by the behavior in the outer disk where grains are covered by water ice. A few species deviate from this, N$_2$H$^+$, O$_2^+$ and SO. In these three cases, the temperature dependent adsorption energies always yield smaller species masses than any of the other two models. This is related to the higher value of N$_2$ adsorption energy in the temperature range $110\! \leq\! T\!\leq\!10$~K \citep[][on water ice]{Girardet2001}, which allows nitrogen to deplete from the gas phase at smaller radii than in the other two models. This impacts many nitrogen bearing species, but also those which form through nitrogen chemistry including for example gas phase water beyond 10~au.

None of the lines in our selection changes by more than a factor two with respect to the bare grain case \citep{Aikawa1996} except the N$_2$H$^+$ line (see Fig.~\ref{fig:Eads-lines-3}). The latter becomes a factor six weaker in the case of temperature-dependent adsorption energies. The largest changes in the chemical composition are seen in the ice reservoir inside 100~au (Fig.~\ref{fig:Eads-ices}); however most lines originate well above the surface ice line.

\subsection{Reactions of excited H$_2$}
\label{Sect:resultsH2EXC}

We discuss the compilation of reaction rates for excited H$_2$ (denoted throughout the rest of the paper as H$_2^*$) in Appendix~\ref{App:chem-not-in-UMIST}, where we assume a representative excitation state of $v\!=\!1$ ($E\!=\!5980$~K). It is assumed that 90\% of the UV absorption leads to excited H$_2$, 10\% to dissociation \citep{Tielens1985}.

We find from our standard model that the molecular ions CH$^+$ and HCO$^+$ are the most affected species. In both cases, the key reaction is 
\begin{equation}
{\rm H}_2(v\!=\!1) + {\rm C^+} \rightarrow {\rm CH^+} + {\rm H} \,\,\, .
\end{equation}
We use here the updated rate from Eq.(\ref{Eq:Zanchet-rate}) as explained in the Appendix. The mass of CH$^+$ increases by a factor $\sim\!10$ due to the presence of excited H$_2$ chemistry. All other species masses change by less than 10~\%. The change in CH$^+$ abundance transmits through various channels into CH$_2^+$, CH$_3^+$ and all those molecular ions have a pathway to form HCO$^+$
\begin{eqnarray}
{\rm CH^+} + {\rm H_2O} & \rightarrow & {\rm HCO^+} + {\rm H_2} \\
{\rm CH_2^+} + {\rm O} & \rightarrow & {\rm HCO^+} + {\rm H} \\
{\rm CH_3^+} + {\rm O} & \rightarrow & {\rm HCO^+} + {\rm H_2} \,\,\, .
\end{eqnarray}
However, there are also many alternative pathways forming and destroying HCO$^+$ that do not involve H$_2^*$. Hence, the species changes much less than CH$^+$.

\begin{figure}[!thbp]
\includegraphics[width=4.4cm]{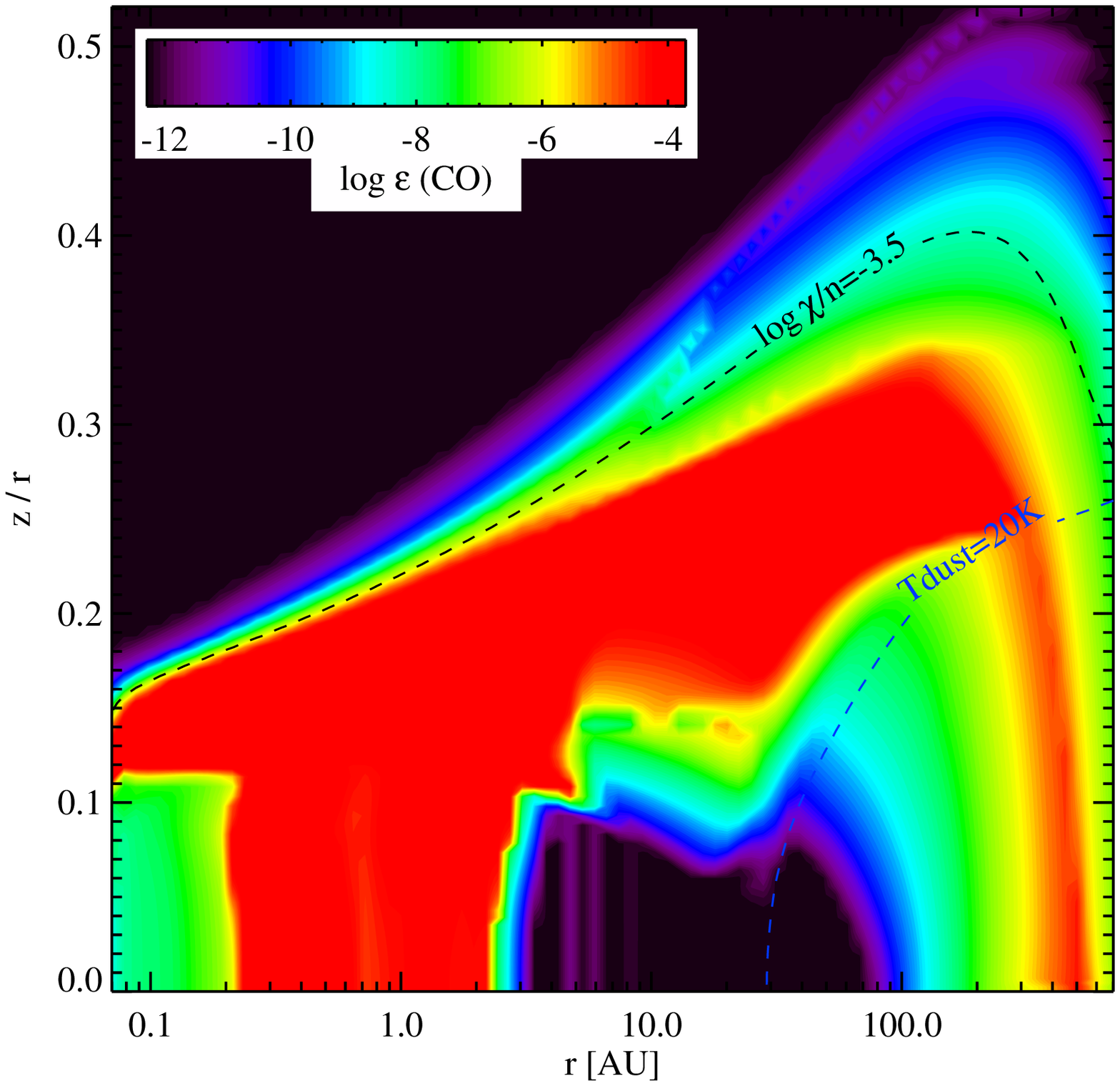}
\includegraphics[width=4.4cm]{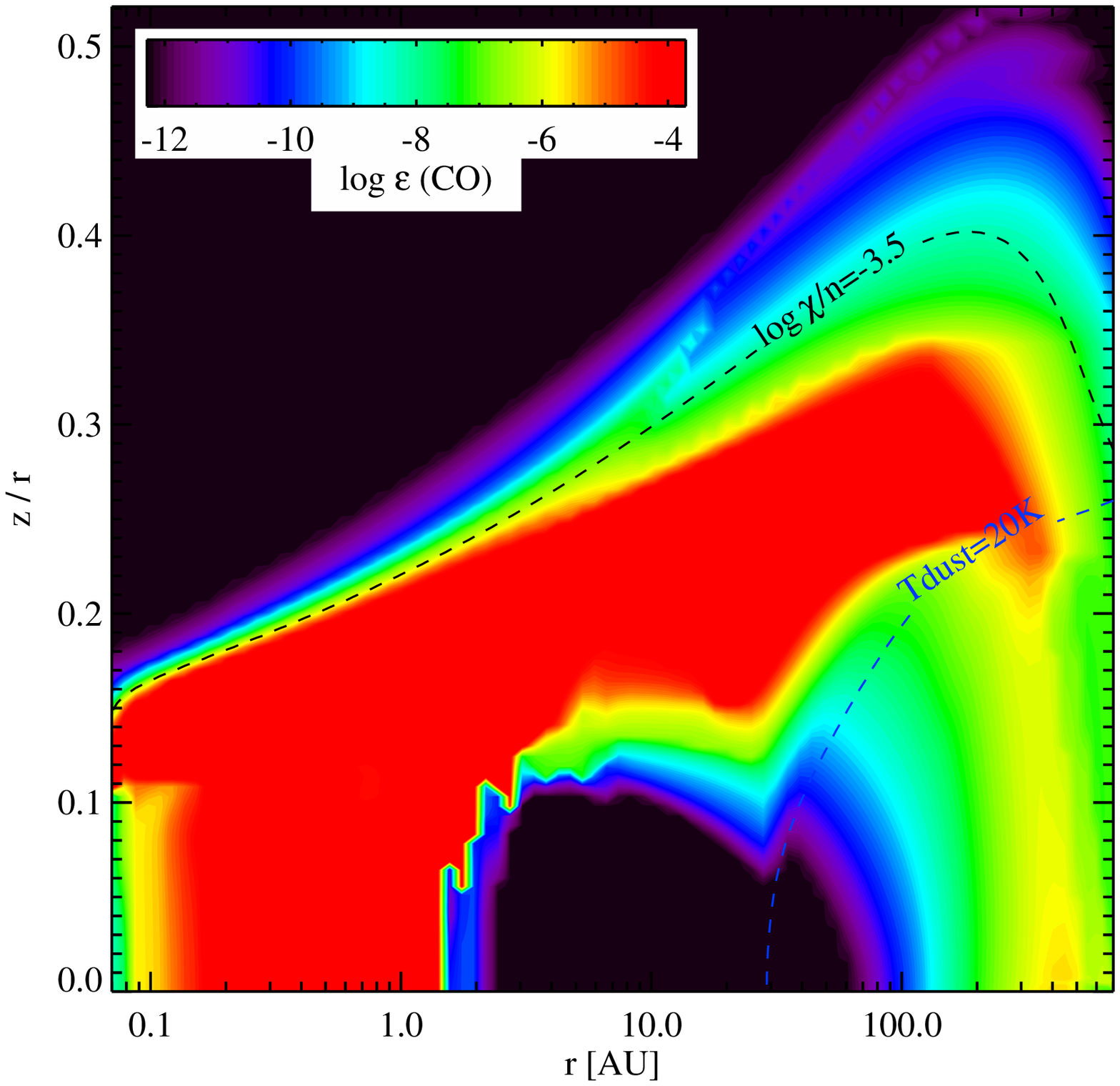}
\includegraphics[width=4.4cm]{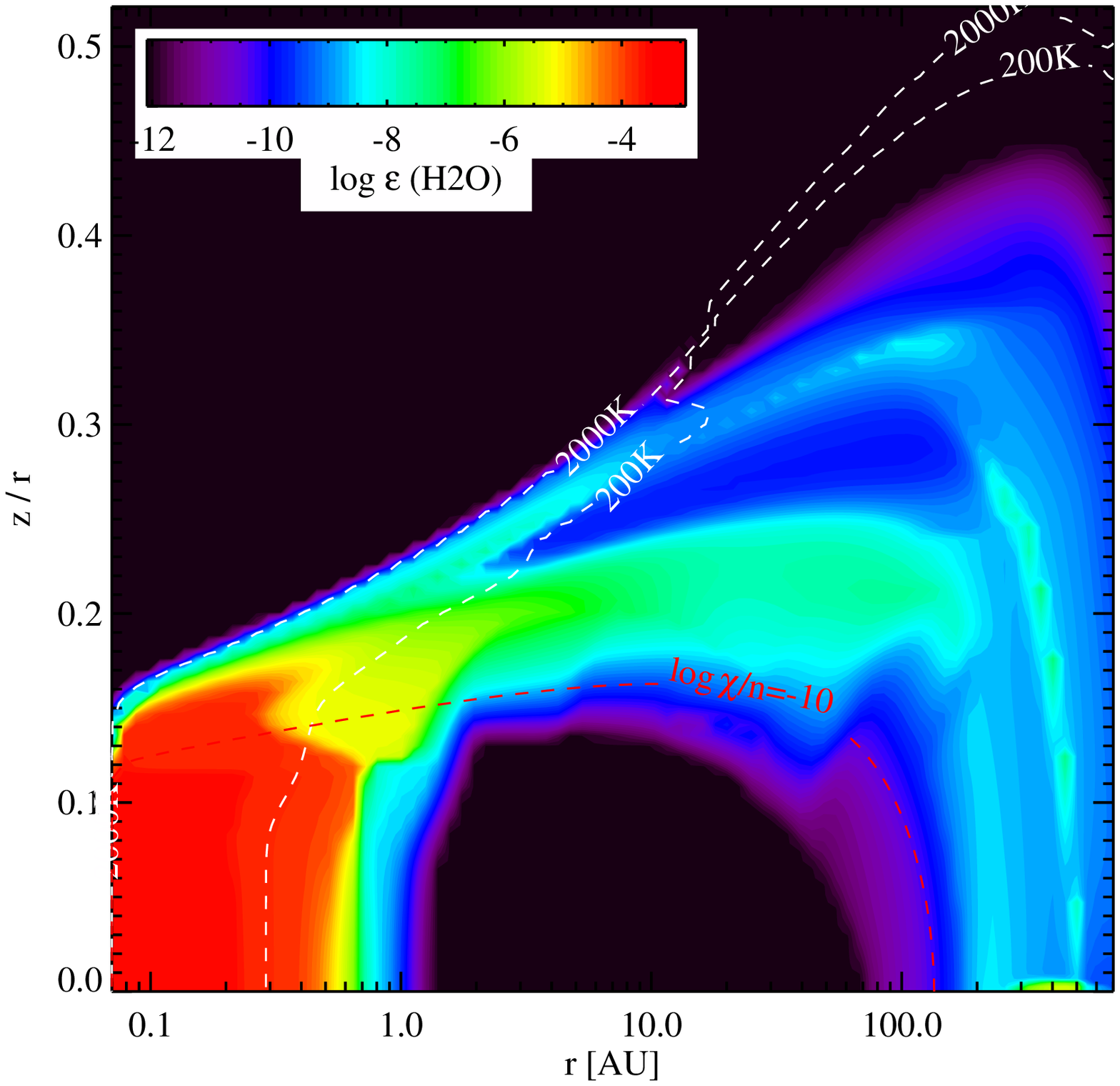}
\includegraphics[width=4.4cm]{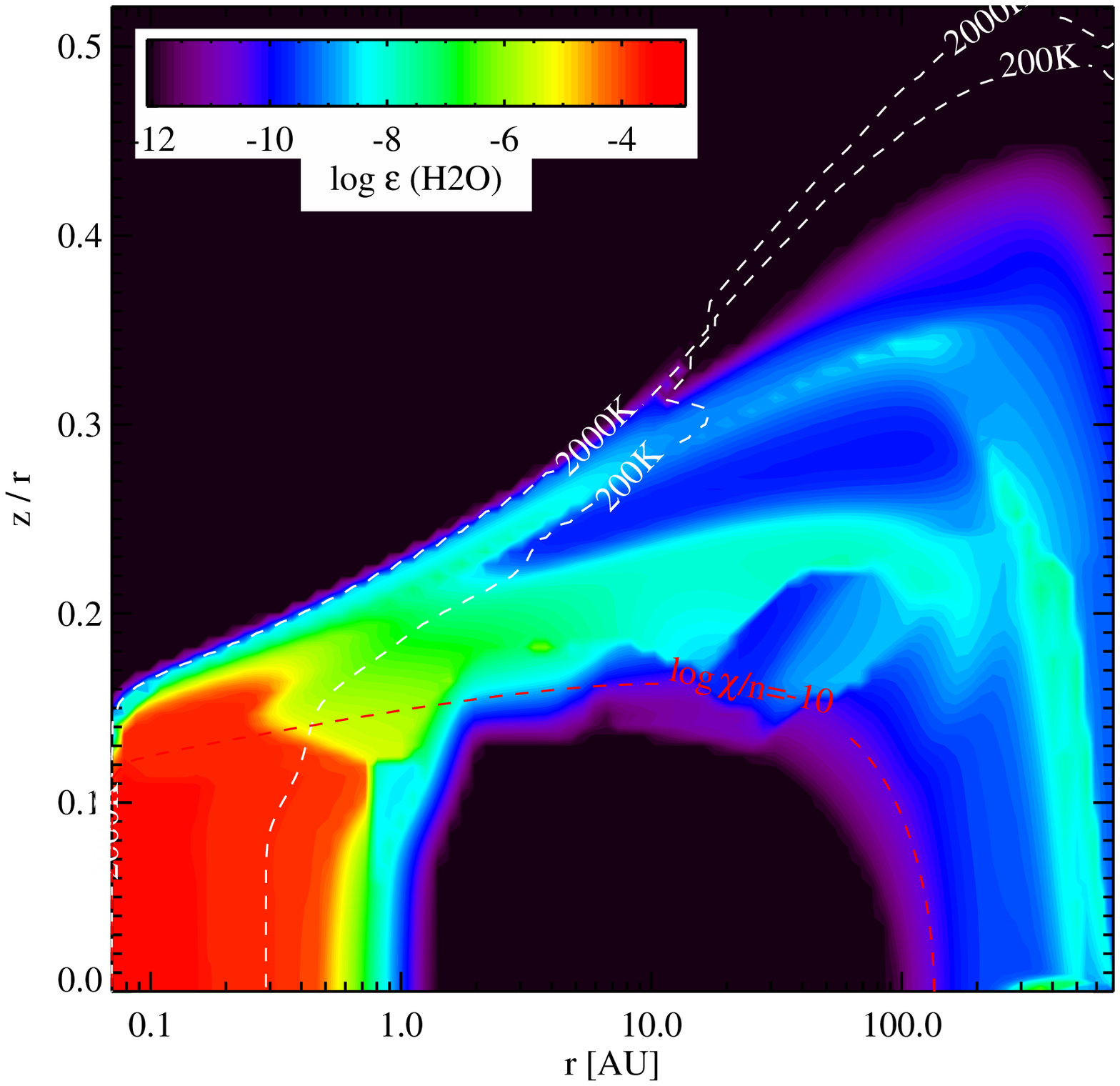}
\includegraphics[width=4.4cm]{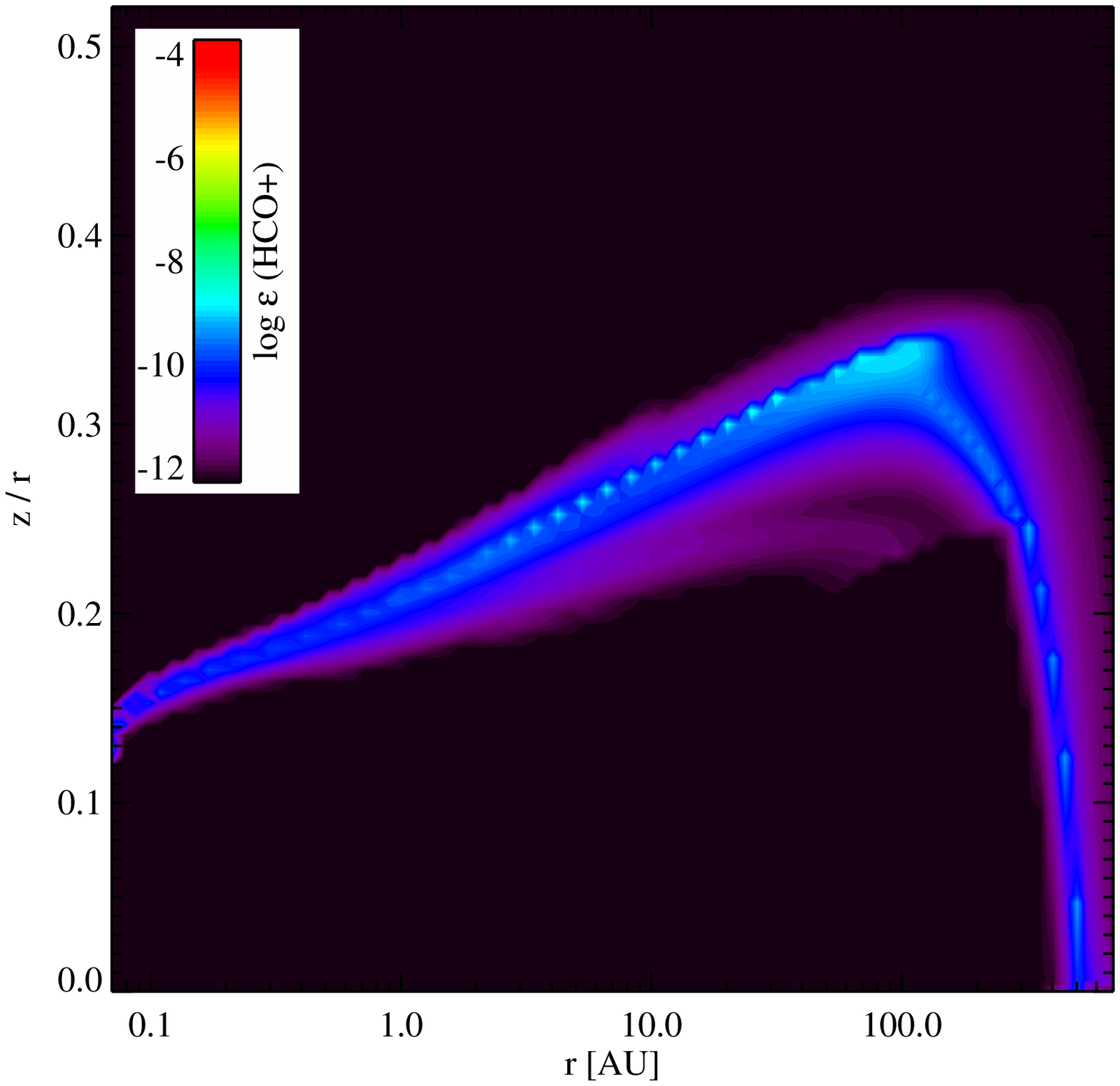}
\includegraphics[width=4.4cm]{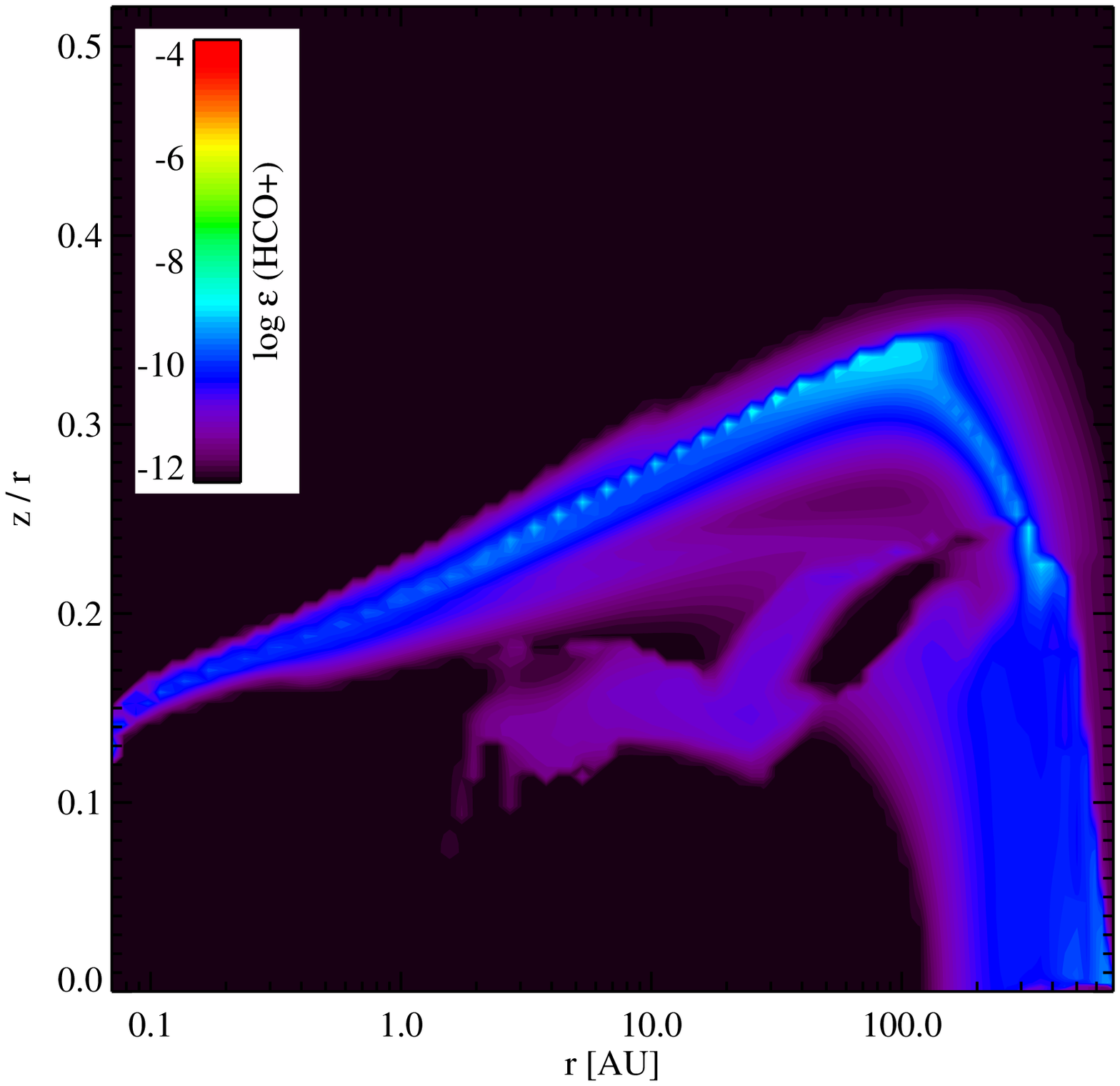}
\includegraphics[width=4.4cm]{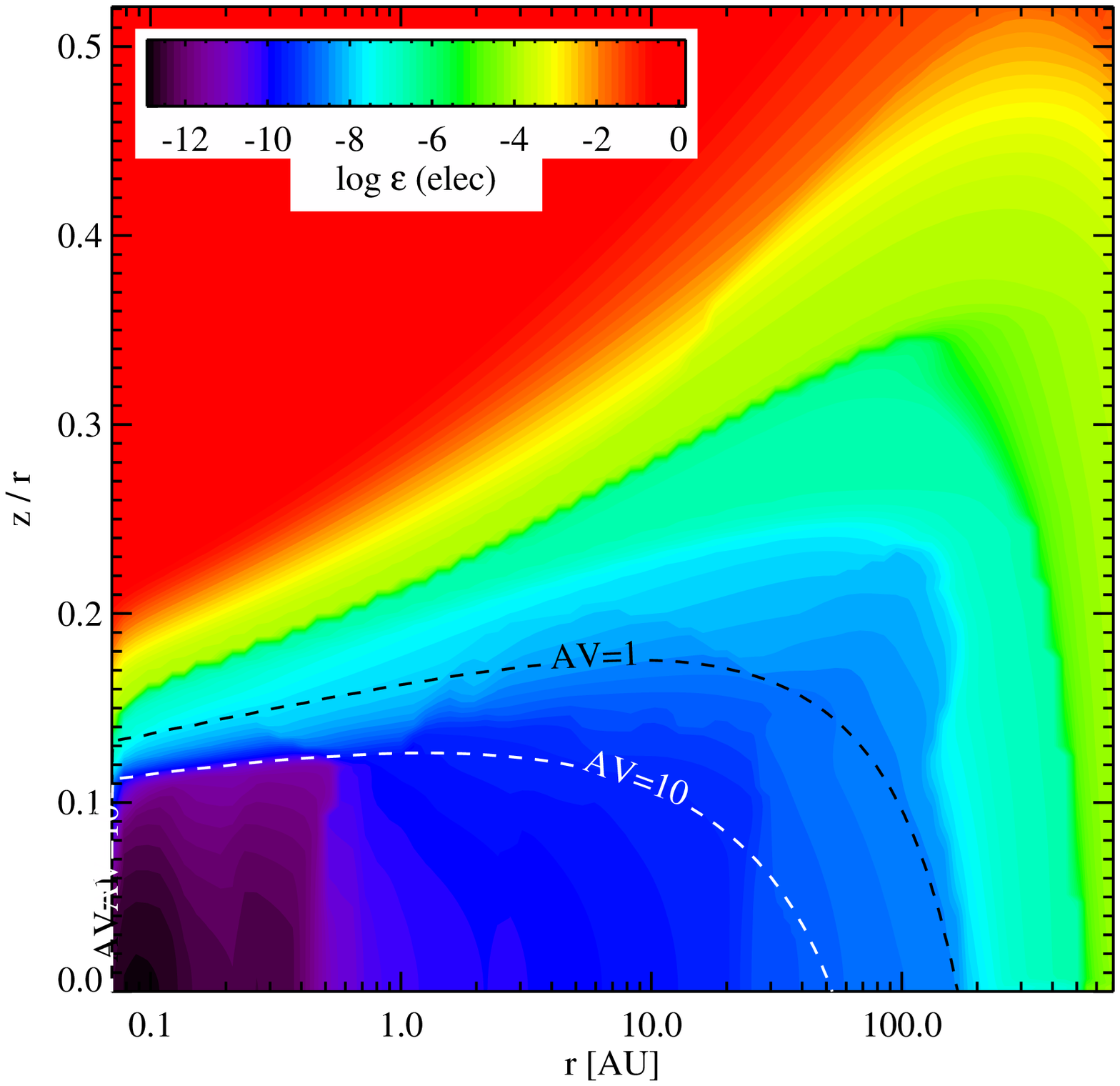}
{\hspace*{2mm}\includegraphics[width=4.4cm]{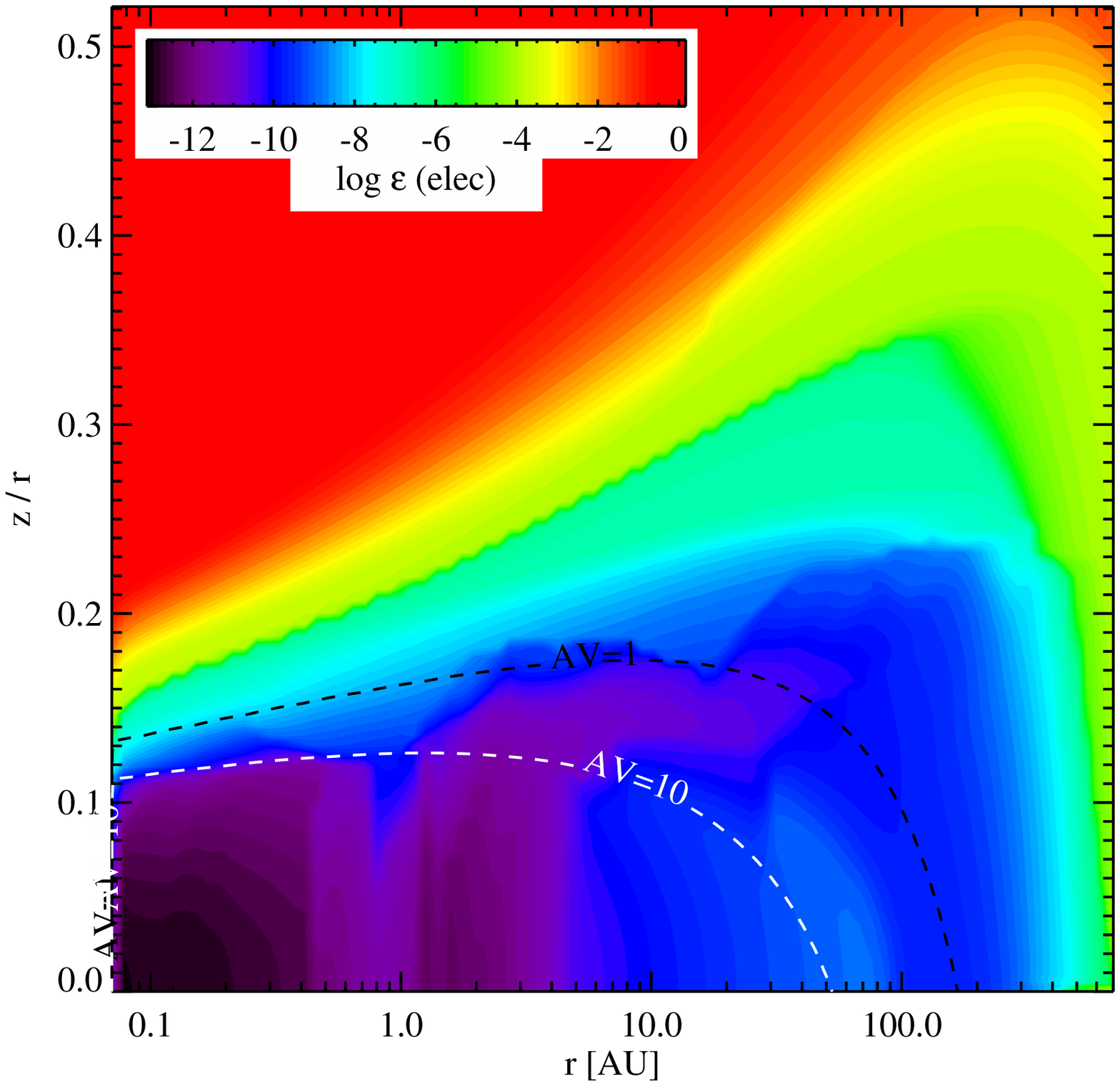}}
\caption{Distribution of key species abundances using the small (left) and large (right) chemical network: CO, H$_2$O, HCO$^+$, and electrons. Contours are the same as Fig.~\ref{fig:base-model}.}
\label{fig:Chemistrylargenetwork1}
\end{figure}

\begin{figure}[!thbp]
\includegraphics[width=4.4cm]{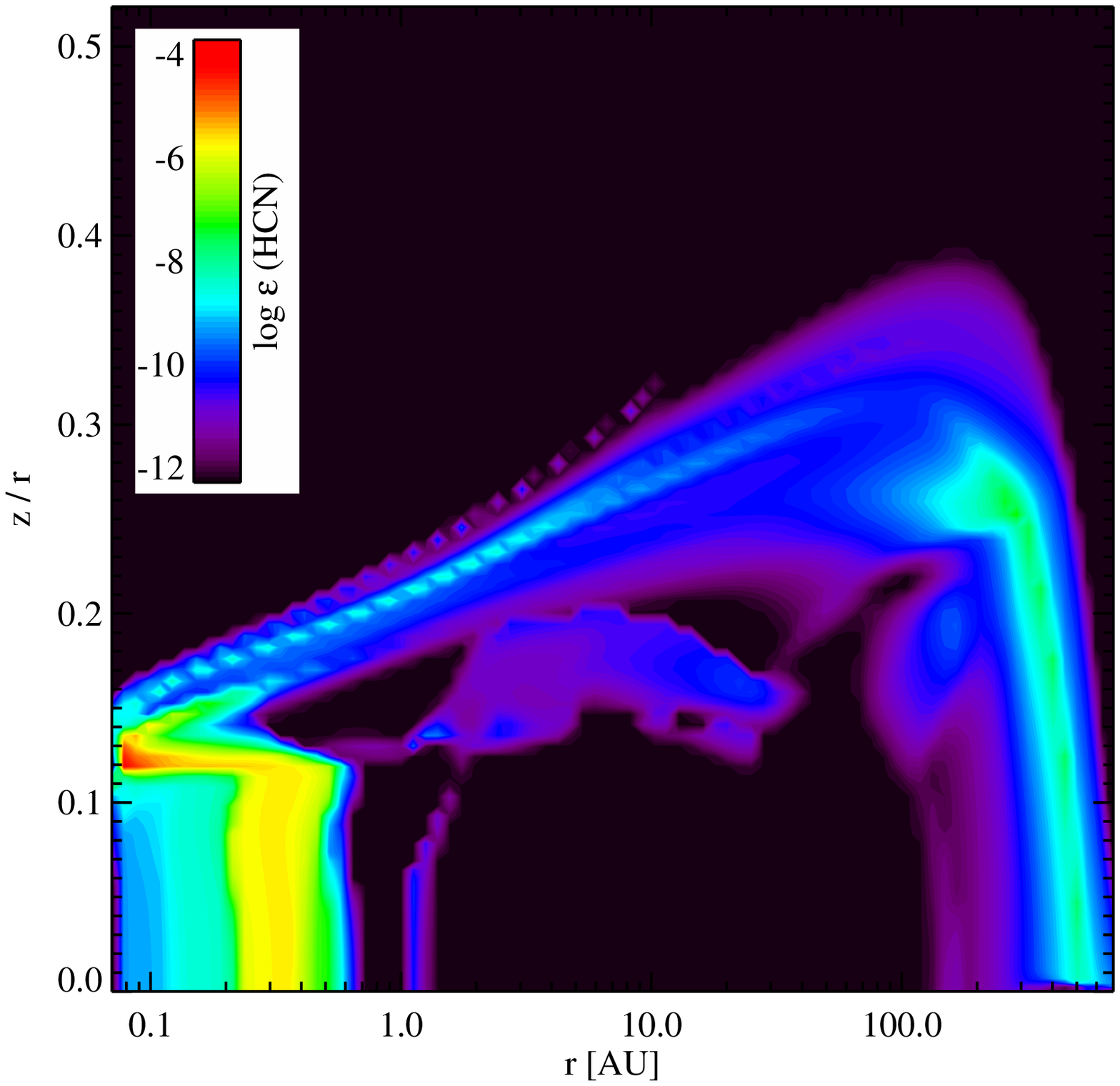}
\includegraphics[width=4.4cm]{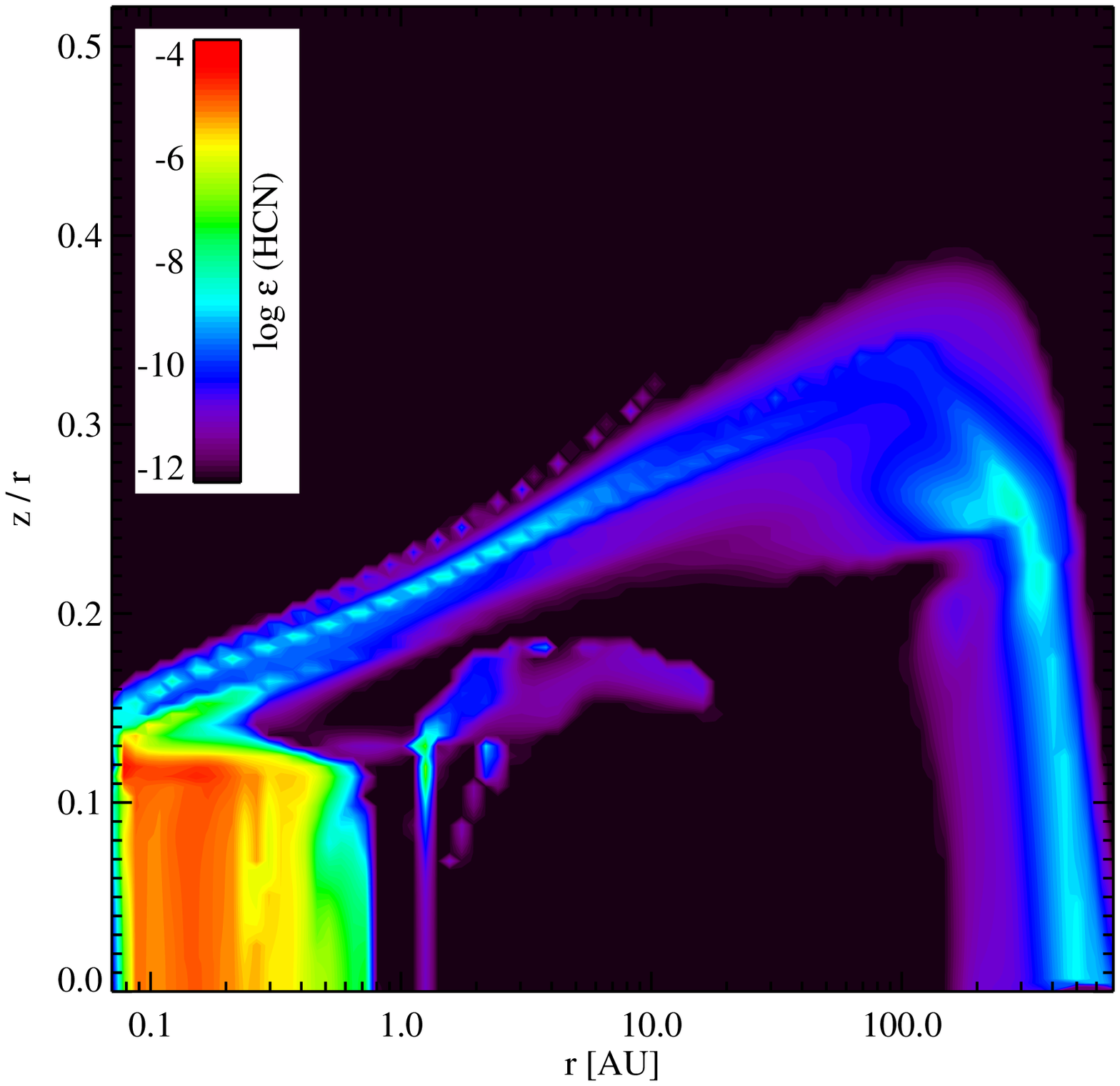}
\includegraphics[width=4.4cm]{model6_icereservoirs_v2.ps}
{\hspace*{2mm}\includegraphics[width=4.4cm]{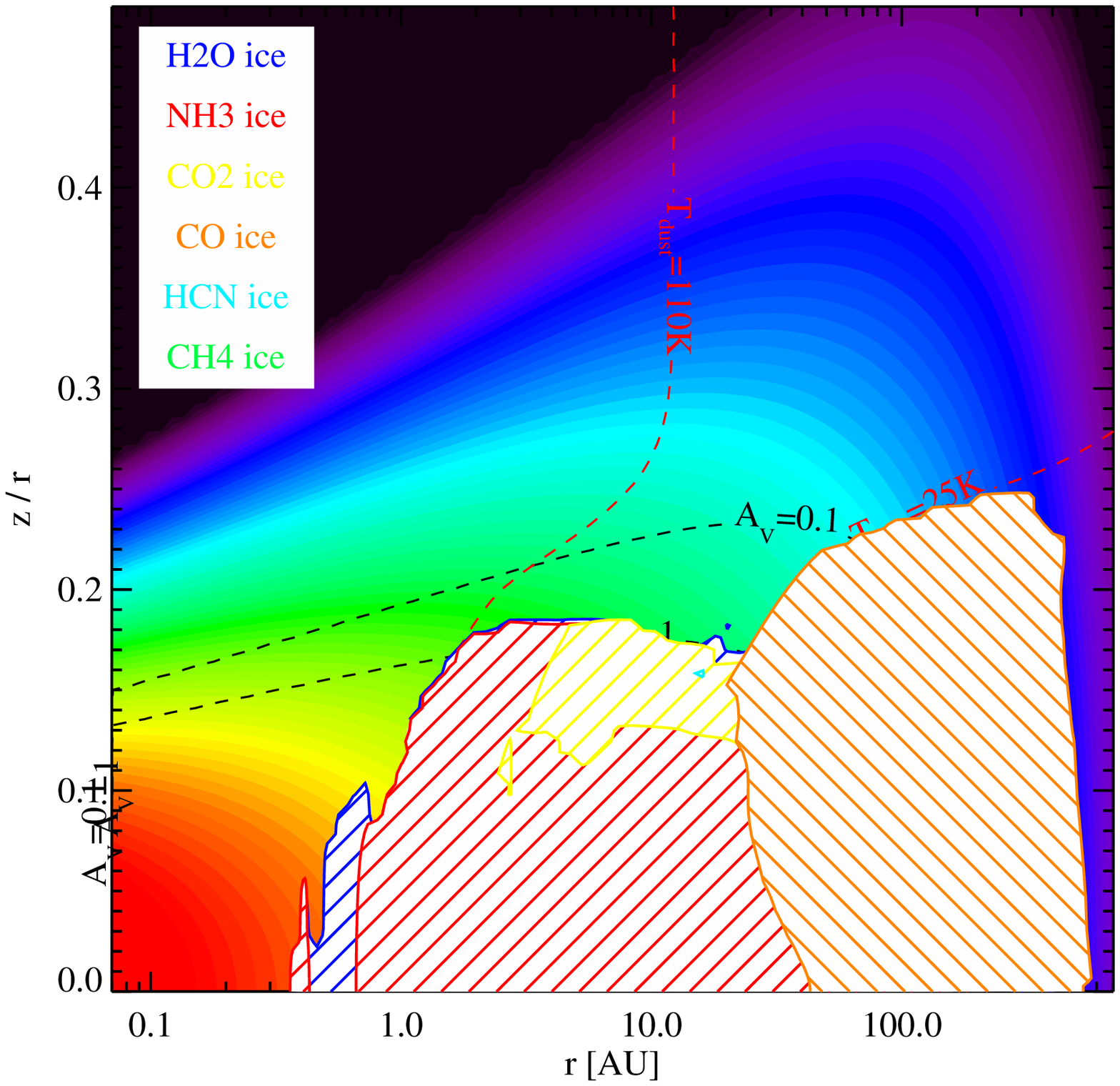}}
\caption{Distribution of key species abundances using the small (left) and large (right) chemical network: HCN and ice reservoirs. Contours and legend are the same as Fig.~\ref{fig:Eads-ices}.}
\label{fig:Chemistrylargenetwork2}
\end{figure}

\subsection{Large versus small networks}
\label{Sect:resultsSizenetwork}

With the advent of ALMA, more complex molecular species and especially molecular ions will be detected in many more disks. Hence, we compare the use of small versus large networks. Again, we keep the disk density and thermal structure fixed and compare model 6 (100 species, 1288 reactions) to model 8 (235 species, 3167 reactions) using the UMIST2012 chemical database and adsorption energies.

\begin{figure*}[!htbp]
\includegraphics[width=18cm]{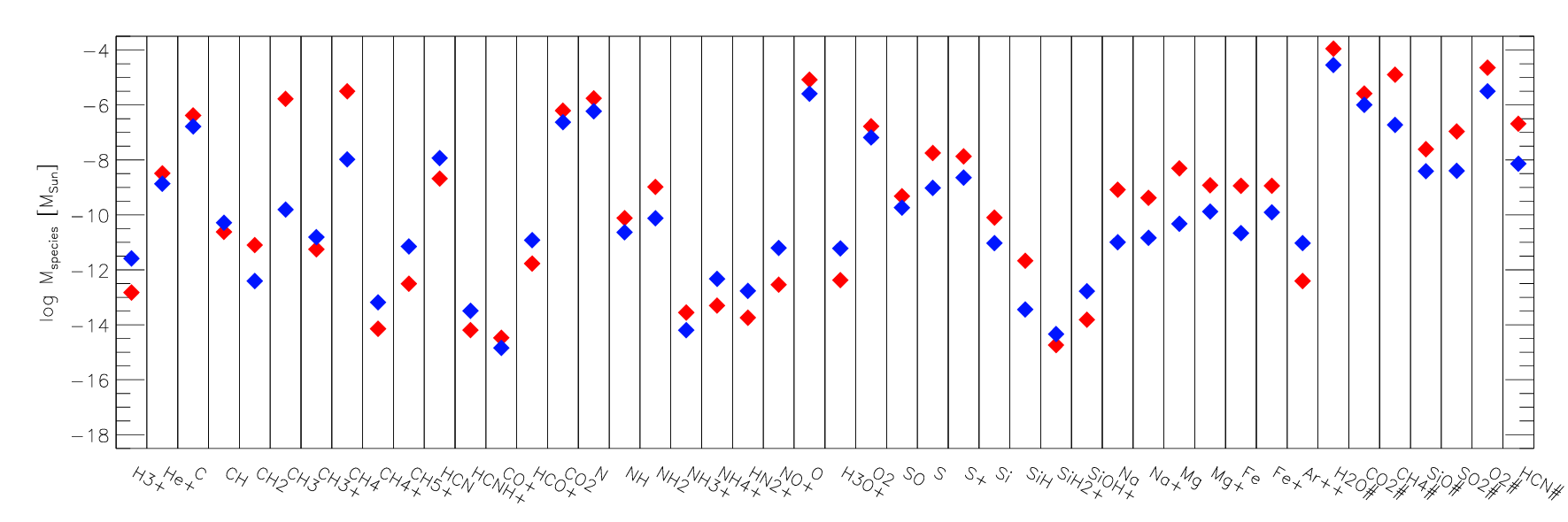}
\caption{Differences in species masses using the UMIST2012 database and its adsorption energies for the small (red, model 6) and large chemical network (blue, model 8).}
\label{fig:largenetwork-change-in-speciesmass}
\end{figure*}

Figures~\ref{fig:Chemistrylargenetwork1} and \ref{fig:Chemistrylargenetwork2} show differences at the outermost radii due to the presence of more complex ices. Those affect also the outer water, HCO$^+$ and HCN reservoirs. Most of these changes come from new branches of chemistry allowed in the larger network such as C-chain chemistry (C$_n$H$_m$), more links between the nitrogen, oxygen and carbon chemistry networks through C-N and N-O bearing species, and sulphur chemistry. In addition, the presence of additional ices and PAHs (with their ice counterpart) changes the electron abundance in the disk.
Hydrocarbons change in some cases by several orders of magnitude in species mass (e.g.\ CH$_3$, CH$_4$ and CH$_5^+$ in Fig.~\ref{fig:largenetwork-change-in-speciesmass}). In addition, many hydronitrogens (azanes) change in mass between a factor three to ten (e.g.\ NH$_2$, NH$_3^+$, NH$_4^+$, N$_2$H$^+$). Differences in molecular species mass beyond a factor 10 are also seen for H$_3^+$ ($\sim\!1.2$~dex), NO$^+$ ($\sim\!1.3$~dex), H$_3$O$^+$ ($\sim\!1$~dex), SiH ($\sim\!1.8$~dex), and SiOH$^+$ ($\sim\!1$~dex). Many metals and metal ions also change their species masses by more than a factor three. For the ices in common, the largest changes are seen in CH$_4$ ice ($\sim\!1.8$~dex), SO$_2$ ice ($\sim\!1.4$~dex) and HCN ice ($\sim\!1.5$~dex).

\begin{figure*}[!htbp]
\begin{center}
\includegraphics[width=16cm]{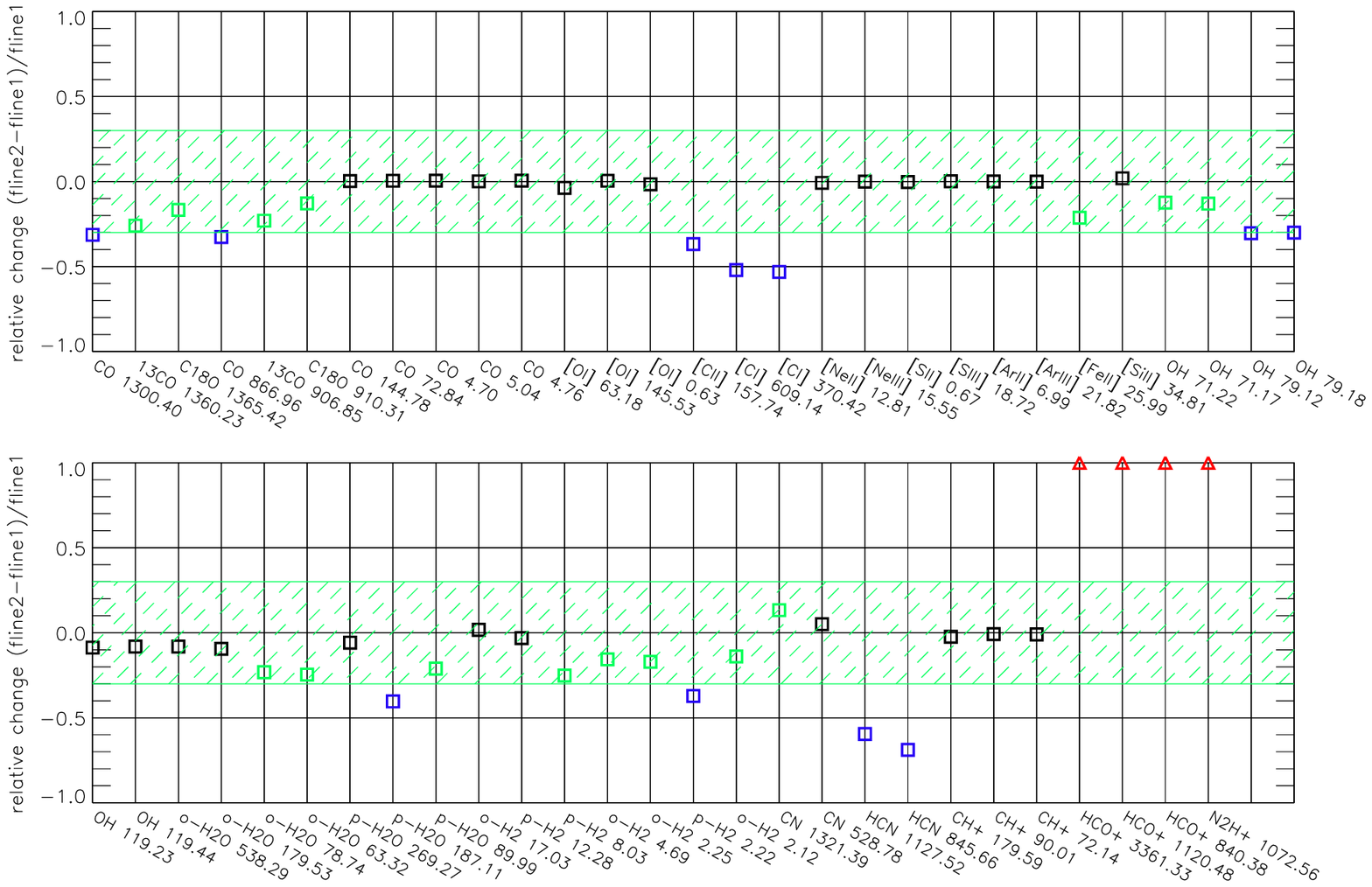}
\caption{Comparison of line fluxes using the UMIST2012 database and its adsorption energies for the small (fline1, model 6) and large chemical network (fline2, model 8). Black and green squares denote differences of less than 25\% and less than a factor two respectively, blue squares and red triangles denote differences larger than a factor three and ten respectively.}
\label{fig:largenetwork-lines}
\end{center}
\end{figure*}

Fig.~\ref{fig:largenetwork-lines} reveals that the majority of lines investigated here do not change when we expand the chemical network to include more complex chemistry. Some lines change within a factor three, something easily buried in uncertainties within other disk input parameters; examples are the fine-structure lines of neutral carbon at 609 and 370~$\mu$m. The sub-mm lines of HCN decrease in the larger network by more than a factor three.

HCO$^+$ and N$_2$H$^+$ lines increase by more than an order of magnitude when the larger network is considered. For HCO$^+$, this is due to a decrease in electron abundance in the regions where this molecule can form (see Fig.~\ref{fig:largenetwork-lines}), especially in the outer disk beyond 100~au. The change in electron abundance (see Fig.~\ref{fig:Chemistrylargenetwork1}) is related to the freeze out of all neutral molecules and atoms (e.g. also sulphur and iron) included in the large network ; the small network comprises only freeze-out of the molecules CO, H$_2$O, CO$_2$, CH$_4$, NH$_3$, SiO, SO$_2$, O$_2$, HCN and N$_2$. More chemical details behind these changes are explained in \citet{Rab2017} with the caveat that they only use the large chemical network.

\section{Discussion}

Most of the results outlined above are not specific to the choice of thermo-chemical disk code. We fixed the disk structure and exploited purely changes related to the choice of chemical database, set of adsorption energies and size of the network. Similar changes would show in any chemical code if it is applied to the large range of physical and irradiation conditions in disks. It has to be kept in mind that the chemical databases used in astrochemistry were originally developed for low density cold environments such as molecular clouds. Networks extending to higher temperatures more appropriate for inner disk regions have been developed \citep[e.g.][]{Agundez2008, Harada2010}, but are not routinely included in disk chemical models. With this work, we thus push the existing widely-used databases into regimes they have originally not been tested for.

Contrary to \citet{McElroy2013b} we find large differences when comparing different chemical databases. We calculate the chemistry in environments of higher densities ($\gtrsim\!10^8$~cm$^{-3}$) and temperatures ($10-5000$~K), while \citet{McElroy2013b} used a dark cloud environment with $n({\rm H_2})\!=\!10^4$~cm$^{-3}$, $T\!=\!10$~K, $A_{\rm V}\!=\!10$~mag. Hence, we find differences both in the spatial distribution of species and also in the resulting line fluxes. The differences for OH and water between UMIST and OSU/KIDA can likely be attributed to collider reactions. In addition, many lines originating from the inner disk show changes larger than a factor two. This indicates differences in the warm chemistry between the networks; abundances of many even simple molecules change in the inner disk where gas temperatures are in excess of a 300~K. These differences would not show up under the low density and low temperature conditions of a dark cloud.

The tests with different sets of adsorption energies shows that most atomic and molecular lines do not depend on these values. Many of these lines are optically thick and hence originate largely in the surface layers well above the ice reservoirs. However, the C$^{18}$O lines are optically thin and therefore directly linked to the size and height of the CO ice reservoir. The fluxes and emission maps of C$^{18}$O will depend on the details of how ices are treated within the chemical network. Another optically thin line is CN 528.78~$\mu$m. If adsorption energies from bare grains are used, the nitrogen reservoir changes significantly and the CN line has an additional contribution from regions inside 100~au. Yet another optically thin line is N$_2$H$^+$. The emitting region and also column densities of this species depend crucially on the choice of N$_2$ adsorption energy and especially also the relative difference between CO and N$_2$ adsorption energies. 

It has been shown by \citet{Agundez2010} that reactions with excited H$_2$ play an important role in the formation of CH$^+$ in diffuse clouds and in Photon Dominated Regions (PDRs). The authors also point out the possible relevance to circumstellar disks. Our tests show now that state-to-state chemical reactions in disks affect indeed mostly CH$^+$; the effect on other molecular ions is minor. Hence for the interpretation of line fluxes and rotational diagrams of CH$^+$ such as presented in \citet{Thi2011b} and \citet{Fedele2013}, it is important to take reactions with excited H$_2$ into account.

\citet{Semenov2004} found that especially the intermediate layers of disks where ion-molecule chemistry is active require larger chemical networks in excess of 100 species. However, they focussed largely on the ionization degree to inform MHD disk models and their model assumes that gas and dust temperatures are equal. The latter assumption leads to colder disk surface layers compared to our model. Many neutral-neutral reactions with barriers become only important for gas temperatures above 300~K. Our comparison between the small (100 species, 1288 reactions) and large (235 species, 3167 reactions) network shows the importance of additional freeze-out due to the presence of more ice species. As \citet{Semenov2004}, we note the importance of carbon chain chemistry. The new chemical pathways opened by connecting C-N, N-O and sulphur chemistry affect the abundance distribution of species even in the outer disk. The emission lines affected by this are mostly HCN, N$_2$H$^+$ and HCO$^+$, while the CO and CN lines stay within a factor $\sim\!2$. Hence, for the interpretation of submm maps and emission lines of HCN, N$_2$H$^+$ and HCO$^+$, we recommend the use of larger chemical networks and a careful treatment of the ionization (metal abundances, freeze-out, charge exchange and grain charging).

The effects outlined above are all related to differences in the chemical input data. It is widely known that many of the rates we use bear large uncertainties and some reaction pathways may be even debated. In addition, we did not even include surface chemistry here, a new layer of complexity with even more unknown parameters. It becomes clear that interpreting absolute column densities of fluxes from molecular lines will be affected by the specific choice of database and/or size of the network used. This poses especially a problem when comparing works from different groups using different chemical input data. It also puts a limit to the quantitative interpretation of individual line observations. A more robust approach could be a differential investigation of the impact of specific disk parameters on key observables, e.g.\ the flaring angle, the gas mass, the amount of irradiation. Even though the absolute column densities of specific species may not be known to better than a factor few, the relative changes should be trustable.

\section{Conclusions}

From the detailed investigation of various chemical databases, different sets of adsorption energies and sizes of chemical networks, we conclude the following key points. 

Many atomic and molecular lines are very robust against changes in the chemical rates and in the size of the network. Caution, however, is required for\\[-6mm]
\begin{itemize}
\item HCN, N$_2$H$^+$ and HCO$^+$ lines (size of the network), 
\item high excitation CO, CN, CH$^+$, H$_2$O, OH lines (database dependency), 
\item CH$^+$ lines (reactions of excited H$_2$),
\item HCO$^+$ lines (UMIST2006 to UMIST2012 update in rates).\\[-6mm]
\end{itemize}

Collider reactions play a major role even in the upper layers of disks. Hence, it would be good to revisit those in experiments. Special attention should be given to checking their low temperature extrapolations.

There is not a single consistent set of adsorption energies to be used for disks. Instead, we recommend a self-consistent approach, where the adsorption energy depends on the nature of the already existing ice, e.g.\ polar or non-polar. This is of minor importance for most of the observed gas lines. However, it will affect the spatial position of ice lines in the disk and thus the emitting region of the rarer CO isotopologues and molecular ions such as HCO$^+$ and N$_2$H$^+$.

For CH$^+$ state-to-state reactions become important in the upper layers of disks. Only very few reactions of excited molecular hydrogen have so far been investigated in detail. The here proposed simplified scheme of using the H$_2$ $v\!=\!1$ state and scaling the known reaction rates for H$_2$ $v\!=\!0$ can only be a first step.

As demonstrated here, the absolute line fluxes can be very sensitive to the specific choice of rate network. However, this will not affect studies where the sensitivity of lines is tested against specific disk parameters using the same chemical network and database. However, discrepancies in disk models for specific objects from groups using different chemical networks should be taken with caution. 

More recent discussions among the disk modelers and the chemical database groups start to diminish some of the discrepancies noted in this work. Hence, we expect that new database releases will bring the results in even closer agreement.

\begin{acknowledgements}
We would like to thank K.\ Oeberg, E.\ Bergin, T.\ Millar and E.\ van Dishoeck for insightful discussions during the development of this work. We also thank the anonymous referee for a careful reading of the manuscript and suggestions that improved the clarity of the figures and text. IK, WFT, CR, and PW acknowledge funding from the EU FP7- 2011 under Grant Agreement nr. 284405. CR also acknowledges funding by the Austrian Science Fund (FWF), project number P24790.
\end{acknowledgements}

\appendix

\section{Chemical reaction rates}
\label{App:diffUMIST}

Next to the standard rates from chemical databases such as UMIST, OSU or KIDA, we use the additional set of reaction rates described below. Some of them are added to the standard set of rates, some overwrite rates from the databases if this option is invoked. This is indicated in the respective subsections. Abbreviations for the references are listed in Table~\ref{Tab:diffUMIST2006}.

\begin{table}[htb]
\caption{List of reference abbreviations}
\begin{tabular}{l|l}
\hline
Abbreviation & Reference \\
\hline
\hline
A93 & \citet{Anicich1993} \\
A96 & \citet{Aikawa1996} \\
AG-estimate & Al Glassgold private communication \\
B06 & \citet{Badnell2006} \\
BC92 & \citet{Baulch1992} \\
EF03 & \citet{Eiteneer2003} \\
FS06 & \citet{Fontijn2006} \\
G01 & \citet{Girardet2001} \\
H17 & \citet{Heays2017} \\
HH93 & \citet{Hasegawa1993a} \\
HM89 & \citet{Hollenbach1989} \\
HM97 & \citet{Hierl1997} \\
HTT91 & \citet{Hollenbach1991} \\
H09 & \citet{Hollenbach2009} \\
JB86 & \citet{Jones1986} \\
JR99 & \citet{Jodkowski1999} \\
KR97 & \citet{Kruse1997} \\
KY90 & \citet{Koshi1990} \\
L88 & \citet{Lennon1988}\\
LF91 & \citet{Landini1991}\\
MM99 & \citet{Mebel1999} \\
TH85 & \citet{Tielens1985} \\
TH86 & \citet{Tsang1986} \\
T87  & \citet{Tsang1987} \\
vD08 & \citet{vanDishoeck2008} \\
VY95 & \citet{Verner1995}\\
ZZ98 & \citet{Zhu1998} \\
\hline
\end{tabular}
\label{Tab:diffUMIST2006}
\end{table}

\subsection{Chemistry different from UMIST}
\label{App:diffUMIST}

A key reaction is the H$_2$ formation which is hardcoded in ProDiMo. The reaction rate and implementation is described in \citet{Woitke2009}.

\subsection{Chemistry added to UMIST}
\label{App:addUMIST}

Reactions which we include in addition to UMIST are given in Table~\ref{tab:Reaction-additions-to-UMIST}. Most of them describe the photodissociation of molecular ions; these latter rates are taken from \citet{Heays2017}. We use the formalism of \citep{Woitke2009} to implement adsorption of gas phase species onto grains and thermal and non-thermal desorption.

\begin{table*}[ht]
\caption{Rate constants for reactions added on top of the UMIST database.}
\begin{tabular}{lrrrll}
\hline
reaction & $A$ & $B$ & $C$ & temperature range & reference \\
\hline
\hline
CH$_5^+$   + $h\nu$          $\rightarrow$ CH$_4^+$   + H                         & 3.00(-11) & 0.0 & 2.0  &  10.0 - 41000.0 & H17\\
CH$_5^+$   + $h\nu$          $\rightarrow$ CH$_3^+$   + H$_2$                        & 3.00(-11) & 0.0 & 2.0  &  10.0 - 41000.0 & H17\\
H$_2$O    + $h\nu$          $\rightarrow$ O      + H$_2$                        & 8.89(-11) & 3.90 & 4.12  &  10.0 - 41000.0 & H17\\
H$_2$O$^+$   + $h\nu$          $\rightarrow$ O      + H$_2^+$                       & 5.00(-13) & 0.0 & 2.0 &   10.0 - 41000.0 & H17\\
H$_2$O$^+$   + $h\nu$          $\rightarrow$ O$^+$     + H$_2$                        & 5.00(-13) & 0.0 & 2.0  &  10.0 - 41000.0 & H17\\
H$_3$O$^+$   + $h\nu$          $\rightarrow$ H$_2$O    + H$^+$                        & 2.00(-11) & 0.0 & 2.0  & 10.0 - 41000.0 & H17\\
H$_3$O$^+$   + $h\nu$          $\rightarrow$ OH     + H$_2^+$                       & 1.00(-11) & 0.0 & 2.0  &  10.0 - 41000.0 & H17\\
H$_3$O$^+$   + $h\nu$          $\rightarrow$ H$_2$O$^+$   + H                         & 2.00(-11) & 0.0 & 2.0   & 10.0 - 41000.0 & H17 \\
H$_3$O$^+$   + $h\nu$          $\rightarrow$ OH$^+$    + H$_2$                        & 2.00(-11) & 0.0 & 2.0 &  10.0 - 41000.0 & H17 \\
SiOH$^+$  + $h\nu$          $\rightarrow$ SiO$^+$   + H                         & 5.00(-11) & 0.0 & 2.0  &  10.0 - 41000.0 & H17 \\
NH$^+$    + $h\nu$          $\rightarrow$ N$^+$     + H                         & 5.40(-11) & 0.0 & 1.64 &  10.0 - 41000.0 & vD08\\
C$^+$     + SiO             $\rightarrow$ SiO$^+$   + C                         & 5.40(-10) & 0.0 & 0.0  &  10.0 - 41000.0 & TH85 \\
H$_2$     + H$_2$              $\rightarrow$ H$_2$     + H      + H                & 2.30(-11) & 1.25 & 65000.0  &  10.0 - 41000.0 & TH85 \\
CO     + H        $\rightarrow$        HCO                                & 5.29(-34) & 0.0 & 370.0 &    10.0 - 41000.0 & NIST \\
\hline
\end{tabular}
\tablefoot{The coefficients $A$, $B$, and $C$ have their usual meaning \citep[see e.g.][]{McElroy2013b}. The notation $x(y)$ denotes $x\,10^y$.}
\label{tab:Reaction-additions-to-UMIST}
\end{table*}

\subsection{Photochemistry}

We calculate the photo rates from using the local radiation field from the 2D continuum radiative transfer \citep{Kamp2010} and the photoionization and -dissociation cross sections from the Leiden database \citep{vanDishoeck2008,Heays2017}. These rates replace the standard UMIST photo rates even if the UMIST database is chosen.

\subsection{Chemistry of excited H$_2$}
\label{App:chem-not-in-UMIST}

Reactions of excited H$_2$ (H2EXC, H$_2^*$) are neither included in UMIST2006 nor in UMIST2012. They can be important in disks, especially at the surface and thus have been included in Reactions.in. The basic assumption is that the vibrational energy of the excited H$_2$ can be entirely used to overcome a potential reaction barrier \citep{Tielens1985}. While this might be a reasonable assumption for low barriers, it may overestimate the rates for reactions with a high barrier. {\em In any case all those rates should be treated as guesses at most.} In the absence of any better rates, the reaction rate of a species with H$_2$ is simply modified by subtracting the energy corresponding to the first vibrational excited level $v\!=\!1$, 5980~K.

The collisional de-excitation rate (in units of cm$^3$~s$^{-1}$) of the pseudo excited state (effective quantum number $v\!=\!6$) H$_2^*$ by collisions with H and H$_2$ is taken from \citet{Tielens1985} as one-sixth of the collisional de-excitation rate from $v\!=\!1$, $\gamma_{10}^{\rm H,H_2}$ 
\begin{eqnarray}
R_{ul}^{\rm H} & = & 2.887\,10^{-12} \left(\frac{T}{300} \right)^{0.5} \exp{(-1000/T)} \\
R_{ul}^{\rm H_2} & = & 4.042\,10^{-12} \left(\frac{T}{300} \right)^{0.5} \exp{(-18100/(T\!+\!1200))}
\end{eqnarray}
Note that \citet{Tielens1985} provide in their Table~9 a de-excitation rate for collisions with atomic hydrogen that is a factor 0.67 smaller than this. The collisional excitation rates (in units of cm$^3$~s$^{-1}$) are the inverse of these de-excitation rates \citep{Woitke2009} using the energy of the pseudo-level for vibrationally excited H$_2$, $E_v\!=\!2.6$~eV, 
\begin{eqnarray}
R_{lu}^{\rm H} & = & R_{ul}^{\rm H} \exp{(-E_v/kT_{\rm gas})} \\
R_{lu}^{\rm H_2} & = & R_{ul}^{\rm H_2} \exp{(-E_v/kT_{\rm gas})}
\end{eqnarray}
Other rates are taken explicitly from their Table~9.  

\begin{figure}[!htbp]
\includegraphics[width=9cm]{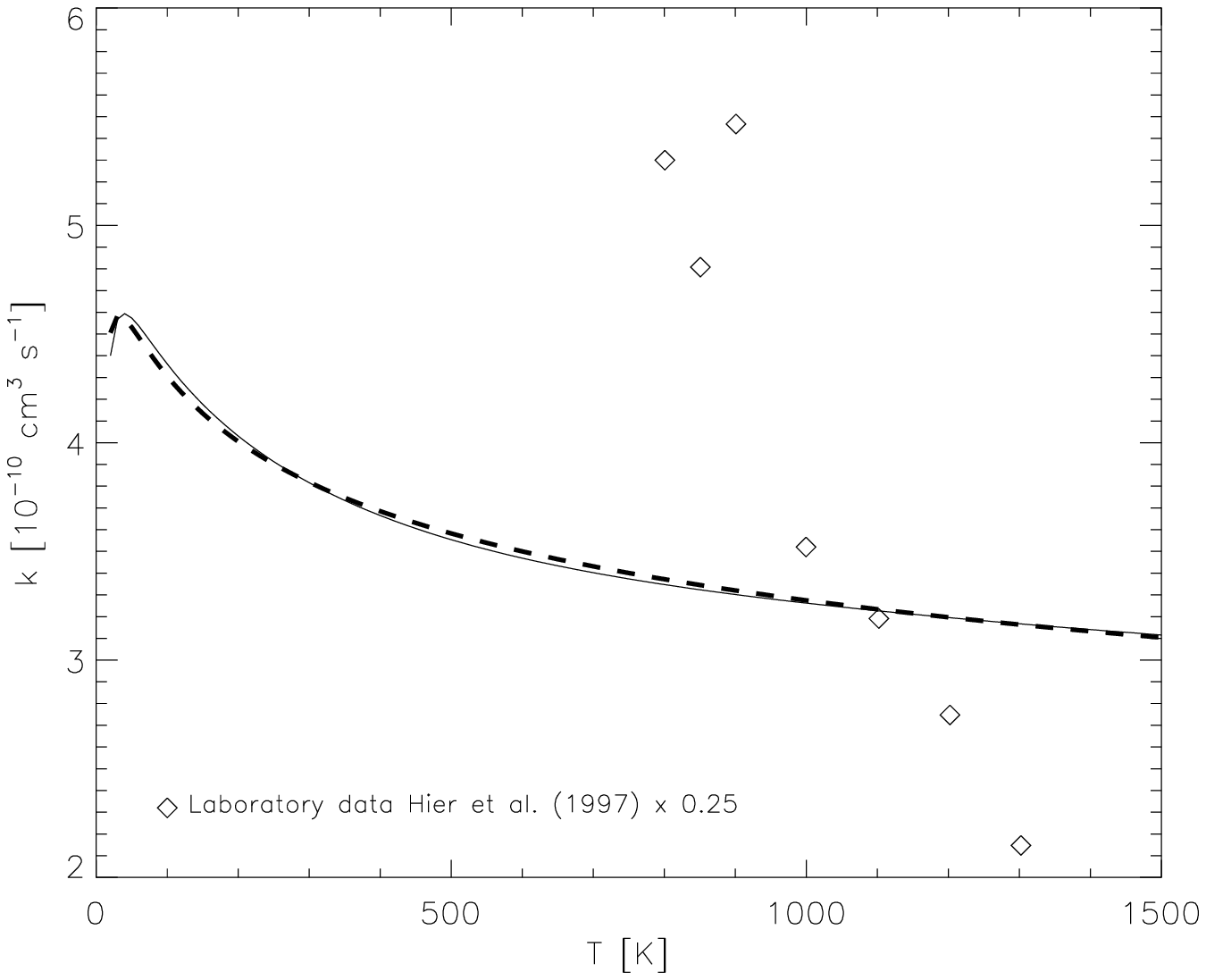}
\caption{Comparison of the theoretically calculated rate A.7 \citep[solid line,][]{Zanchet2013b}, the laboratory data scaled by a factor 0.25 \citep[diamonds,][]{Hierl1997} and the fit presented here (dashed line).}
\label{fig:H2excCplus}
\end{figure}

\citet{Agundez2010} use for the reaction H$_2^*\,+\,$C$^+\,\rightarrow\,$CH$^+\,+\,$H a constant Langevin rate coefficient of $1.6 \, 10^{-9}$~cm$^3$~s$^{-1}$. Fig.10 of \citet{Hierl1997} suggest a possible weak temperature dependence. \citet{Zanchet2013a} performed quantum calculations to derive state-to-state reaction rates for the system C$^+$ and H$_2$. In the following, we neglect the formation excitation of CH$^+ $ and use the parametrized rate from \citet{Zanchet2013b} for H$_2 (v\!=\!1)$ (Table~3). To bring it into the usual form, we present here a simple Arrhenius fit 
\begin{equation}
R = 3.87\,10^{-10} \left(\frac{T}{300}\right)^{-0.136} \exp{(-4.33/T)}~~{\rm cm^3~s^{-1}}
\label{Eq:Zanchet-rate}
\end{equation}
The theoretical calculations are a factor 4 lower than the laboratory work from \citet{Hierl1997} (Fig.~\ref{fig:H2excCplus}). The newer rate is only used in Sect.~\ref{Sect:resultsH2EXC} discussing the effect of excited H$_2$ on the chemistry.

In addition to those reactions, we add
\begin{eqnarray}
{\rm H_2^*} + {\rm CN} & \rightarrow & {\rm HCN} + {\rm H} \\
{\rm H_2^*} + {\rm He^+} & \rightarrow & {\rm H^+} + {\rm H} + {\rm He} 
\end{eqnarray}
taken from \citet{Zhu1998} and \citet{Jones1986}, respectively. In the latter case, $R=0.18-1.8\,10^{-9}$~cm$^3$/s and we approximate that with $0.5\,10^{-9}$~cm$^3$/s.
\begin{eqnarray}
{\rm H_2^*} + {\rm C_2} & \rightarrow & {\rm C_2H} + {\rm H} \\
{\rm H_2^*} + {\rm C_2H} & \rightarrow & {\rm C_2H_2} + {\rm H} 
\end{eqnarray}
taken from NIST \citep{Kruse1997,Eiteneer2003}. The activation barriers of $4000$~K, $460.6$~K (respectively) are set to zero.
\begin{eqnarray}
{\rm H_2^*} + {\rm CH_2} & \rightarrow & {\rm CH_3} + {\rm H} \\
{\rm H_2^*} + {\rm CH_3} & \rightarrow & {\rm CH_4} + {\rm H} 
\end{eqnarray}
taken from NIST \citep{Tsang1986,Baulch1992}. The first one has only an upper limit and we use that for the rate and set the activation barrier of 6400~K to 420~K. The activation barrier of $4740$~K for the second one is set to zero.
\begin{eqnarray}
{\rm H_2^*} + {\rm CH_3O} & \rightarrow & {\rm CH_3OH} + {\rm H} \\
{\rm H_2^*} + {\rm CH_2OH} & \rightarrow & {\rm CH_3OH} + {\rm H} 
\end{eqnarray}
taken from NIST \citep{Jodkowski1999,Tsang1987}. The activation barriers of $2470$~K, $6720$~K (respectively) are set to zero and 740~K respectively.
\begin{eqnarray}
{\rm H_2^*} + {\rm N} & \rightarrow & {\rm NH} + {\rm H} \\
{\rm H_2^*} + {\rm NH} & \rightarrow & {\rm NH_2} + {\rm H} \\ 
{\rm H_2^*} + {\rm NH_2} & \rightarrow & {\rm NH_3} + {\rm H} 
\end{eqnarray}
taken from NIST \citep{Koshi1990,Fontijn2006,Mebel1999}. The activation barriers of $16600$~K, $7760$~K, $3610$~K (respectively) are set to 12115~K, 1802~K and zero respectively.
\begin{eqnarray}
{\rm H_2^*} + {\rm C} & \rightarrow & {\rm CH} + {\rm H} \\
{\rm H_2^*} + {\rm CH} & \rightarrow & {\rm CH_2} + {\rm H} 
\end{eqnarray}
are taken from UMIST2006 and their activation barriers of 11700~K and 1943~K are set to 5720~K and zero respectively.
\begin{eqnarray}
{\rm H_2^*} + {\rm O} & \rightarrow & {\rm OH} + {\rm H} \\ 
{\rm H_2^*} + {\rm OH} & \rightarrow & {\rm H_2O} + {\rm H} \\ 
{\rm H_2^*} + {\rm O_2} & \rightarrow & {\rm OH} + {\rm OH} 
\end{eqnarray}
are taken from UMIST2006 and their activation barriers of 3150~K, 1736~K, 21890~K are set to zero, zero and 15910~K respectively.
\begin{eqnarray}
{\rm H_2^*} + {\rm H} & \rightarrow & {\rm H} + {\rm H} + {\rm H}\\ 
{\rm H_2^*} + {\rm H_2} & \rightarrow & {\rm H_2} + {\rm H} + {\rm H}\\ 
{\rm H_2^*} + {\rm H_2^*} & \rightarrow & {\rm H_2} + {\rm H} + {\rm H}
\end{eqnarray}
are taken from UMIST2006 and their activation barriers of 55000~K, 84100~K, 84100~K are set to 49020~K, 78120~K, and 72260~K.

\subsection{Collider reactions}

Even though UMIST2006 contains collider (three-body reactions, CL), the UMIST2012 rate file does not. There are 32 collider reactions in UMIST2006 ($\#4552-\#4583$). After discussion with T.\ Millar, we decided to append them to the UMIST2012 data file.

\subsection{PAH chemistry}

PAHs participate in charge exchange reactions with other gas-phase species or the ionised PAHs can recombine with electrons. Therefore PAHs are a key species to the modelling of the gas heating and key to the determination of the ionisation fraction in disks.  An example of the role of PAHs is for example that negative PAHs can lock a large fraction of the free electrons, preventing them to recombine with HCO$^+$ or N$_2$H$^+$, two commonly detected species in protoplanetary disks. The PAH reactions are described in details in Thi et al., (in prep.). PAHs can adsorb onto the surfaces of grains and desorb through photons or thermal heating depending on their sizes. PAH freeze-out can modify the heating-cooling balance as well as the ionisation fraction in disks. The ionisation balance is calculated using the local UV radiation field \citep{Woitke2011}. The models described in this paper use circumcoronene (C$_{54}$H$_{18}$) as the representative PAH. Circumcoronene is a large and compact PAH (pericondensed, superaromatic) and the electron delocalization adds to its stability against photodissociation \citep{Tielens1987ASIC..191..273T,Visser2007,DeBecker2013}.

\subsubsection{PAH freeze-out and desorption}

PAHs can adsorb onto grain surfaces coated with water ice or on bare silicate grains. In dense molecular clouds, weak absorption features have been attributed to PAHs frozen in water ice mantles \citep{Bouwman2011A&A2...529A..46B}. \citet{Bouwman2011a_A&A...525A..93B} estimated that the PAH$_{\mathrm{ice}}$/H$_2$O$_{\mathrm{ice}}$ ratio can reach as much as 2\%. Assuming a water ice abundance of $\sim\!10^{-5}$, the PAH\# (PAHs adsorbed on dust surfaces) abundance is $\sim\!2\!\times\!10^{-7}$, which suggests that all the PAHs are removed from the gas phase (the total PAH abundance is estimated to be $\sim\!3\!\times\!10^{-7}$). The standard PAH used in the models presented here is circumcoronene although one can choose also larger condensed PAHs such as circumcircumcoronene (C$_{96}$H$_{24}$), which are even more stable against photodissociation than circumcoronene.

In the absence of measured desorption rates for all individual PAHs, we assume a linear dependance of the desorption energy on the number of carbon and hydrogen atoms of the PAH based on the method of fragment constant. The linear behaviour is consistent with the additivity of van der Waals interactions \citep{Bjork2011_doi:10.1021/ja205857a}. Adsorption (desorption) energies $E_{\mathrm{ads}}$ measured in the laboratory vary from $E_{\mathrm{ads}}/k\!\simeq\!5600$~K for benzene C$_6$H$_6$ \citep{Thrower2009MNRAS.394.1510T} up to $E_{\mathrm{ads}}/k\!\simeq\!18900$~K for pentacenene C$_{22}$H$_{14}$ \citep{Oja1998_doi:10.1021/je970222l}. Table~\ref{tab_PAH_Edes} summarises the adsorption (desorption) energies for several PAHs. It should be noted that the surface on which the PAHs are adsorbed varies in the different studies. We plotted the adsorption energy normalised by the number of carbon atoms $E_{\mathrm{des}}/N_{\mathrm{C}}$ as a function of $N_{\mathrm{H}}/N_{\mathrm{C}}$ in Fig.~\ref{fig_PAH_normalize_binding_energy}. The adopted law is
\begin{equation}  
  E_{\mathrm{ads}}/k = (482\times (N_{\mathrm{C}}-N_{\mathrm{H})}) +  (946\times N_{\mathrm{H}})\,\,\, \mathrm{K}\,\,\,,\label{desoprtion_formula}
\end{equation}
where $N_{\mathrm{C}}$ is the number of carbon atoms of the PAH and $N_{\mathrm{H}}$ is the number of hydrogen. $E_{\mathrm{CC}}/k$ ($=\!482$~K) is the fitted desorption energy per graphene-like carbon, i.e. carbon atoms not attached to  hydrogen and $E_{\mathrm{CH}}/k$ ($=\!946$~K) is the fitted energy per benzene-like carbon and its adjoining H-atom. Graphene-like carbons are C-atoms with three covalent bonds with carbons, whereas benzene-like carbon have two covalent bonds with carbons and one bond with a hydrogen atom. Graphenes are PAHs where all the hydrogens have been stripped away. The value for $E_{\mathrm{CC}}/k$ is consistent with the measured range of adsorption energies of graphene on amorphous SiO$_2$, which lies between 450 and 1685~K per carbon \citep{Thrower2013_doi:10.1021/jp404240h,Kumar2016}. The adsorption energies are also consistent with the theoretical values of \citet{Meszar2013doi:10.1021/jp401532x}. 
\begin{figure*}[!ht]
\centering  
\includegraphics[angle=90,scale=0.32]{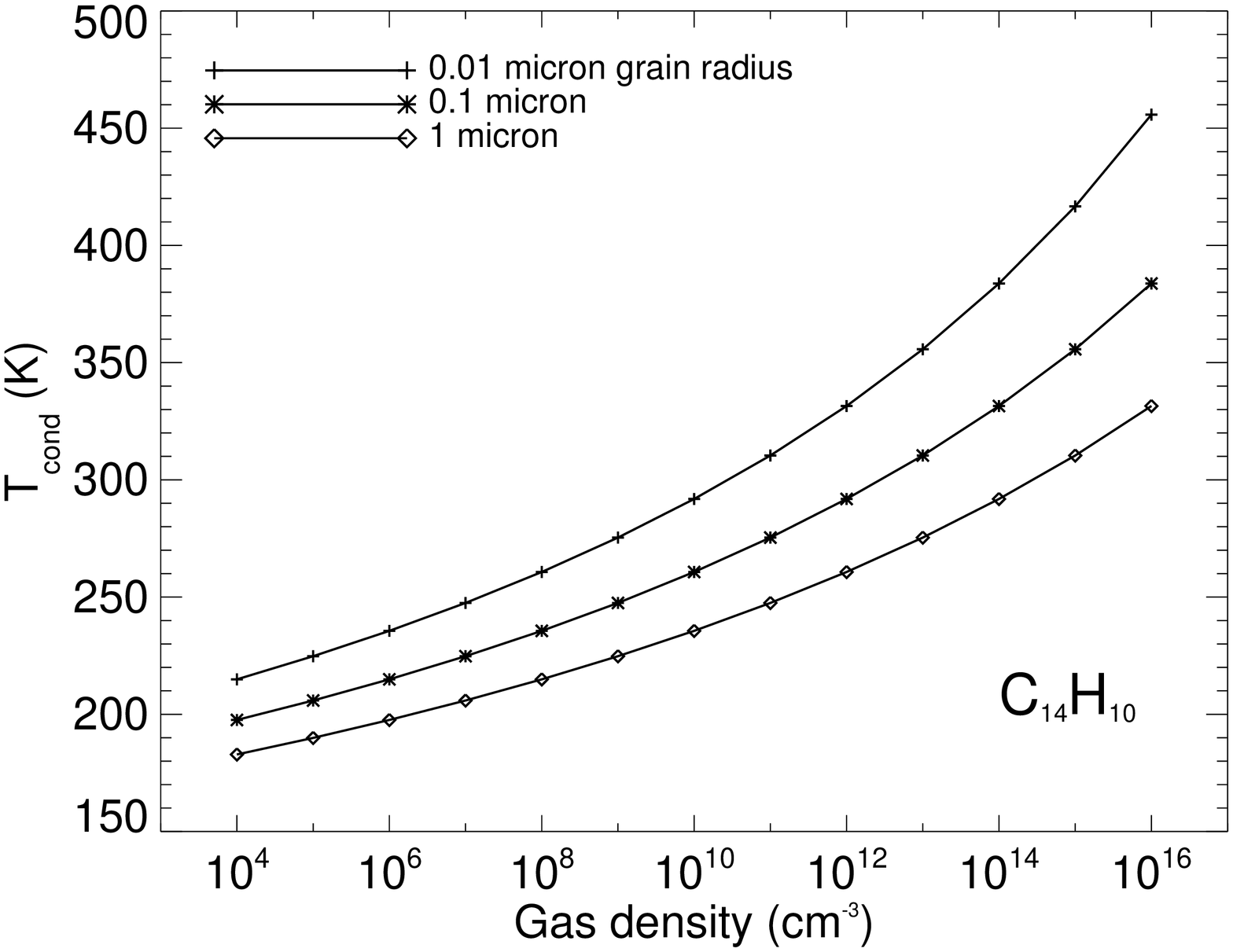}
\includegraphics[angle=90,scale=0.32]{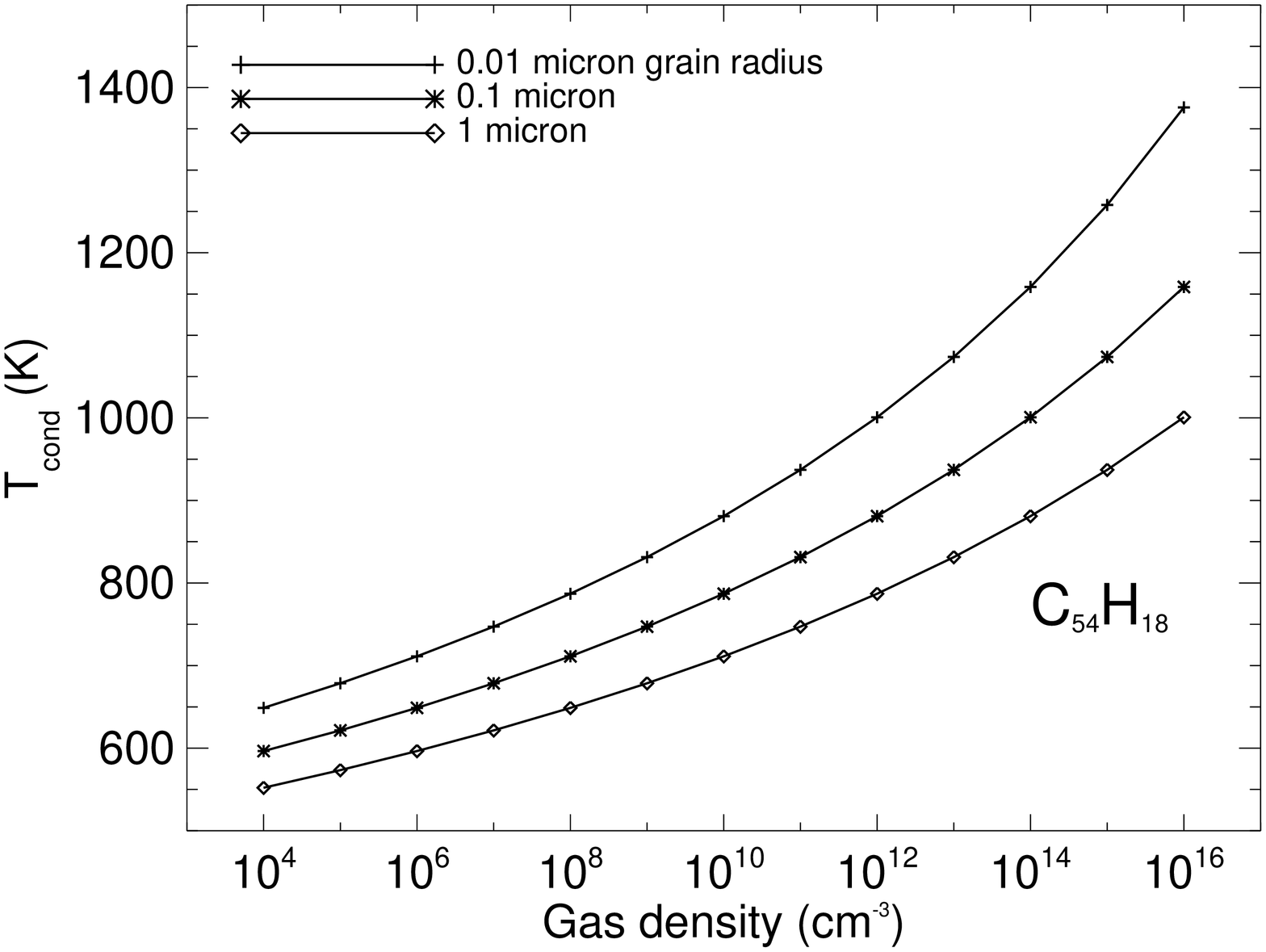}
\caption{PAH condensation temperature for anthracene (C$_{14}$H$_{10}$) and circumcoronene (C$_{54}$H$_{18}$) assuming three different grain radii.}  
  \label{fig_PAH_Tcond}          
\end{figure*}    

\begin{figure}[!ht]
\centering
\resizebox{\hsize}{!}{\includegraphics[angle=90]{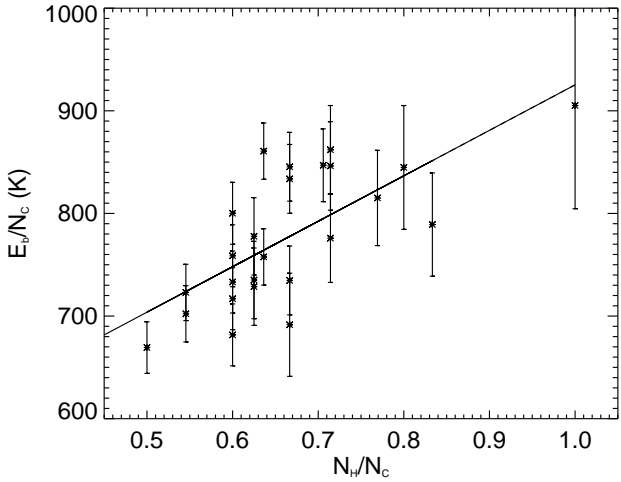}}
\caption{PAH binding energy normalized to the number of carbon atoms.}  
  \label{fig_PAH_normalize_binding_energy}          
\end{figure}    

\begin{figure}[!ht]
\centering  
\resizebox{\hsize}{!}{\includegraphics[angle=90]{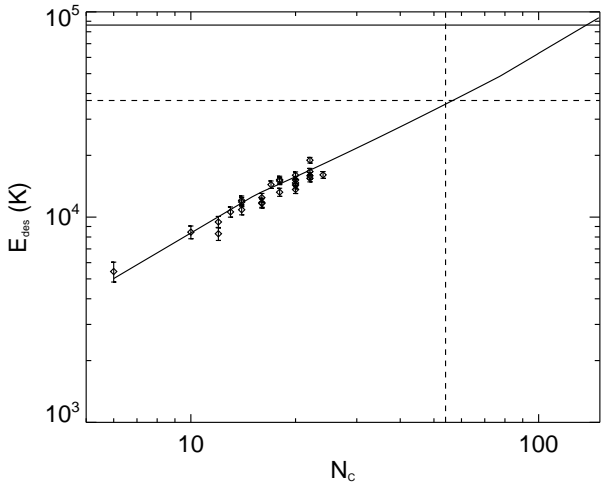}}
\caption{PAH desorption energy. The horizontal line corresponds to the limit imposed by the heat of vaporisation of graphite. The dashed-lines designate the desorption energy for circumcorene ($N_{\rm C}\!=\!54$).}  
  \label{fig_PAH_desorption_energy_as_function_of_the_number_of_carbon_atoms_in_the_PAH}          
\end{figure}    

The fitting formula is valid for PAH made of up to 100 carbon atoms. For bigger PAHs, the formula gives values that are larger than the heat of vaporization of graphite of $\Delta H_{\mathrm{f}}/k\!=\!86240$~K, Fig.~\ref{fig_PAH_desorption_energy_as_function_of_the_number_of_carbon_atoms_in_the_PAH} \citep{Pierson1993}. For circumcoronene, the estimated desorption energy is 34380~K.  A simple way to estimate the condensation temperature $T_{\mathrm{cond}}$ is to balance the adsorption with the thermal desorption
\begin{eqnarray}
\sigma_{\mathrm{dust}}n_{\mathrm{dust}}n_{\mathrm{PAH}} S_{\mathrm{PAH}}\sqrt{\frac{8kT_{\mathrm{cond}}}{\pi m_{\mathrm{PAH}}}}\\ \nonumber
= n_{\mathrm{PAH\#}}\nu_{\mathrm{osc}}\exp{(-E{_\mathrm{des}}/kT_{\mathrm{cond}})}\,\,\,,
\end{eqnarray}
where the exponential prefactor is \citep{Aikawa1996ApJ...467..684A}
\begin{equation}
\nu_{\mathrm{osc}} = \sqrt{\frac{2 n_{\mathrm{surf}}  E_{\mathrm{des}}}{\pi^2 m_{\mathrm{PAH}}}}\ \,\,\mathrm{s}^{-1}\,\,\,. 
\end{equation}
$S_{\rm PAH}$ is the sticking coefficient for PAH molecules and $n_{\rm PAH}$ and $n_{\rm PAH\#}$ are the number density of PAH in the gas phase and adsorbed on surfaces respectively. The number of surface adsorption sites is assumed to be $n_{\mathrm{surf}}\!=\!1.5 \times10^{15}$~cm$^{-2}$ for grains with a radius of $0.1~\mu$m. The average number of dust grains in the disk is
\begin{eqnarray}\label{eq_ndust}
n_{\mathrm{d}} & = & \frac{2.2 \,\mathrm{amu}\ n_{\mathrm{H}}}{(4/3)\pi \rho_{\mathrm{d}}<\!a^3\!> \delta} \\ \nonumber
 & \approx & 2.907\times 10^{-15} n_{\mathrm{H}} \left(\frac{\mathrm{\mu m^3}}{<\!a^3\!>}\right)\left(\frac{100}{\delta}\right)\,\,\,,\label{nd_formula}
\end{eqnarray}
where $\delta$ is the gas-to-dust mass ratio which is assumed to be 100. We have assumed a silicate mass density of 3.0~g~cm$^{-3}$. The balance between adsorption and thermal desorption for a gas density of $10^{10}$~cm$^{-3}$, a grain radius of $1~\mu$m, a gas-to-dust mass ratio of 100, and assuming $n_{\mathrm{PAH}}\!=\!n_{\mathrm{PAH\#}}$ and a sticking coefficient of unity, leads to a condensation temperature for circumcoronene of $\sim\!710$~K. In disks, PAHs should stick onto silicate grains and to each other. The condensation temperature for anthracene and circumcoronene is shown in Fig.~\ref{fig_PAH_Tcond}. The desorption of PAHs occurs thermally, induced by cosmic ray hits on the grains, and by absorption of a UV photon. We assumed a yield of 10$^{-3}$ per absorbed photon for the photodesorption and a standard cosmic-ray induced desorption similar to the other ice species. The encounter between two PAHs can also result in the formation of clusters, which may subsequently grow by further accretion of PAHs. However, this mechanism is not accounted for in our current model.

\begin{table*}
  \caption{Measured and computed (M\'{e}sz\'{a}r) PAH desorption energies.\label{tab_PAH_Edes}}          
\begin{tabular}{lllllllll}     
\hline
\noalign{\smallskip}        
\multicolumn{1}{c}{Name} & \multicolumn{1}{c}{Formula} & \multicolumn{1}{c}{Oja98} & \multicolumn{1}{c}{Thower09} & \multicolumn{1}{c}{Domine07} & \multicolumn{1}{c}{Fu11} &  \multicolumn{1}{c}{Goldfarb08} & \multicolumn{1}{c}{Meszar13} & \multicolumn{1}{c}{Thower13} \\
 & & \multicolumn{6}{c}{$E_{\mathrm{des}}$ [K]} \\[2mm]
\hline
\hline
\noalign{\smallskip}   
 benzene       & C$_{6}$H$_{6}$ & & 5431 & & & & &  \\
  naphtalene    & C$_{10}$H$_{8}$ & & & & & & 8448 & \\
acenaphthylene  &   C$_{12}$H$_{8}$   & & & & & 8298~$\pm$~989 & & \\               
 acenaphthene    &   C$_{12}$H$_{10}$   & & & & & 9470~$\pm$~269 & & \\
 fluorene        &   C$_{13}$H$_{10}$   & & & & & 1059~$\pm$~229 & & \\
 anthracene    & C$_{14}$H$_{10}$ & 12068 & & & & 11848~$\pm$~399 & & \\
 phenanthrene  & C$_{14}$H$_{10}$ & 10860 & & 10862 & & & &  \\
 fluoranthene  & C$_{16}$H$_{10}$ & & & & & 11658~$\pm$~332  & & \\
 pyrene            & C$_{16}$H$_{10}$ & 12442 & & & & 11762~$\pm$~392 & & \\
 2,3-benzofluorene  & C$_{17}$H$_{12}$  & 14397 & & & & & & \\
 naphthacene        & C$_{18}$H$_{12}$  & 15218 & & & & 15006~$\pm$~350 & & \\ 
 perylene           & C$_{20}$H$_{12}$  & 16003 & & & & 15175~$\pm$~98 & & \\
 benzo[a]phenanthrene  &   C$_{18}$H$_{12}$ & & & & & 13224~$\pm$~430 & & \\ 
 benzo[a]pyrene     & C$_{20}$H$_{12}$ & & & & & 13633~$\pm$~375 & & \\
 benzo[b]fluoranthene    & C$_{20}$H$_{12}$  & &&&14337   & & & \\                             
 benzo[k]fluoranthene    & C$_{20}$H$_{12}$  & &&&14663   & & & \\                             
 benzo[ghi]perylene      & C$_{22}$H$_{12}$  & &&&15448   & & & \\                            
 indeno[1,2,3-cd]pyrene  & C$_{22}$H$_{12}$  & &&&15906   & & & \\                           
 dibenz[a,h]anthracene   & C$_{22}$H$_{14}$  & &&&16667   & & & \\
 pentacenene        & C$_{22}$H$_{14}$   & 18936 & & & & & & \\ 
 coronene           & C$_{24}$H$_{12}$   & 16063 & & & & & & 19727\\    
 \hline                          
\noalign{\smallskip}
\end{tabular}
\tablefoot{Oja98: \citet{Oja1998_doi:10.1021/je970222l}; Thower09: \citet{Thrower2009JChPh.131x4711T}; Domine07: \citet{Domine2007doi:10.1021/es0706798}; Fu11: \citet{Fu20111_660}; Goldfarb08: \citet{Goldfarb2008_doi:10.1021/je7005133}; Meszar13: \citet{Meszar2013doi:10.1021/jp401532x}; Thrower13: \citet{Thrower2013_doi:10.1021/jp404240h}}  
\end{table*}  

\subsection{X-ray chemistry}
\label{App:X-rays}

The X-ray chemistry in ProDiMo is described in detail in the appendix of \citet{Meijerink2012}. Abbreviations for the references are listed in Table~\ref{Tab:diffUMIST2006}.

The individual atomic ionization cross sections are used to derive the molecular dissociation rate from the individual ionization rates using the relative weight difference. If the difference in atomic weight between the components of the molecule is large, the cross section of the heavier one is used. If the difference is small, the cross sections are combined \citep{Aresu2011}. An example is the reaction CO\,$+$\,XPHOT which can lead to C$^{2+}\,+\,$O, C$^+\,+\,$O$^+$, or C$\,+\,$O$^{2+}$, so that the total cross section is the sum 
\begin{equation}
\sigma_{\rm CO}=\frac{1}{3}\sigma_{\rm C} + \frac{1}{6}(\sigma_{\rm C}+\sigma_{\rm O}) + \frac{1}{3}\sigma_{\rm O}
\end{equation}
where $\sigma_{El}$ is the ionization cross sections of the element, in this case C and O.

More detailed explanations/discussions of rates can be found in \citet{Adamkovics2011}. The di-electronic recombination rates are from tables of \citet{Landini1991}, \citet{Badnell2003}, \citet{Badnell2006}, \citet{Zatsarinny2003} and subsequent papers of this series. In case of X-rays, the high temperature di-electronic recombination rates get added to the radiative recombination rates taken from UMIST. The two elements Ne and Ar have only di-electronic recombination rates currently available. We include here also charge exchange reaction of Ar and Ne with water, O$_2$, CO, N$_2$, CH$_4$, NH$_3$, CO$_2$, and NO. In addition we consider charge exchange between He$^+$ and Ne.

\section{How $E_{\rm ads}$ affects line fluxes}

\begin{figure*}[!htbp]
\begin{center}
\includegraphics[width=16cm]{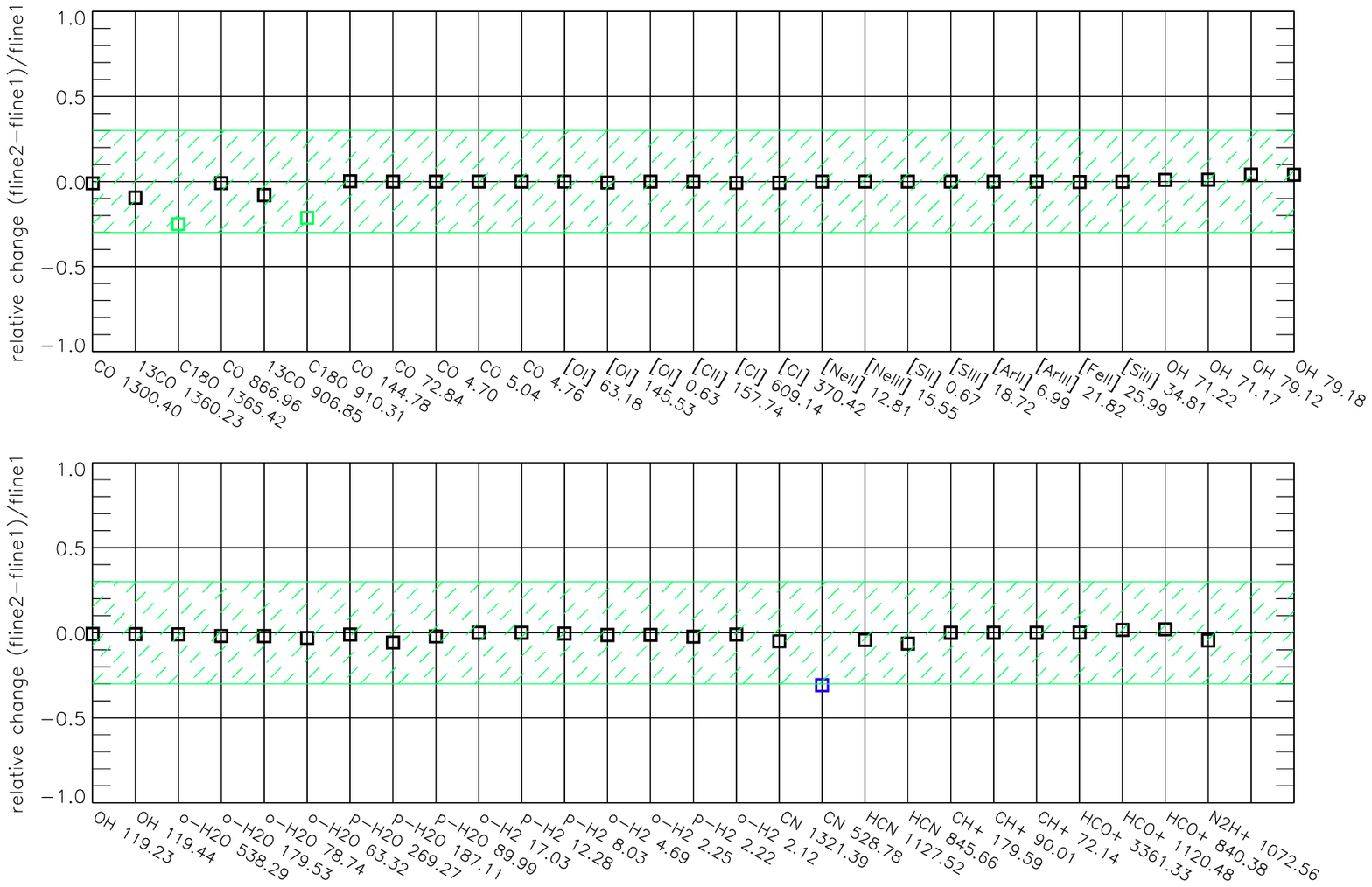}
\caption{Comparison of line fluxes for two sets of adsorption energies: Aikawa (fline1, model 1) and GH06 (fline2, model 5). Black and green squares denote differences of less than 25\% and less than a factor two respectively, blue squares and red triangles denote differences larger than a factor three and ten respectively.}
\label{fig:Eads-lines-1}
\end{center}
\end{figure*}

\begin{figure*}[!htbp]
\begin{center}
\includegraphics[width=16cm]{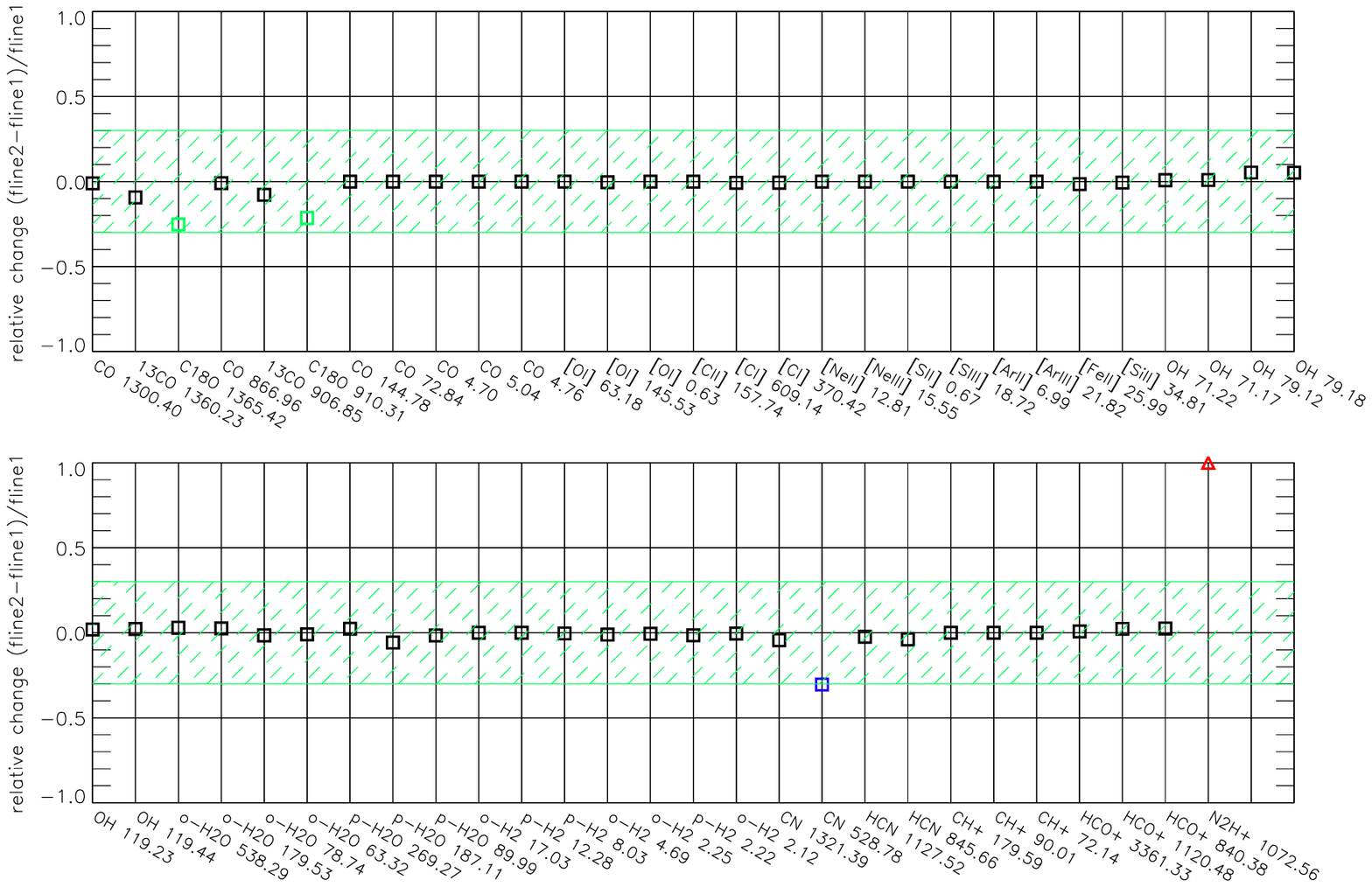}
\caption{Comparison of line fluxes for two sets of adsorption energies: Aikawa (fline1, model 1) and UMIST2012 (fline2, model 6). Black and green squares denote differences of less than 25\% and less than a factor two respectively, blue squares and red triangles denote differences larger than a factor three and ten respectively.}
\label{fig:Eads-lines-2}
\end{center}
\end{figure*}

\begin{figure*}[!htbp]
\begin{center}
\includegraphics[width=16cm]{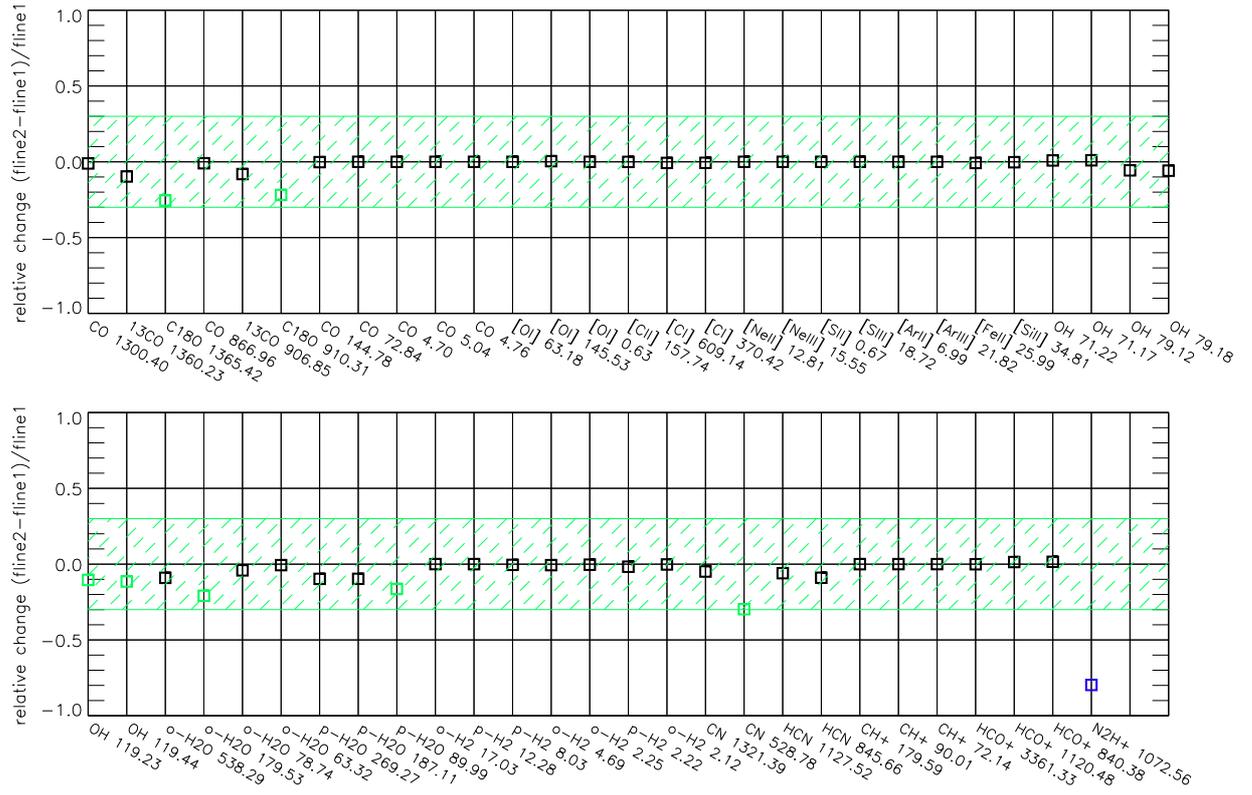}
\caption{Comparison of line fluxes for two sets of adsorption energies: Aikawa (fline1, model 1) and $T$-dependent adsorption energies (fline2, model 7). Black and green squares denote differences of less than 25\% and less than a factor two respectively, blue squares and red triangles denote differences larger than a factor three and ten respectively.}
\label{fig:Eads-lines-3}
\end{center}
\end{figure*}

We show in Figs.~\ref{fig:Eads-lines-1} to \ref{fig:Eads-lines-3} how models with different sets of adsorption energies (Aikawa, GH06, UMIST2012 and temperature-dependent $E_{\rm ads}$) affect our selection of line fluxes.

\bibliography{chemistry-reference,mri_bib}

\end{document}